\documentclass[aps,footinbib, notitlepage,superscriptaddress, longbibliography,eqsecnum]{revtex4-2}

\usepackage{silence}
\usepackage[english]{babel}
\usepackage[utf8]{inputenc}
\usepackage{amsmath}
\usepackage{amsthm}
\usepackage{braket}
\usepackage{amssymb}
\usepackage{graphicx}
\usepackage{hyperref}
\usepackage{dsfont}
\usepackage{xcolor}
\usepackage{geometry}
\usepackage{mathtools}
\usepackage{bbold}
\usepackage{upgreek}
\usepackage{bbm}
\usepackage{float}
\usepackage[capitalize]{cleveref}

\usepackage{tikz}

\newcommand{\panelimg}[3][]{%
  \begin{tikzpicture}
    \node[inner sep=0pt, anchor=south west] (img) at (0,0)
      {\includegraphics[#1]{#3}};
    \node[
      anchor=north west,
      fill=none,
      inner sep=1pt,
      font=\footnotesize
    ] at ([xshift=-6pt,yshift=-0pt]img.north west) {(#2)};
  \end{tikzpicture}%
}

\newcommand{\panelimgshift}[5][]{%
  \begin{tikzpicture}
    \node[inner sep=0pt, anchor=south west] (img) at (0,0)
      {\includegraphics[#1]{#4}};
    \node[
      anchor=north west,
      fill=none,
      inner sep=1pt,
      font=\footnotesize
    ] at ([xshift=#2,yshift=#3]img.north west) {(#5)};
  \end{tikzpicture}%
}

\DeclareFontFamily{OT1}{pzc}{}
\DeclareFontShape{OT1}{pzc}{m}{it}{<-> s * [1.10] pzcmi7t}{}
\DeclareMathAlphabet{\mathpzc}{OT1}{pzc}{m}{it}

\definecolor{deepchampagne}{rgb}{0.98, 0.84, 0.65}

\providecommand{\st}[1]{_{\text{#1}}}

\providecommand{\ut}[1]{^{\text{#1}}}
\def\LRA{\Leftrightarrow}

\def\bra{\ensuremath{\langle}}
\def\ket{\ensuremath{\rangle}}
\def\pd{\partial}
\def\tr{\mathrm{tr}}
\def\nullsp{\mathrm{null}}
\def\rowsp{\mathrm{row}}
\def\Imat{\mathbbm{I}}
\def\im{\mathrm{i}}

\def\av{\bv{a}}

\def\uv{\bv{u}}
\def\ev{\bv{e}}
\def\gv{\bv{g}}

\def\Fv{\bv{F}}
\def\fv{\bv{f}}
\def\mv{\bv{m}}

\def\yv{\bv{y}}
\def\zv{\bv{z}}

\def\vv{\bv{v}}

\def\sv{\bv{s}}

\def\xv{\bv{x}}
\def\wv{\bv{w}}

\def\b0{\bv{0}}

\def\Bcal{\mathcal{B}}
\def\Ecal{\mathcal{E}}

\def\Hcal{\mathcal{H}}

\def\Jcal{\mathcal{J}}

\def\Dcal{\mathcal{D}}
\def\Kcal{\mathcal{K}}
\def\Lcal{\mathcal{L}}
\def\Mcal{\mathcal{M}}
\def\Ncal{\mathcal{N}}

\def\Ocal{\mathcal{O}}

\def\Rcal{\mathcal{R}}
\def\Scal{\mathcal{S}}

\def\Vcal{\mathcal{V}}

\def\phiv{\boldsymbol{\upphi}}
\def\psiv{\boldsymbol{\uppsi}}

\def\rank{\mathrm{rank}}

\def\reals{\mathbb{R}}

\def\enc{\st{enc}}
\def\Herm{\mathrm{Herm}}
\def\E{\mathbb{E}}

\def\numM{{m}}		
\def\numMtot{{B}}		
\def\numF{{q}}		
\def\numTM{{L}}		

\newcommand{\beq}{\begin{equation}}
\newcommand{\eeq}{\end{equation}}
\newcommand{\beqn}{\begin{equation*}}
\newcommand{\eeqn}{\end{equation*}}
\newcommand{\bv}[1]{\mathbf{#1}}

\begin{document}

\title{Theory and interpretability of Quantum Extreme Learning Machines:\\ a Pauli transfer matrix approach}

\author{Markus Gross}
\email{markus.gross@dlr.de}
\author{Hans-Martin Rieser}

\affiliation{Institute for AI Safety and Security, German Aerospace Center (DLR),\\ Sankt Augustin and Ulm, Germany}

\date{\today}

\begin{abstract} 
Quantum reservoir computers (QRCs) have emerged as a promising approach to quantum machine learning, since they utilize the natural dynamics of quantum systems for data processing and are simple to train. 
Here, we consider $n$-qubit quantum extreme learning machines (QELMs) with initial-state encoding and continuous-time reservoir dynamics.
Applying the Pauli transfer matrix (PTM) formalism reveals the encoding-generated features and their linear transformation by quantum channels, and thereby provides a nonlinear vector autoregression model as an interpretable classical representation of a QELM.
We show that the nonlinear information processing capacity of a QELM factorizes into the ability of the encoding to represent a given target function and the decodability of the relevant Pauli features after the effective PTM. 
Comparing geometric and statistical decodability reveals features that are algebraically hidden yet recoverable through data correlations.
We relate the decodability of Pauli features to the spreading of measurement operators under Heisenberg evolution (Krylov growth).
Focusing on the unitary channel, we discuss how structured Hamiltonians, when paired with certain observables, can impair model performance, as reflected in a low rank of the readout and the effective PTMs.
We trace this effect to Hamiltonian symmetries and derive asymptotic rank estimates for symmetry-resolved observable families. 
Our decodability analysis disentangles the contributions of encoding, reservoir dynamics, and measurement observables, and thus provides interpretable diagnostics that can inform the design of QELMs and reduce reliance on exhaustive hyperparameter search.
As a specific application, we consider forecasting nonlinear dynamical systems and show that the readout of a trained QELM closely matches the expression of the underlying flow map at leading orders.
\end{abstract}

\maketitle

\section{Introduction}
QRCs have enjoyed significant attention since their introduction \cite{fujii_harnessing_2017}, featuring applications in various domains, such as image classification \cite{sakurai_quantum_2022,kornjaca_largescale_2024,sakurai_simple_2025,lorenzis_harnessing_2025}, time series forecasting \cite{fujii_harnessing_2017,chen_temporal_2020,suzuki_natural_2022,mujal_time_2023,ahmed_prediction_2024,ahmed_optimal_2025}, fluid dynamics \cite{pfeffer_hybrid_2022,pfeffer_reducedorder_2023}, or post-processing of quantum experiments \cite{suprano_experimental_2024,vetrano_state_2025,zia_quantum_2025,assil_entanglement_2025}.
A major advantage of QRCs is their suitability for noisy intermediate-scale quantum hardware, primarily because the training process is restricted to the classical readout layer, thereby rendering the training problem convex and avoiding the barren plateau problem associated with variational quantum algorithms \cite{mcclean_barren_2018,larocca_barren_2025}. 
QRCs are often resilient to, and can even benefit from, environmental noise \cite{domingo_taking_2023,fry_optimizing_2023,sannia_dissipation_2024,monzani_nonunital_2024}.
The nonlinear processing of input by QRCs has been analyzed in \cite{mujal_analytical_2021,nokkala_gaussian_2021,govia_nonlinear_2022,innocenti_potential_2023}, finding that the encoding is often the main supplier of nonlinearity. Universal function approximation properties have been discussed in \cite{chen_learning_2019,chen_temporal_2020,goto_universal_2021,martinez-pena_quantum_2023,monzani_universality_2024,gonon_universal_2025}.
Notably, QRCs are closely related to quantum kernel methods \cite{schuld_supervised_2021}---a fact that can be utilized to find optimal measurement operators \cite{gross_kernelbased_2026} and harness insights from the classical domain \cite{canatar_bandwidth_2022,gross_expressivity_2025,kempkes_double_2026}.
For general overviews on QRCs, we refer to \cite{fujii_quantum_2020,mujal_opportunities_2021}.

A specific QRC variant without any coherently evolving memory is the QELM \cite{wilson_quantum_2019,fujii_quantum_2020,mujal_opportunities_2021,sakurai_quantum_2022}. 
Its classical counterpart, the `extreme learning machine', is essentially a feedforward network with fixed internal architecture \cite{huang_extreme_2006}, similar to a random-feature model \cite{rahimi_random_2007}.
QELMs are capable of various ML tasks, such as image classification and time series forecasting.
A typical QELM differs from a conventional quantum neural network with fixed internal parameters in two important ways: first, instead of gate-based architectures, it employs the natural dynamics of a physical quantum system (a reservoir). Second, instead of optimizing a measurement output for a certain task, one usually collects multiple unspecific measurements into a large readout vector and feeds it into a classical regression or classification algorithm. 
The main task in optimizing QELMs thus consists in tuning the quantum properties of the reservoir such that information processing is maximized and relevant features can be properly decoded from the measurements \cite{martinez-pena_dynamical_2021,domingo_optimal_2022,hayashi_impact_2023,gotting_exploring_2023,innocenti_potential_2023,kobayashi_coherence_2024,cindrak_engineering_2025,schutte_expressivity_2025,martinez-pena_inputdependence_2025,lorenzis_entanglement_2025,vetrano_state_2025}.
Identifying possible quantum advantages, e.g., related to the exponentially large Hilbert space into which data is encoded, is an active research topic \cite{schuld_effect_2021,mujal_opportunities_2021,shin_exponential_2023,gotting_exploring_2023,kornjaca_largescale_2024,schutte_expressivity_2025}.
However, the exponential concentration effect emerging for large qubit numbers is a phenomenon conceptually related to barren plateaus and can strongly reduce QELM performance  \cite{xiong_fundamental_2025}. 

The present study aims at an interpretable framework that can support systematic design of QELMs and reduce the need for expensive training-based hyperparameter tuning. 
To this end, we will discuss temporal multiplexing \cite{fujii_harnessing_2017} as well as information processing capacity \cite{dambre_information_2012,martinez-pena_information_2023} in the light of the PTM formalism, based on the notion of feature \emph{decodability}.
This allows us to systematically address which parts of the model expressivity stem from the reservoir channel (such as the unitary) and which from the encoding. Moreover, our decodability framework can identify features algebraically hidden but still recovered via correlations in the data.
We analyze the role of Hamiltonian symmetries for the rank of the readout matrix, which is a crucial measure of expressivity in reservoir computing.
These symmetries partition the operator space under Krylov evolution, which allows one to identify suitable observable families for a given Hamiltonian.
Chaotic dynamical systems are used as benchmarks due to their beneficial properties for system identification \cite{shumaylov_when_2025}.
The present formalism applies to $n$-qubit QELMs in which the encoding is time-independent and separated from reservoir evolution. 
Our findings also hold implications for other QML models that fall into this category.
However, models with dynamical input injection \cite{mccaul_minimal_2025,vetrano_state_2025} or dissipatively driven reservoirs \cite{sannia_dissipation_2024} are not covered by the present formalism.

\section{Model}
\label{sec_model}
We briefly present the formalism of QELMs and the specific implementation used in this work \cite{fujii_harnessing_2017,fujii_quantum_2020,mujal_opportunities_2021}.
Let $\Herm(\Hcal)$ denote the space of $d\times d$ Hermitian matrices (real vector space of dimension $d^2$) on the $d$-dimensional Hilbert space $\Hcal$, where $d=2^n$ in terms of the number of qubits $n$.
Given classical inputs $\xv\in\reals^D$, a QELM is defined by the output function 
\beq f(\xv) = \sum_{k=1}^\numM w_k \tr[M_k \rho(\xv)] ,
\label{eq_QELM}\eeq 
where $\{M_k\}_{k=1}^\numM$ are a set of measurement operators in $\Herm(\Hcal)$, and $\rho(\xv)\in \Herm(\Hcal)$ is the density matrix of the quantum system, obtained after applying encoding and other quantum channels to an initial state, $\rho(\xv)=\Ecal(\cdots \Ecal\ut{enc}_\xv(\rho_0)\cdots)$.
The trainable weights $w_k$ are determined by minimizing the empirical risk:
 \beq \min_{\mathbf{w}\in\reals^\numM} \sum_{i=1}^P \Lcal \left( f(\xv_i),\, y_i \right) + \frac{\lambda}{2} \|\mathbf{w}\|^2_2 ,
 \label{eq_QELM_opt}\eeq
where $\Lcal$ is a loss function, $\lambda$ is a regularization parameter, and $\{(\xv_i,y_i)\}_{i=1}^P$ are $P$ training data points.
Typical loss functions are the least-squares loss $\Lcal(\hat y,y)=\frac{1}{2}(\hat y-y)^2$ for regression (including time-series prediction), and the hinge loss $\Lcal\st{hinge}(\hat y,y)=\max(0,1-y\hat y)$, with $y\in\{-1,+1\}$ for classification.
In the case of the least-squares loss, the optimal weights are given by
\beq \hat{\mathbf{w}} = (G G^\top + \lambda \Imat_\numM)^{-1} G \mathbf{y}, \qquad G_{ki} = \tr(M_k \rho(\xv_i)),
 \label{eq_ols_weights}\eeq
where $\mathbf{y}\in\reals^P$ is the sample vector and $G\in\reals^{\numM\times P}$ is the readout design matrix.

When optimizing over a complete basis of measurement operators $\{M_k\}_{k=1}^{d^2}$, the QELM takes the form of a quantum kernel model (see \cref{app_kernel} for a summary). In the present work, we do not use that formalism, but instead work with a preselected observable set and optimize only over the associated weights $w_k$.
Our notation is summarized in \cref{tab:notation}.

\subsection{Feature map}
\label{sec_qelm_feature_map}

We consider pure states, i.e., $\rho(\xv)=|\psi(\xv)\ket\bra \psi(\xv)|$, with the quantum state $|\psi(\xv)\ket \in \Hcal$ given by 
\beq |\psi(\xv)\ket = U_R(t) U\st{enc}(\xv) |\Phi_0\ket.
\label{eq_q_feature_map}\eeq 
Here, $U\st{enc}(\xv)$ is a data \emph{encoding} unitary, $U_R$ is a \emph{reservoir} unitary, and we take the \emph{initial state} as $|\Phi_0\ket=|0\ket^{\otimes n}$.
The unitary $U_R(t)=\exp(-\im t H_R)$ is defined by time evolution under a reservoir Hamiltonian $H_R$. 
Specifically, we consider random Hamiltonians $H_R$ as well as two variants of a transverse field Ising model (TFIM)
\begin{subequations}\label{eq_H_TFIM}
\begin{align}
H\st{TFIM}\ut{zz-x} &= - J \sum_{j=1}^{n-1} Z_j Z_{j+1} - h \sum_{j=1}^n X_j, \label{eq_H_TFIM_zzx}\\
H\st{TFIM}\ut{xx-z} &= - J \sum_{j=1}^{n-1} X_j X_{j+1} - h \sum_{j=1}^n Z_j, \label{eq_H_TFIM_xxz}
\end{align}
\end{subequations}
with couplings $J,h\in\reals$.
Here and in the following, $\{X_j, Y_j, Z_j\}$ denote single-qubit Pauli operators acting on the $j$-th qubit and $Z_j Z_{j+1}$ is a shorthand for the usual $n$-qubit tensor product (with identity operators acting on the remaining qubits).
The two TFIM variants in \cref{eq_H_TFIM} are related by a global Hadamard transform $U\st{H}=\mathrm{H}^{\otimes n}$: 
\beq U\st{H} X_j U\st{H}^\dagger = Z_j,\quad U\st{H} Y_j U\st{H}^\dagger = -Y_j,\quad U\st{H} Z_j U\st{H}^\dagger = X_j,\quad U\st{H} H\st{TFIM}\ut{zz-x} U\st{H}^\dagger = H\st{TFIM}\ut{xx-z}.
\label{eq_Hadamard_equiv}\eeq

\subsection{Encoding schemes}
\label{sec_encodings}

We consider time-independent (parallel) product state encodings with $D=n$, producing the state 
\beq \rho\st{enc}(\xv) = U\st{enc}(\xv)\rho_0 U\st{enc}(\xv)^\dagger = \bigotimes_{j=1}^n |\psi\st{enc}(x_j)\ket\bra \psi\st{enc}(x_j)|,\qquad U\st{enc}(\xv)=\bigotimes_{j=1}^n U_{\text{enc},j}(x_j),
\label{eq_encoding_Uenc}\eeq
where the unitaries $U_{\text{enc},j}$ act on the $j$-th qubit.
In our numerical experiments we employ amplitude and rotational encoding schemes \cite{schuld_machine_2021}. The former comes in two variants, where one either encodes data on half a great circle of the Bloch sphere:
\beq
|\psi\st{enc}(x)\ket = \sqrt{x}|0\ket + \sqrt{1-x}|1\ket,\qquad x\in[0,1],
\label{eq_enc_amplsqrt}\eeq
or on the full great circle:
\beq
|\psi\st{enc}(x)\ket = x|0\ket+\sqrt{1-x^{2}}|1\ket,\qquad x\in[-1,1].
\label{eq_enc_amplsq}\eeq
Rotational encoding is defined by the unitary 
\beq
U_{\text{enc},j}(x_j) =\exp(-\im \pi x_j  \sigma^{p_j}) =R_{p_j}(2\pi x_j),\qquad x_j \in [0, 1],
\label{eq_enc_rot}\eeq
where $\sigma^p\in \{\Imat, X, Y, Z\}$ are single-qubit Pauli matrices, $p\in \{0,x,y,z\}$, and $R_{p_j}$ denotes rotation gates. 
In order to avoid irrelevant phases or trivial states, we exclusively use $R_Y$ operators, which yield the single-qubit state
\beq
|\psi\st{enc}(x_j)\ket= R_Y(2\pi x_j)|0\ket=\cos(\pi x_j)|0\ket+\sin(\pi x_j)|1\ket .
\label{eq_state_enc_rot}
\eeq

For rotational encoding, the two TFIM-based QELMs are equivalent under $x_j\mapsto \tfrac14-x_j\pmod 1$ provided the observables are also Hadamard-conjugated, exchanging $Z/ZZ$ with $X/XX$ readout. This does not imply equivalence at fixed $Z$-type readout.


\section{Pauli transfer formulation of a QELM}

QML models, including QELMs, have often been analyzed from the Fourier perspective, which is natural for Hamiltonian-based encodings \cite{schuld_effect_2021,xiong_fundamental_2025,shin_exponential_2023}. We instead employ the Pauli transfer matrix (PTM) formalism \cite{caro_learning_2024,hantzko_fast_2025,angrisani_classically_2025,angrisani_simulating_2025,rudolph_pauli_2025}, which allows a more direct identification of the relevant features for models with initial-state encoding, and has been previously utilized to study QRCs from the state-space perspective \cite{fujii_quantum_2020,martinez-pena_quantum_2023,kobayashi_coherence_2024,martinez-pena_inputdependence_2025}.
We focus in the following on this classical feature-based representation of a QELM to analyze its readout expressivity and feature decodability.

\subsection{Pauli transfer matrix}
\label{sec_qelm_ptm}

In the PTM formalism, we expand states and operators in the $n$-qubit Pauli basis
\beq
\{P_k=\sigma^{a_1}\otimes\cdots\otimes\sigma^{a_n}:\ a_i\in\{0,x,y,z\}\}_{k=0}^{d^2-1},
\eeq
which consists of $d^2$ mutually orthogonal Pauli strings $P_k$, with $P_0=\Imat$, $d=2^n$, and normalization $\tr(P_k P_l) = d\delta_{kl}$.
Every state admits the expansion
\beq
\rho(\xv)=\frac{1}{2^n}\sum_{k=0}^{d^2-1} \phi_k(\xv) P_k,
\qquad
\phi_k(\xv)=\tr(P_k \rho(\xv)),
\label{eq_pauli_feature}\eeq
giving rise to a vector $\phiv(\xv)\in\mathbb{R}^{d^2}$ of \emph{Pauli features}, which are the fundamental input-dependent objects in our study.
They have a straightforward expression in the case of product-state encodings (see \cref{sec_feat_enc}).
Any quantum channel $\Ecal$ acts linearly on this vector, $\phiv' = T_{\Ecal}\, \phiv$, via the Pauli transfer matrix:
\beq (T_{\Ecal})_{kl}=\frac{1}{d}\,\tr \big( P_k\, \Ecal(P_l)\big).
\label{eq_PTM}\eeq

Consider the QELM in \cref{eq_QELM} with a subset $\Scal$ of measured Pauli operators $\{P_k\}_{k\in\Scal}$, $\numM \equiv |\Scal|\leq d^2$. 
This selection can be represented by a matrix $S\in\reals^{\numM\times d^2}$ (a subset of rows of the identity matrix) that extracts the corresponding rows from $T_{\Ecal}$.
The readout vector $\Fv(\xv)\in\mathbb{R}^{\numM}$ contains their expectation values $F_k(\xv)=\tr\bigl[P_k\, \Ecal(\rho\st{enc}(\xv))\bigr]$.
Expanding $\rho\st{enc}$ as in \cref{eq_pauli_feature} and using \cref{eq_PTM} gives the model output function
\beq
  f(\xv) = \wv^\top\,\Fv(\xv) = \wv^\top \,T_{\Ecal}^{(\numM)}\,\phiv(\xv) =  \sum_{k\in\Scal} w_k \sum_{j=0}^{d^2-1} (S T_\Ecal)_{kj} \phi_j(\xv),
\label{eq_qelm_ptm}\eeq
where $\wv\in\mathbb{R}^{\numM}$ are classical weights and $T_{\Ecal}^{(\numM)} \equiv S T_\Ecal$ is the $\numM\times d^2$ block of $T_{\Ecal}$ restricted to the measured observables.
This formulation explicitly shows that $f(\xv)$ is a linear function of the features $\phiv(\xv)$ generated solely by the encoding and mixed by the reservoir channel $T_{\Ecal}$.

If the quantum channel $\Ecal$ is completely positive and trace-preserving (CPTP), it admits a Kraus representation 
\beq \Ecal(\rho)=\sum_{\ell=1}^\mu K_\ell \rho K_\ell^\dagger
\label{eq_Kraus_rep}\eeq 
with $\sum_{\ell=1}^\mu K_\ell^\dagger K_\ell = \Imat$ and $\mu\leq d^2$ (dependent on the channel, e.g., $\mu=1$ for a unitary channel, see \cref{app_ptm} for further discussion).
With this representation, we can write
\beq (T_{\Ecal}^{(\numM)})_{kj} = \frac{1}{d}\sum_{\ell=1}^\mu \tr\bigl(P_k\, K_\ell P_j K_\ell^\dagger\bigr) = \frac{1}{d} \tr\bigl(P_k\ut{(H)} P_j\bigr),
\label{eq_Kraus_rep_T}\eeq
where the last expression introduces the generalized Heisenberg representation of an observable $O$, 
\beq O\ut{(H)} = \sum_{\ell=1}^\mu K_\ell^\dagger O K_\ell.
\label{eq_heisenberg_rep}\eeq
Specifically, for the unitary channel, we have $K=U(t)$ ($\mu=1$), such that $T_{\Ecal}$ becomes an orthogonal matrix $V(t)\in\mathrm{SO}(d^2)$ with elements 
\beq V_{kj}(t)=\frac{1}{d}\tr\bigl[P_k\, U(t) P_j U(t)^{\dagger}\bigr],
\label{eq_PTM_uni}\eeq 
and $V_{kj}(0) = \delta_{kj}$.
For general CPTP maps, $T_{\Ecal}$ can also induce contraction and translation of the feature vector (see \cref{sec:general-noise}).

\paragraph*{Temporal multiplexing.} In temporal multiplexing \cite{fujii_harnessing_2017,cindrak_enhancing_2024,steinegger_predicting_2025}, instead of working at a fixed evolution time $t$ and a potentially large set of observables $P_k$, one selects a small subset $\Scal$ of size $\numM$ [see \cref{eq_qelm_ptm}] and evaluates them at several times $t_1,  \ldots, t_\numTM$. 
One then stacks the corresponding rows of $V(t)$ into a $\numM \numTM\times d^2$ observability matrix 
\beq R_\numTM = \begin{pmatrix}
  S V(t_1) \\
  S V(t_2) \\
  \vdots\\
  S V(t_\numTM)
\end{pmatrix}.
\label{eq_tmux}\eeq 
We will simply call this the \emph{effective} PTM, denoted by $R$.
In the case without temporal multiplexing ($\numTM=1$), the full unitary PTM $R=V(t)$ is recovered for $S=\Imat_{d^2}$.

\subsection{Feature encoding}
\label{sec_feat_enc}

The structure of the Pauli feature vector $\phiv(\xv)$ can be made explicit for product-state encodings. Using \cref{eq_encoding_Uenc} and an initial product state, the encoded state before the reservoir unitary is
\beq
\rho\st{enc}(\xv)=U\st{enc}(\xv)\,|0\ket\bra 0|^{\otimes n}\,U\st{enc}(\xv)^\dagger=\bigotimes_{j=1}^n \rho_j(x_j),\qquad \rho_j(x_j)=U_{\text{enc},j}(x_j)|0\ket\bra 0|U_{\text{enc},j}(x_j)^\dagger.
\eeq
It is convenient to introduce the single-qubit Pauli features
\beq
\phiv^{(j)}(x_j)=\bigl(1,\phi_x^{(j)}(x_j),\phi_y^{(j)}(x_j),\phi_z^{(j)}(x_j)\bigr)^\top,\qquad \phi_a^{(j)}(x_j)=\tr\bigl[\sigma^a\,\rho_j(x_j)\bigr],
\label{eq_single_qubit_phi}
\eeq
with $\phi_0^{(j)} = \tr(\rho_j) = 1$ (normalization), $\sigma^0\equiv\Imat$ and $\sigma^{x,y,z}\equiv X,Y,Z$ as before.
The three nontrivial components of $\phiv^{(j)}$ define the Bloch vector, which has unit magnitude for pure states (see \cref{app_ptm} for further discussion).
Since both the state and the Pauli basis factorize across qubits, the $n$-qubit feature vector is  
\beq
\phiv(\xv) = \bigotimes_{j=1}^n \phiv^{(j)}(x_j),
\label{eq_phi_tensor}\eeq
and its components are given by
\beq
\phi_{a_1\ldots a_n}(\xv)= \tr\!\Bigl[\Bigl(\bigotimes_{j=1}^n \sigma^{a_j}\Bigr)\,\rho\st{enc}(\xv)\Bigr] =\prod_{j=1}^n \phi_{a_j}^{(j)}(x_j),\qquad a_j\in\{0,x,y,z\}.
\label{eq_phi_nqubit}\eeq
The subscript $a_1 \ldots a_n$ corresponds to the explicit site representation of the index $k$ in \cref{eq_pauli_feature}.
The $\phi_k$ thus run over \emph{all possible combinations} of the single-qubit features in \cref{eq_single_qubit_phi}: for each qubit one chooses one of the four factors $\{1,\phi^{(j)}_x,\phi^{(j)}_y,\phi^{(j)}_z\}$, yielding a total of at most $d^2=4^n$ features.
The readout features resulting from some simple encoding schemes are summarized in \cref{tab:enc_proj}. Effects of feature mixing by a unitary channel are discussed in the next section.

\begin{table}[tb!]
    \centering
    \renewcommand\arraystretch{1.3}
    \begin{tabular}{c|c|c|c}
        \hline
         & \multicolumn{3}{c}{\textbf{Feature} $\phi_{a_k}^{(k)}(u_k)$} \\
        \textbf{Factor in observable $P$} & \textbf{Amplitude on $[0,1]$} & \textbf{Amplitude on $[-1,1]$} & \textbf{Rotational} \\
        \hline
        $\cdots\otimes X_k\otimes\cdots$ & $2\sqrt{(1-u_k)u_k}$ & $2u_k\sqrt{1-u_k^2}$ & $\sin(2\pi u_k)$ \\
        $\cdots\otimes Y_k\otimes\cdots$ & $0$ & $0$ & $0$ \\
        $\cdots\otimes Z_k\otimes\cdots$ & $2u_k-1$ & $2u_k^2-1$ & $\cos(2\pi u_k)$ \\
        $\cdots\otimes\tfrac12(\Imat+Z_k)\otimes\cdots$ & $u_k$ & $u_k^2$ & $\cos^2(\pi u_k)$ \\
        \hline
    \end{tabular}
    \caption{Pauli-basis projections $\bra \psi\st{enc}|P|\psi\st{enc}\ket$ of an encoded product state $\rho\st{enc}(\uv)$ for amplitude encoding \eqref{eq_enc_amplsqrt} or \eqref{eq_enc_amplsq} as well as rotational encoding \eqref{eq_enc_rot} with $R_Y$-gates. Here $P=\bigotimes_{k=1}^{n}\sigma^{a_k}$ denotes a Pauli operator ($\sigma^{a_k}\in\{\Imat,X,Y,Z\}$) and the classical datum entering qubit $k$ is $u_k$. For convenience, we have included the projection operator $\frac{1}{2}(\Imat+Z_k)$. One observes that each Pauli factor $\sigma^{a_k}$ (1st column) in an observable $P$ produces the corresponding single-qubit feature $\phi_{a_k}^{(k)}(u_k):=\tr\!\big(\sigma^{a_k}\,\rho_k(u_k)\big)$ (2nd--4th columns). For instance, the $X_k$ row gives $\phi_x^{(k)}(u_k)$ and the $Z_k$ row gives $\phi_z^{(k)}(u_k)$. }
    \label{tab:enc_proj}
\end{table}

\subsection{Classical representation of a QELM}
\label{sec_classical}

Let $\numF\leq d^2$ be the effective number of features (excluding features identically vanishing due to the encoding), and $\numMtot$ be the total number of measurements, e.g. $\numMtot=\numM \numTM$.
According to \cref{eq_qelm_ptm}, a QELM has an equivalent classical representation as a PTM-weighted nonlinear function library,
\beq f(\uv) = \wv^\top \Fv(\uv),\qquad  \Fv(\uv) \equiv R \phiv(\uv),
\label{eq_qelm_classical}\eeq
where $R \in \mathbb{R}^{\numMtot\times \numF}$ is the effective PTM \eqref{eq_tmux} (including any row-selection and temporal multiplexing) and $F_k = \sum_{j=1}^q R_{kj}\phi_j$ are the Pauli features available at the readout. 

If $R$ has full column rank, it is left-invertible \footnote{This allows for the existence of a pseudoinverse such that $R^+ R = \Imat_\numF$.} and thus the regressor can always find weights $\wv\propto (R^+)^\top \ev_r$ that reconstruct any desired element $\phi_r$.
Since one usually trains with arbitrary targets $\yv$ instead of $\phi_r$, the actual weights [\cref{eq_ols_weights}] contain projections from the Pauli feature space to the data space:  
\beq \hat{\wv} = (G G^\top +\lambda \Imat_\numMtot)^{-1} G \yv = R (C R^\top R + \lambda \Imat_\numF)^{-1} \Phi \yv, 
\label{eq_ols_weights_red}\eeq
where we introduced the feature design and Gram matrices
\beq \Phi \equiv [\phiv(\uv_1)\cdots \phiv(\uv_P)]\in \reals^{\numF\times P},\quad C = \Phi \Phi^\top\in \reals^{\numF\times \numF},
\label{eq_gram_matrix}\eeq
and the corresponding readout design and Gram matrices
\beq G\equiv [\Fv(\uv_1)\cdots \Fv(\uv_P)] = R\Phi \in \reals^{\numMtot\times P},\quad  G G^\top = R C R^\top \in \reals^{\numMtot\times \numMtot}.
\label{eq_gram_matrix_readout}\eeq
The latter are typically computed in a QELM implementation, whereas $R$ and $\Phi$ are theoretical objects.
We will analyze in the following subsections the requirements on the encoding and reservoir unitary to improve trainability and avoid expensive hyperparameter search.

The last equality in \eqref{eq_ols_weights_red} (which follows from the ``push-through'' identity) implies that for the full unitary PTM $R=V$ [\cref{eq_PTM_uni}], the weights are simply rotated versions of the ones without PTM mixing:
\beq \hat{\wv} = R(C+\lambda\Imat_\numF)^{-1} \Phi \yv .
\label{eq_ols_weights_unitary}\eeq
Moreover, since $V$ is orthogonal, it drops out from the predictor \eqref{eq_qelm_classical}, consistent with the absence of a reservoir unitary in the kernel-representation \eqref{eq_QELM_kernel}.
Such a QELM is then simply a regressor acting on the full $d^2$-dimensional feature library $\phiv$. 
In the context of time series processing (especially when using some inputs $u_j$ to encode delay states) this reduces to a ``next-generation'' reservoir computer (NG-RC) \cite{gauthier_next_2021} or, equivalently, a nonlinear vector autoregressive process (NVAR).
When $R$ merely has full column rank, it can be interpreted to induce a regressor operating on the mixed feature set $\phiv' = R\phiv$.

A coarse measure for expressivity is the rank of the readout Gram matrix in \cref{eq_ols_weights_red}, which can be bounded according to
\beq \rank(G)\leq \min(\rank(R),\rank(\Phi)) \leq \min(\numMtot,P,\numF).
\label{eq_gram_rank}\eeq
More precisely, $\rank(G)=\rank(R)-\dim\!\left(\rowsp(R)\cap\nullsp(\Phi^\top)\right)$.
Hence $\rank(G)=\rank(R)$ whenever the sampled feature matrix is non-degenerate on the row space of $R$. This condition is satisfied, for example, if $\Phi$ has full row rank, and is expected generically once $\rank(\Phi)\ge \rank(R)$ and the data are sufficiently diverse. Accidental data degeneracies or numerical ill-conditioning, since the features $\phi_k$ are in general not orthogonal, can reduce the actual rank. Accordingly, in the data-rich generic regime,
\beq
\rank(G)\simeq \rank(R).
\label{eq_gram_ptm_rank}\eeq 
For random Hamiltonians, $\rank(R)$ typically grows proportionally to the measurement budget $\numMtot$ and then saturates near $\numF$ (see \cref{fig:feat_iso} below). 
By contrast, structured Hamiltonians like the TFIM are not fully scrambling, resulting in $\rank(R)\ll \numF$.
The achieved ranks indicate the quality of the measurement strategy (see \cref{sec_Pauli_decoding}) and thus of the resulting training accuracy (see also \cref{fig:tmux_training} below).

\section{PTM analysis for a unitary quantum channel}
\label{sec_PTM_analysis}

For separable (product) quantum channels, the PTM factorizes as $T=\bigotimes_{j=1}^n T_j$ into single-qubit PTMs $T_j$. The action of a local unitary is discussed in \cref{app_PTM_unitary}.
To understand and optimize a QELM with general nonlocal quantum channels, we must analyze the full PTM [\cref{eq_qelm_ptm,eq_Kraus_rep_T}]. We focus in the following on the unitary case.
The optimization can be split into several subtasks that define the fundamental reconstruction problem in (quantum) reservoir computing with explicit feature maps \cite{innocenti_potential_2023,martinez-pena_inputdependence_2025,cindrak_engineering_2025,vetrano_state_2025}:
\begin{enumerate}
  \item Decide which types of features $\{\phi_r\}$ are best suited to learn the dataset. For example, non-Markovian time series (including scalar harmonic signals) require delay coordinates \cite{so_linear_2005,vaseghi_advanced_2008}, whereas Markovian dynamical systems can already be modeled by instantaneous monomial features \cite{gross_flow_2026}. While commonly used encodings support universal function approximation \cite{chen_learning_2019,chen_temporal_2020,goto_universal_2021,martinez-pena_quantum_2023,monzani_universality_2024,gonon_universal_2025}, they can differ in their approximation errors, generalization behavior \cite{caro_encodingdependent_2021,schuld_effect_2021,peters_generalization_2022}, or robustness during autoregressive rollout. As shown in \cite{gross_flow_2026}, the training error shows strongest reduction with time resolution when the feature map matches the low-order Lie coefficients of the flow map. This implies that some feature maps are more suitable for learning time series originating from physical models. 
  \item Select suitable measurement observables $\{P_k\}$. Ideally, the Heisenberg-evolved observables $P_k\ut{(H)}(t)$ have strong overlap with the observable $P_r$ generating the desired feature $\phi_r$ [see \cref{eq_Kraus_rep_T}]. One possibility to achieve this is to determine an optimal measurement operator $M^*$ via primal or dual optimization, which can then be further expanded into a set of rotated Pauli-Z operators \cite{gross_kernelbased_2026}. Another possibility is temporal multiplexing \cite{appeltant_information_2011,fujii_harnessing_2017,cindrak_enhancing_2024,cindrak_krylov_2025,steinegger_predicting_2025}, discussed further in the PTM framework below (see \cref{sec_op_spread}).
  \item When the available measurement operations are restricted, one must tune the quantum channel $\Ecal$ and the readout layer such that the desired features $\phi_r$ become \emph{decodable}. Strict (geometric) decodability means that, for each such $\phi_r$, there exists a weight vector $\wv_r$ such that $\wv_r^\top T_\Ecal \phiv \propto \phi_r$ (see \cref{sec_Pauli_decoding}). In practice, decodability in the statistical sense (as a correlation) and within a certain error tolerance is often sufficient, since the features needed for typical training tasks can usually be approximated on limited domains from the available Pauli features (see \cref{sec_input_decod}). We will show that temporal multiplexing directly enhances decodability by increasing the operator support. 
\end{enumerate}

\subsection{Krylov growth and Pauli weight spreading}
\label{sec_op_spread}

Temporal multiplexing enables the regressor to ``reverse'' the information scrambling caused by quantum evolution. The latter is captured by the Krylov growth of an operator under conjugation \cite{parker_universal_2019,cindrak_engineering_2025,cindrak_krylov_2025}, which we summarize here in our setting.
Specifically, the unitary Heisenberg-evolution [\cref{eq_heisenberg_rep}] of a given Pauli observable $P_k$ satisfies
\beq
P_k\ut{(H)}(t) \equiv U(t)^\dag P_k U(t)
= e^{\im Ht} P_k e^{-\im Ht}
= e^{t\Lcal}(P_k)
= \sum_{\ell=0}^\infty \frac{t^\ell}{\ell!}\,\Lcal^\ell(P_k),
\eeq
where $\Lcal^\ell$ denotes $\ell$ repeated applications of the Liouvillian $\Lcal(O)\equiv \im[H,O]$. The time-evolved observable therefore generates the (increasing) Krylov subspace
\beq
\Kcal_\ell(P_k)=\mathrm{span}\bigl\{P_k,\, \Lcal(P_k),\, \ldots,\, \Lcal^{\ell-1}(P_k)\bigr\}.
\eeq
While there is no strict one-to-one identification of the integer Krylov order $\ell$ with the continuous time $t$, $\ell$ can be understood as the depth of a truncation of the nested-commutator series. Due to the suppression by the factorial, a rough estimate is $\ell\sim t\,\|H\|$ in case of local Hamiltonians.
It is useful to note that for the TFIM models in \cref{eq_H_TFIM}, the Hadamard equivalence \eqref{eq_Hadamard_equiv} implies that 
\beq (Z_{j_1}\cdots Z_{j_k})\ut{(H)}(t)\big|\st{zz-x} = U\st{H}^\dagger (X_{j_1}\cdots X_{j_k})\ut{(H)}(t)\big|\st{xx-z} U\st{H},
\eeq 
i.e., Pauli-$Z$ observables evolve under $H\st{TFIM}\ut{zz-x}$ equivalently to Pauli-$X$ observables under $H\st{TFIM}\ut{xx-z}$ up to a global unitary. 

\begin{figure}[tb!]
  \centering
  \panelimgshift[width=0.26\textwidth]{-12pt}{0pt}{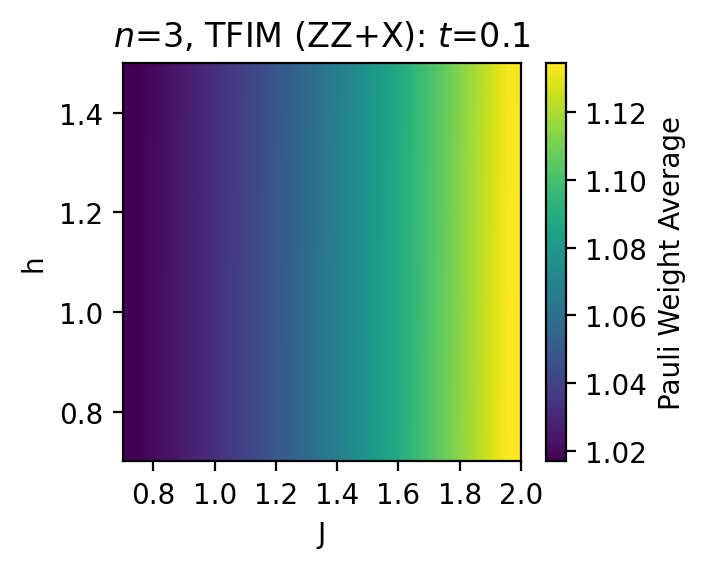}{a}\qquad
  \panelimg[width=0.26\textwidth]{b}{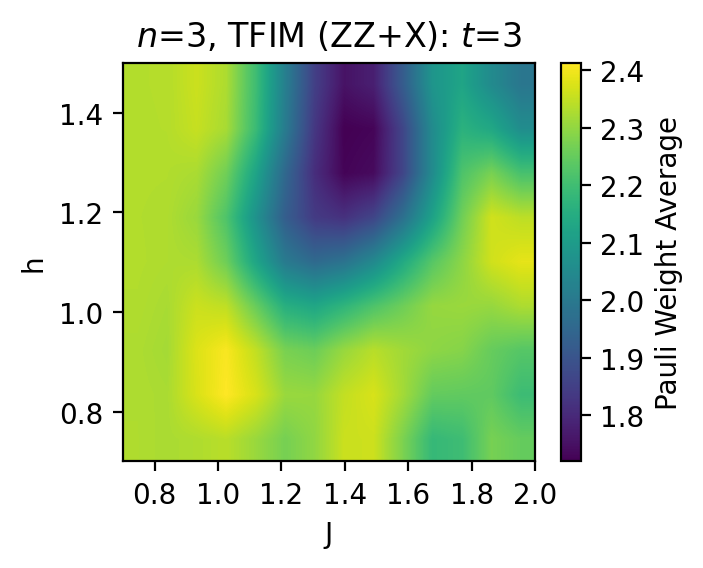}\qquad
  \panelimg[width=0.26\textwidth]{c}{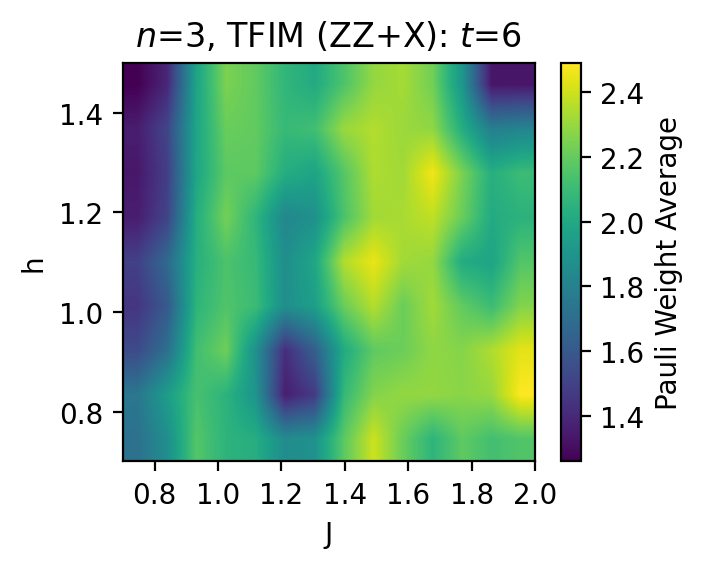}
   \caption{Time evolution of the Pauli weight average $\bar\nu_k$ [\cref{eq_pauli_wt_avg}], averaged over all single-site Paulis $P_k\in\{X_j,Y_j,Z_j\}_{j=1}^n$, in the $h$--$J$ coupling parameter space of the TFIM \eqref{eq_H_TFIM_zzx} for $n=3$ qubits.}
  \label{fig:qrc_pauliwt}
\end{figure}

In the PTM approach, Krylov growth can be represented in terms of \emph{operator spreading} in the Pauli basis \cite{nahum_operator_2018}. 
Accordingly, expanding the time-evolved observable $P_k\ut{(H)}(t)$ gives
\beq P_k\ut{(H)}(t)=\sum_{j=0}^{d^2-1} V_{kj}(t) P_j,
\label{eq_pauli_op_spread}\eeq
where we used cyclicity of the trace to introduce $V(t)$ from \cref{eq_PTM_uni}.
Thus, for non-Clifford $U(t)$, an initial weight fully localized on $P_k$ can spread to other (higher and lower order) Pauli strings as $t$ increases. A convenient diagnostic is the Pauli (Hamming) weight $\nu(P_j)\in\{0,\ldots,n\}$, i.e., the number of non-identity factors in $P_j$, and its coefficient-weighted average
\beq
\bar\nu_k(t)=\sum_{j=0}^{d^2-1} \nu(P_j)\, V_{kj}(t)^2.
\label{eq_pauli_wt_avg}\eeq
For local (including nonintegrable/chaotic ones) Hamiltonians, the Lieb--Robinson bound \cite{lieb_finite_1972,nachtergaele_propagation_2006,anthonychen_speed_2023,mahoney_liebrobinson_2024,lorenzis_entanglement_2025} heuristically implies that an initially local operator can only develop support within distance $\lesssim v_{\mathrm{LR}} t$ until finite-size saturation: 
\beq
\bar\nu_k(t) \sim \begin{cases}
1 + \mathrm{const}\cdot v_{\mathrm{LR}} t & \text{(intermediate times)} \\
\mathcal{O}(n) & \text{(late times)}
\end{cases}
\label{eq_pauli_wt_growth}\eeq
with a typical late-time value $\bar\nu_k\approx \tfrac{3}{4}n$, corresponding to a uniform distribution of coefficients $V_{kj}^2$ over Pauli strings.
The late-time limit is to be understood here and in the following in the sense of typical coverage, since finite-dimensional unitary evolution is generally quasiperiodic.
From the definition \eqref{eq_heisenberg_rep} and the expansion \eqref{eq_pauli_op_spread}, it straightforwardly follows that $\|P_k\ut{(H)}\|^2=d=d \sum_{j=0}^{d^2-1} V_{kj}^2$, and thus $\sum_{j=0}^{d^2-1} V_{kj}^2=1$ for all $t$.
Since, for a Haar-random unitary, a non-identity Pauli operator is distributed approximately uniformly over the traceless Pauli subspace of dimension $d^2-1$, one expects a typical magnitude
\beq \E_U[V_{kj}^2] = \frac{1}{d^2-1}\sim 4^{-n}\qquad (j,k\neq 0).
\label{eq_pauli_spread_latetime}\eeq
This is the origin of the exponential concentration effect, which makes accessing information about the input infeasible for large qubit numbers $n\to\infty$ \cite{xiong_fundamental_2025,sannia_exponential_2025}.
Integrable models (such as the integrable TFIM used here) typically show more structured behavior due to stable quasiparticle modes.
An extreme case is a Clifford unitary, for which $V_{kj}$ is a signed permutation matrix (no spreading).
By contrast, a Haar-random unitary on the full Hilbert space can generate rapid delocalization \cite{gopalakrishnan_hydrodynamics_2018,avdoshkin_euclidean_2020,noh_operator_2021,kaneko_dynamics_2023,sahu_quantifying_2023}.
The evolution of the Pauli weight average is illustrated in \cref{fig:qrc_pauliwt}.
As shown in \cref{sec_tfim_heisenberg_sectors}, the interplay of Hamiltonian structure and observables can have significant impact on the Heisenberg evolution and thus on the resulting Krylov spans.

\begin{figure}[tb!]
  \centering
  \panelimg[width=0.48\textwidth]{a}{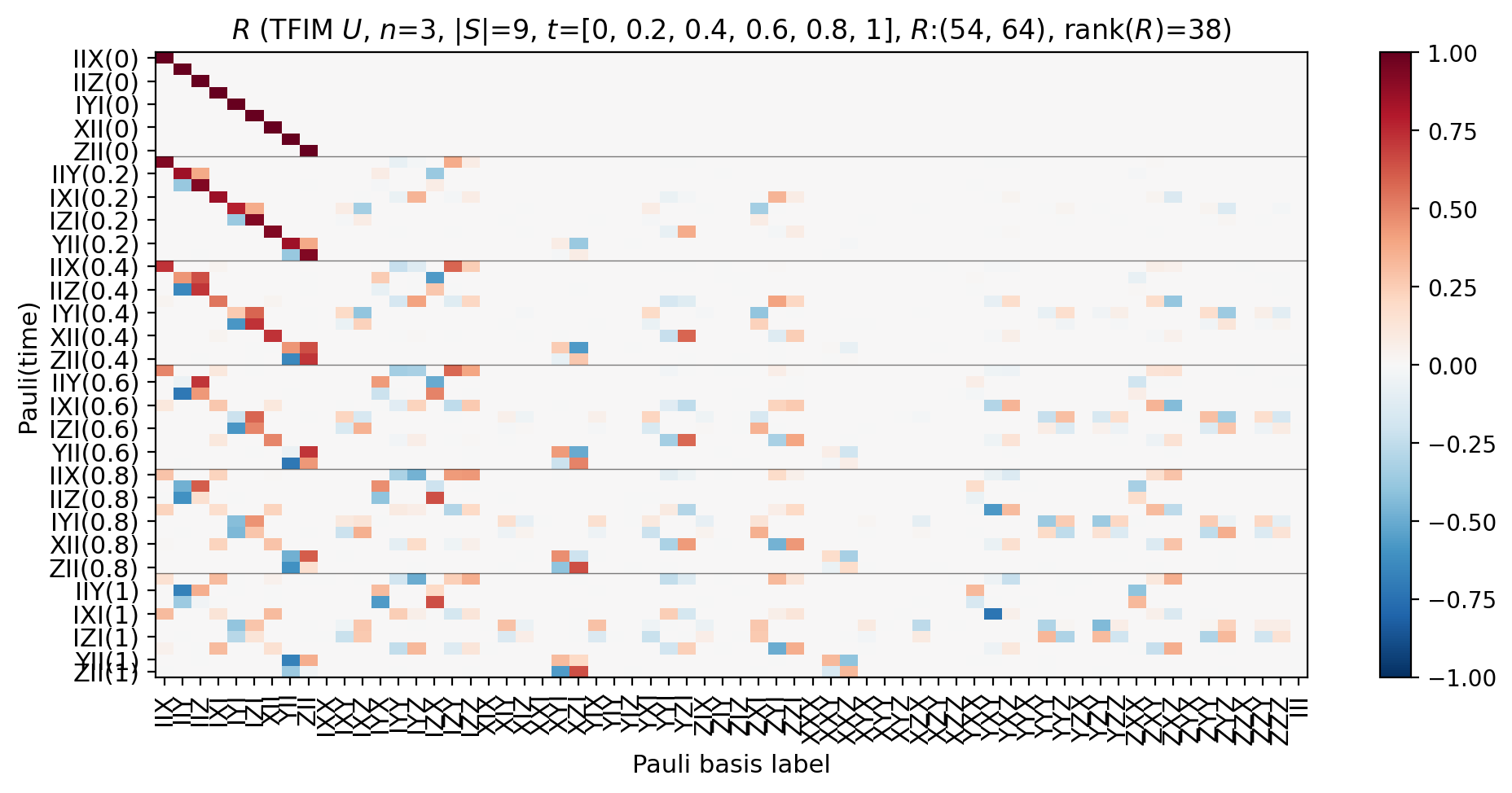}
  \panelimg[width=0.48\textwidth]{b}{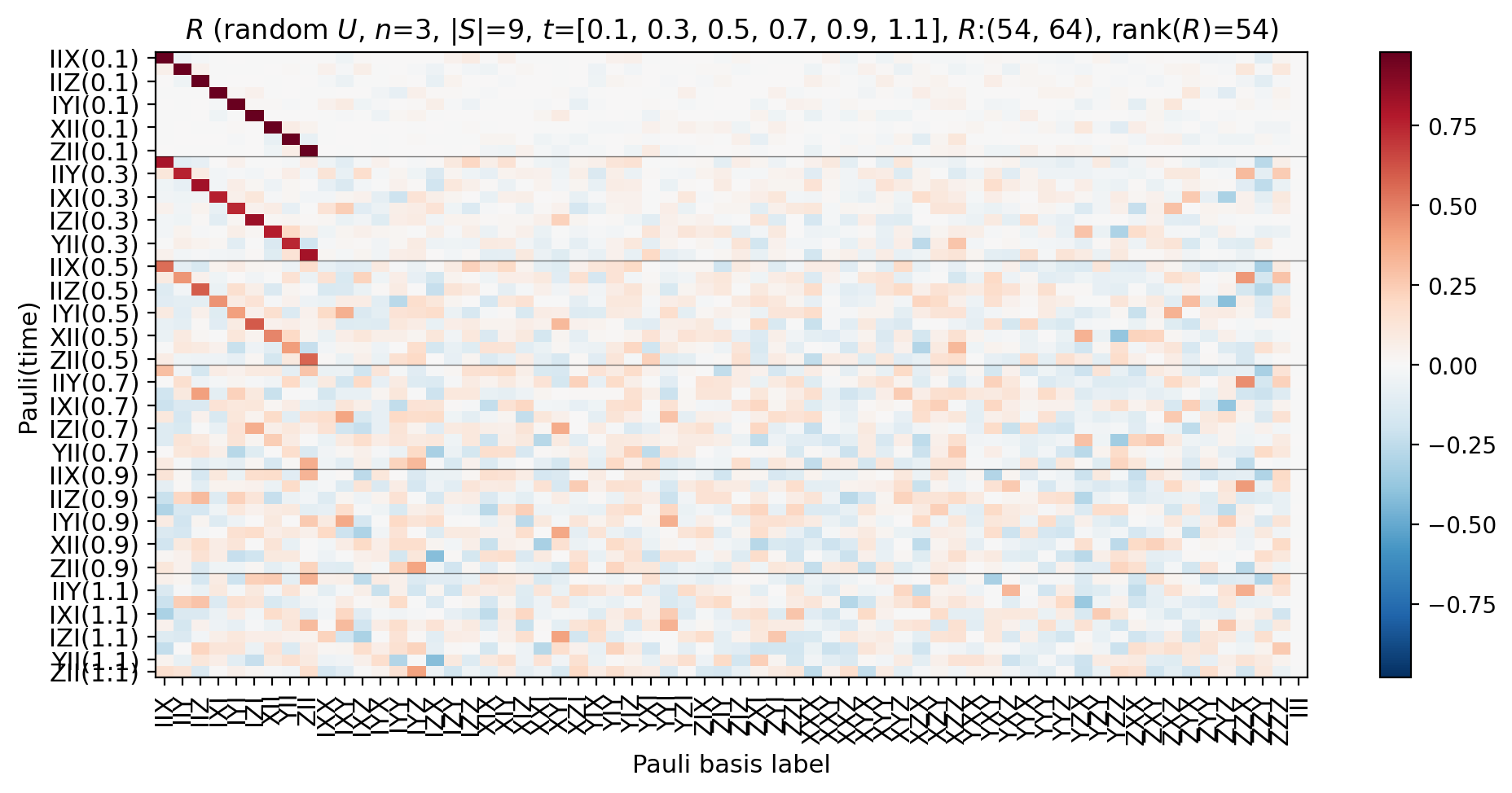}
  \caption{Visualization of the operator spreading \eqref{eq_pauli_op_spread} via the effective PTM $R$ [\cref{eq_tmux}] constructed in the temporal multiplexing scheme for the TFIM (a) and random unitary (b). The selected observable set $\Scal$ are all weight-1 Pauli operators of an $n=3$ qubit system. With increasing evolution time (from top to bottom), these initial operators develop nonzero projection onto the full $d^2$-dimensional Pauli basis [\cref{eq_pauli_op_spread}]. In the TFIM case, the supporting basis exhibits a sparse pattern, whereas it is dense for a random unitary.}
  \label{fig:tmux_eff_ptm}
\end{figure}

\subsection{Pauli feature decoding}
\label{sec_Pauli_decoding}

\begin{figure}[tb!]
  \centering
  \panelimg[width=0.31\textwidth]{a}{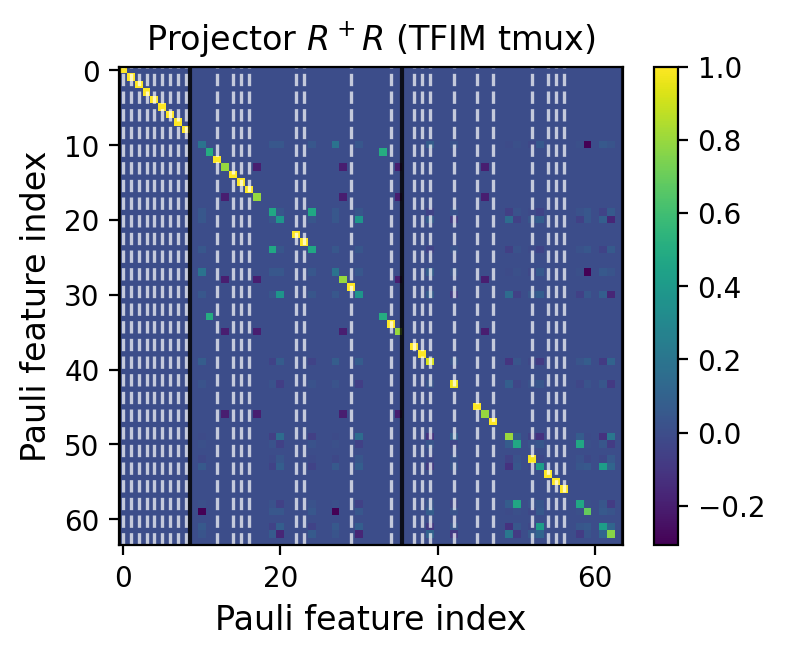}
  \panelimg[width=0.31\textwidth]{b}{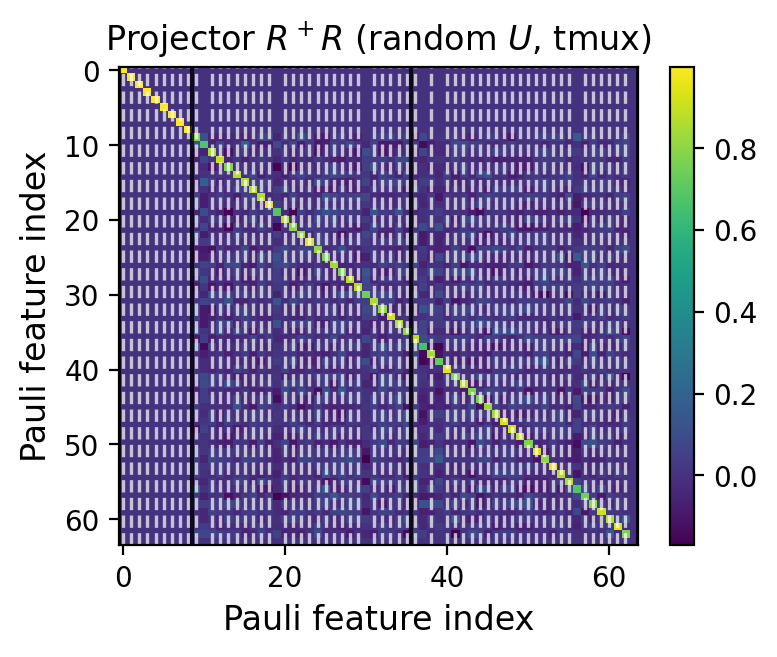}
  \panelimg[width=0.31\textwidth]{c}{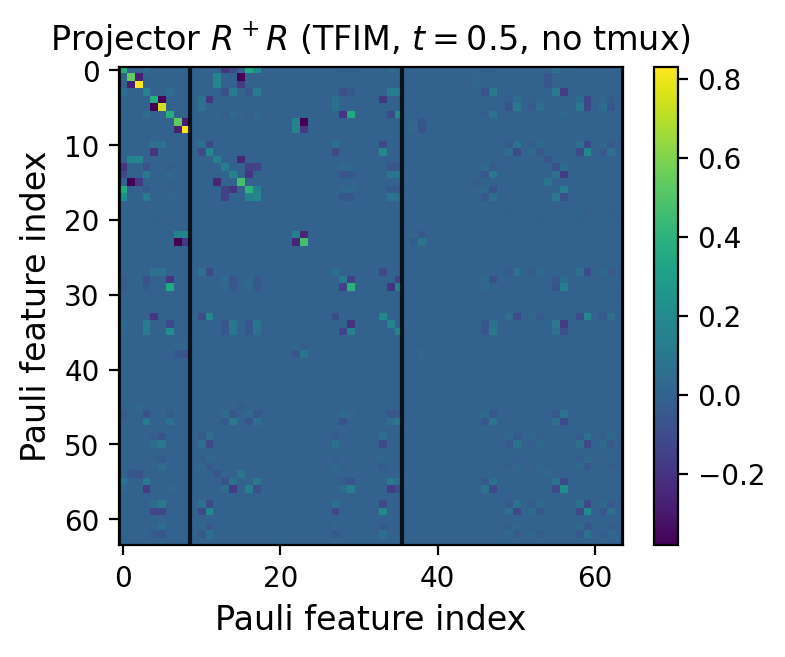}
  \caption{Projector $R^+R$ of the effective PTM [\cref{eq_tmux}] for different unitary types and a complete weight-1 Pauli observable set. Panels (a,b) directly correspond to the $R$ shown in \cref{fig:tmux_eff_ptm}, which are constructed by temporal multiplexing. Panel (c) shows the projector based on $R=SV(t=0.5)$ (no temporal multiplexing). The dashed white lines indicate features for which $\gamma_j^2>0.5$, while the black lines indicate Pauli weight sector boundaries. Due to operator spreading over a time $t=0.5$, decodability in (c) has significantly decreased.}
  \label{fig:proj_heatmap}
\end{figure}

Following \cref{eq_qelm_classical}, we denote the set of Pauli features available at the readout by $F_k = \sum_j R_{kj}\phi_j$ ($k=1,\ldots,\numMtot$), which emerge by linear transformations of the input features $\phi_j$ ($j=1,\ldots,\numF\leq d^2$) via the effective PTM $R\in \reals^{\numMtot\times \numF}$ [\cref{eq_tmux}], possibly including temporal multiplexing.
We now ask for conditions that enable the regressor to reconstruct a given input feature $\phi_r(\xv)$ by finding weights $\wv$ such that 
\beq \wv\cdot \Fv \simeq \phi_r \quad \LRA\quad  \wv^\top R \simeq \ev_r^\top,
\label{eq_feat_decod}\eeq 
where $\ev_r$ is the $r$-th basis vector of $\reals^{\numF}$.
The least-squares solution to this problem is discussed in \cref{app_ols_decoding}, showing that one can distinguish between \emph{geometric} and \emph{statistical} decodability: 
the former indicates whether an exact algebraic solution to \eqref{eq_feat_decod} can be found; in the latter, a solution can additionally make use of correlations in the data. 
We focus in this section on geometric decodability, as it is determined by PTM structure alone, independently of the data. Statistical decodability is then addressed in \cref{sec_input_decod,sec_cap_factorization}.

If $R$ corresponds to the full unitary channel ($\numMtot=\numF=d^2$), then $R=V(t)\in \reals^{\numF\times \numF}$ is orthogonal and any feature can be geometrically decoded by 
\beq \wv =  V(t)\ev_r.\qquad  \text{(unitary channel)}
\label{eq_feat_decod_unitary}\eeq
If we select subsets of observables, then \cref{eq_feat_decod} can in general \emph{not} be satisfied for arbitrary $r$ since features get mixed under a quantum channel.
We thus introduce the geometric decodability score of a Pauli feature (see \cref{app_ols_decoding})
\beq \gamma_r^2 \equiv (R^+ R)_{rr}, \qquad \sum_{r=1}^{\numF} \gamma_r^2 = \rank(R)\leq \numF,
\label{eq_decod_score}\eeq 
and its associated decoding weights
\beq \wv =  \left((R^+)_{r,:}\right)^\top,
\eeq
where $R^+$ is the Moore--Penrose pseudoinverse.
Exact decodability of the $r$-th feature is indicated by $\gamma_r^2=1$.
A value $\gamma_r^2<1$ implies that the feature can only be reconstructed up to an error $\sim |\phi_r-\wv_r^{*\top} R\phiv|\leq \sqrt{1-\gamma_r^2}\|\phiv\|_2$.
We take $\gamma_r^2\gtrsim 0.5$ as a heuristic decodability threshold. 
The score $\gamma_r^2$ is a purely geometric, data-independent notion. 
Statistical decodability scores formulated in \cref{sec_input_decod,app_decodability} can exceed it (via correlations), but can also fall below it (for strongly anisotropic feature distributions). 

Both decodability score and weights preserve tensor product structure of the general PTM $R$: if $R=\otimes_{i=1}^n R_i$ then $\gamma_r^2=\prod_{i=1}^n \gamma_{r,i}^2$ and $\wv=\otimes_{i=1}^n \wv_i$.
For a single evolution time and before feature clipping ($\numF=d^2$), $R=SV(t)$ has orthonormal rows and one obtains $\gamma_r^2 = \sum_{k\in \Scal} V(t)^2_{kr}$, showing that feature $r$ is decodable only to the extent that the ``spread'' of that Pauli under $V(t)$ lands in the measured subset $\Scal$.\footnote{If $R$ contains a $t=0$ block, then since $V(0)=\Imat_{d^2}$ the corresponding rows of $R$ include $\ev_j^\top$ for all $j\in\Scal$ (up to basis ordering), hence $\ev_j\in\mathrm{row}(R)$. Since $R^+R$ is the orthogonal projector onto $\mathrm{row}(R)$, $(R^+R)\ev_j=\ev_j$ and thus $\gamma_j^2=(R^+R)_{jj}=1$.}
Specifically, for Haar-random unitaries, typical decodability decays exponentially with qubit number $n$,
\beq \E_U[\gamma_r^2]  \sim |\Scal| 4^{-n},
\label{eq_decod_haar}
\eeq
which follows from \cref{eq_pauli_spread_latetime}.
Noisy channels can have invertible PTMs [see \cref{sec:general-noise}] and then do not strictly reduce the accessible feature subspace. If they are contracting, they effectively suppress high-weight Pauli features [see \cref{eq_noise_pauli_suppress}], but require large weight norms $\|\wv\|$ which degrades conditioning. Such maps are unstable under finite sampling and thus effectively lead to information loss.

Operator spreading under temporal multiplexing is visualized in \cref{fig:tmux_eff_ptm}, where the effective PTM $R$ is plotted for the case of a TFIM and a random unitary. The related projector matrices $R^+ R$ are shown in \cref{fig:proj_heatmap}(a,b).
When including $t=0$ in the multiplexing scheme, the first $|\Scal|=9$ rows of $R$ have 1 on the diagonal (no spreading), leading to perfect decodability of the weight-1 Pauli sector.
For the Hamiltonian and parameters in \cref{fig:proj_heatmap}(c), no perfectly decodable feature is observed at the chosen nonzero time.

\begin{figure}[tb!]
  \centering
  \panelimg[width=0.4\textwidth]{a}{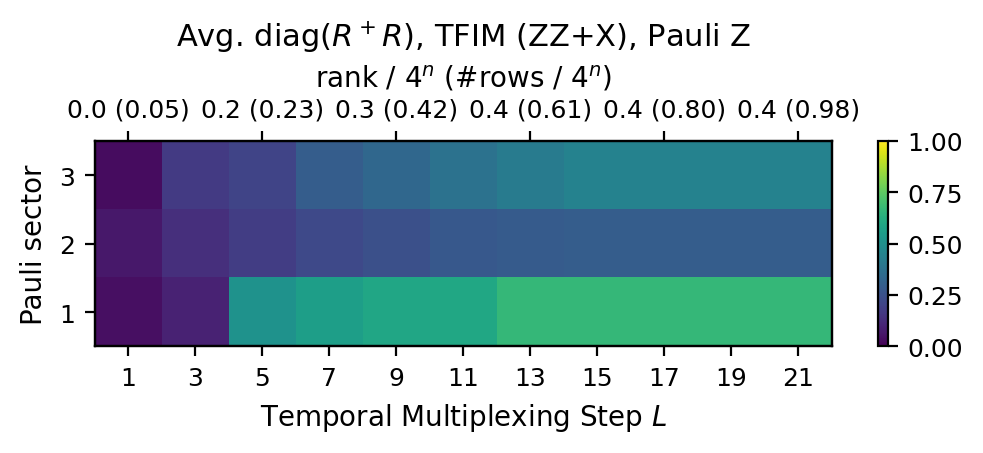}\qquad
  \panelimg[width=0.4\textwidth]{b}{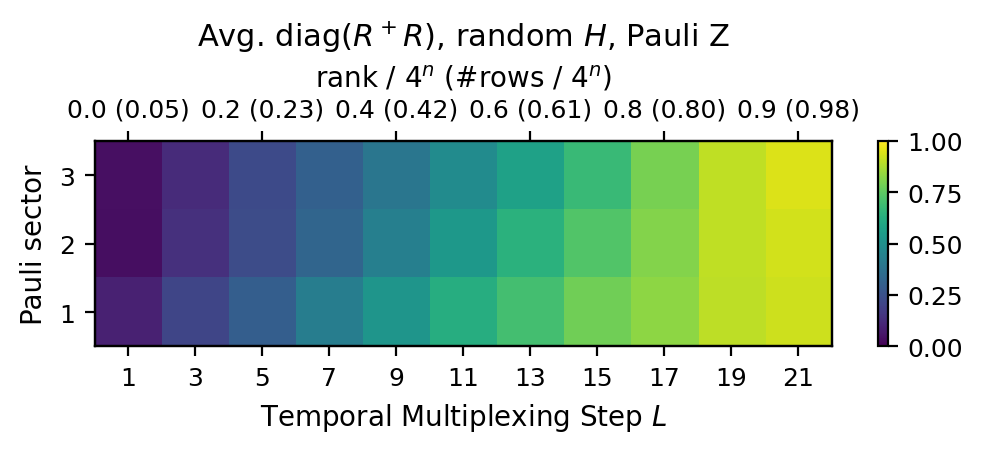}\\
  \panelimg[width=0.4\textwidth]{c}{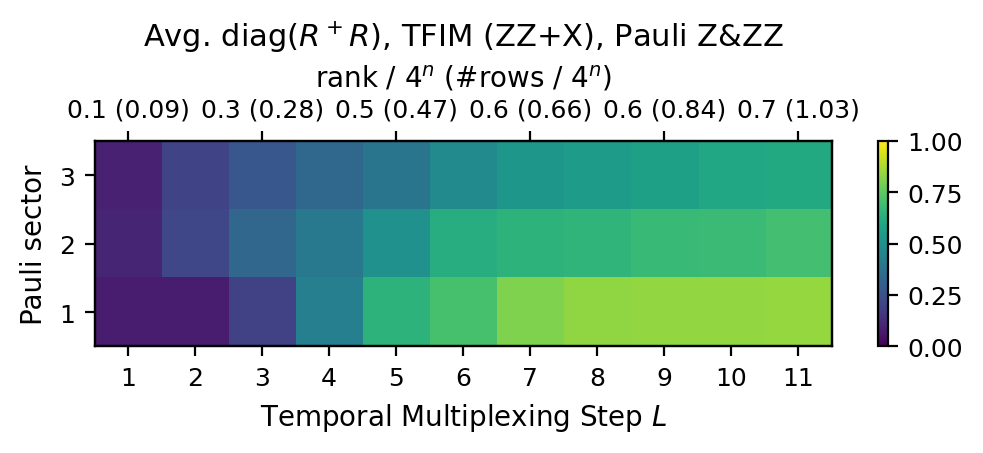}\qquad
  \panelimg[width=0.4\textwidth]{d}{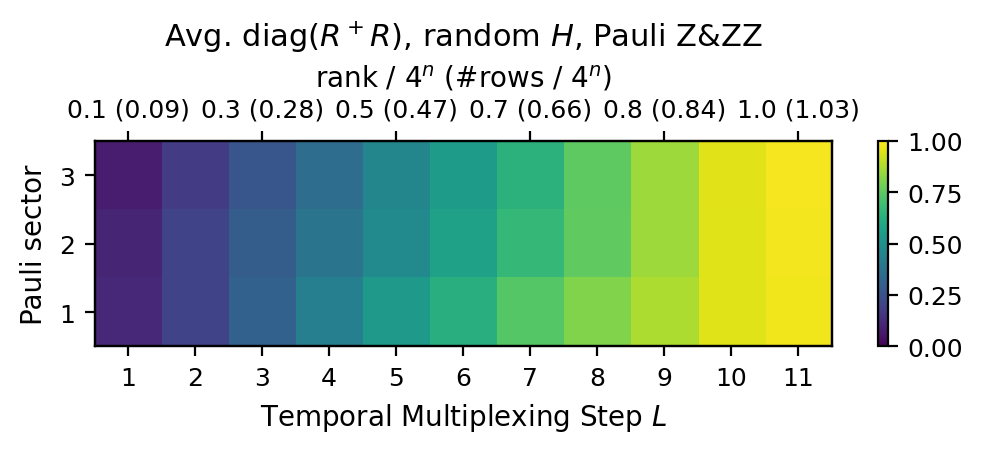}\\
  \panelimg[width=0.4\textwidth]{e}{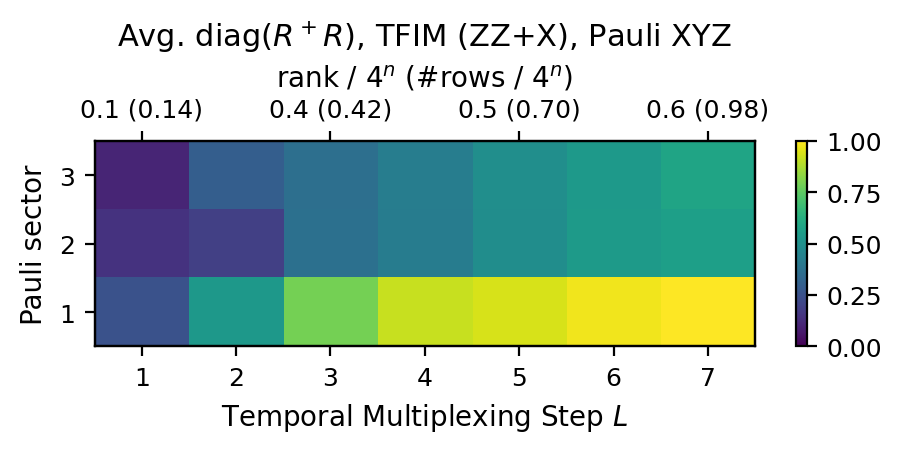}\qquad 
  \panelimg[width=0.4\textwidth]{f}{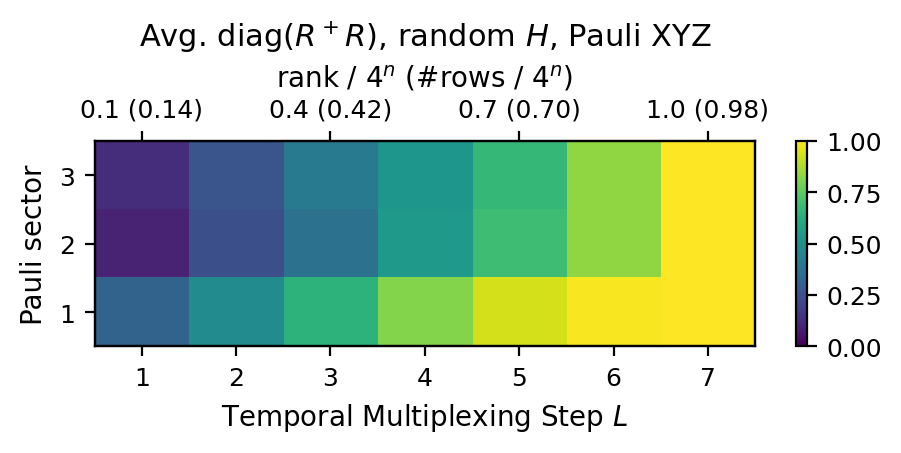}
  \panelimg[width=0.4\textwidth]{g}{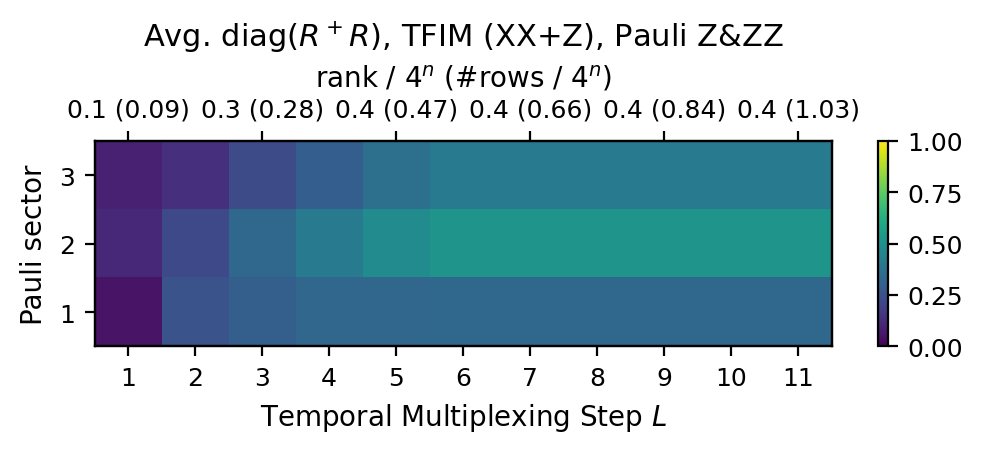}
  \caption{Feature decodability score $\bar\gamma_j^2$ obtained from the effective PTM $R$ [\cref{eq_decod_score}], averaged over each Pauli weight sector, of an $n=3$ qubit model as a function of temporal multiplexing iteration length $\numTM$ [\cref{eq_tmux}]. The unitary evolution time is given by $t_\numTM=\numTM$ (i.e., initial time $t_1=1$). The corresponding rank and number of rows of $R$, representing the measurement budget ($\numMtot = \numM \numTM$), are shown on the top axis (as a fraction of the total operator space dimension $d^2=64$). We consider the TFIM \eqref{eq_H_TFIM_zzx} in (a,c,e), the TFIM \eqref{eq_H_TFIM_xxz} in (g), and random Hamiltonians in (b,d,f). The observable set $\Scal$ is either only $Z$-Paulis (a,b), $Z$- and $ZZ$-Paulis (c,d,g), or all weight-1 Paulis (e,f).} 
  \label{fig:feat_iso}
\end{figure}

The evolution with multiplexing steps $\numTM$ of the mean decodability $\bar\gamma_r^2$ of each Pauli sector is further illustrated in \cref{fig:feat_iso}.
Since we now start with $t=1$, the selected low-weight Paulis are already spread and initial decodability is low on average \footnote{Including $t\approx 0$ in the temporal multiplexing scheme improves decodability at early iterations, but does not affect the final decodability for sufficiently large $\numTM$.}.
Structured Hamiltonians, like the TFIM [\cref{eq_H_TFIM}], can have strong impact on the achievable decodability score depending on the variant or the specific observables used.
Comparing \cref{fig:feat_iso}(c) and (g), Pauli-$Z$'s evolving under the Hamiltonian $H\st{TFIM}\ut{xx-z}$ render only weak scores compared to the Hamiltonian $H\st{TFIM}\ut{zz-x}$ \footnote{By global Hadamard equivalence, the statement flips for Pauli-$X$ as initial operators.}.

\begin{figure}[tb]
  \centering
  \panelimg[width=0.45\textwidth]{a}{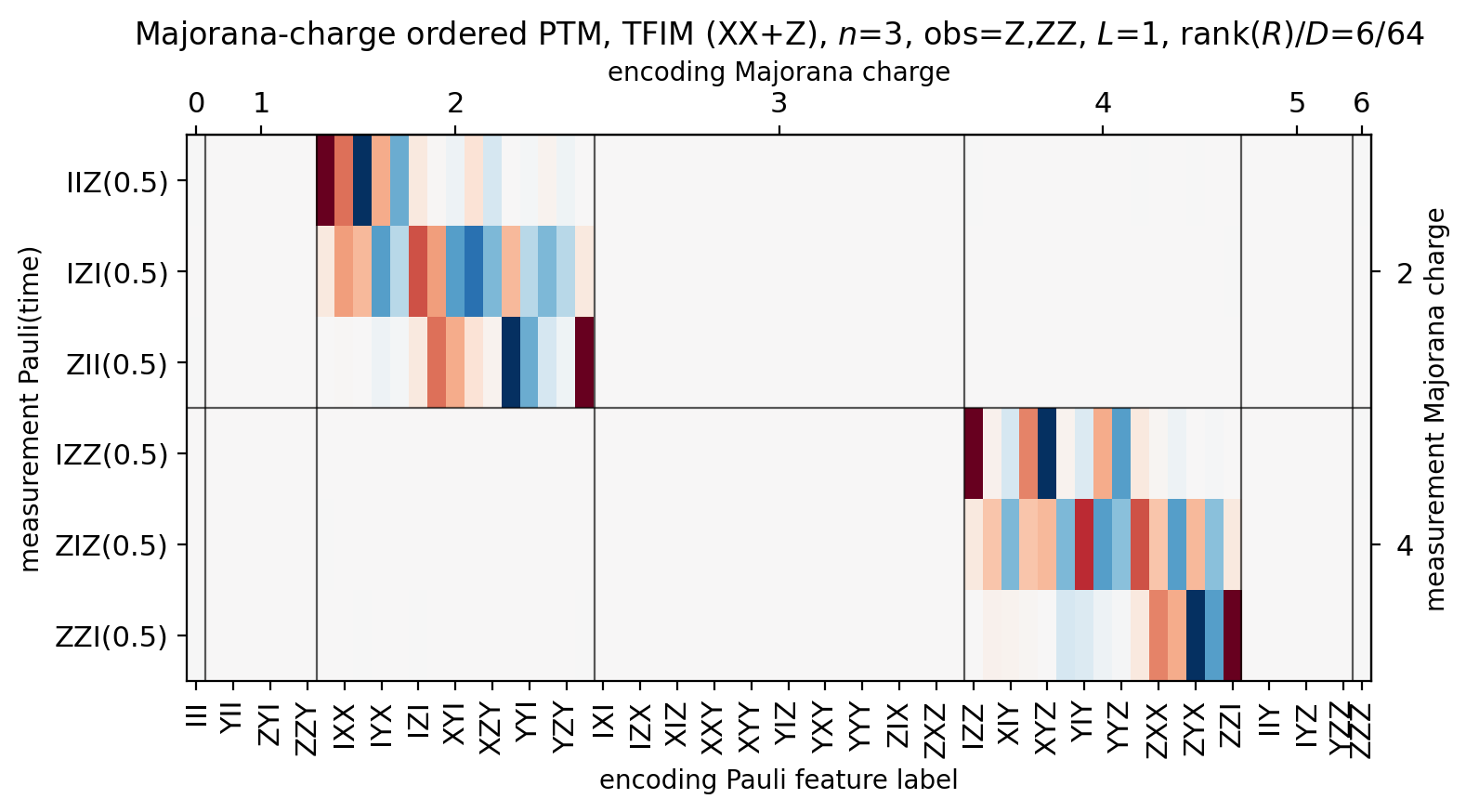}\qquad
  \panelimg[width=0.45\textwidth]{b}{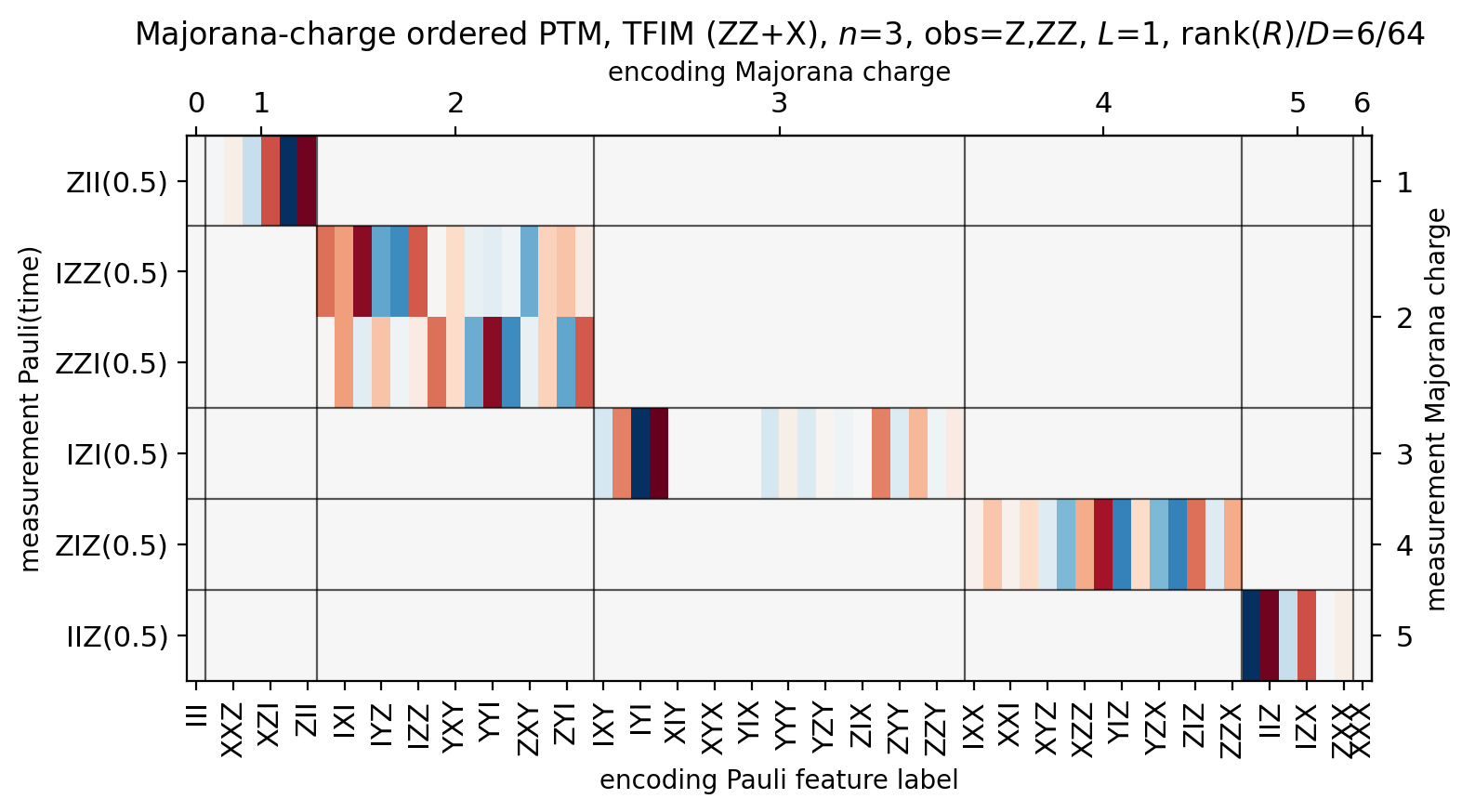}
  \panelimg[width=0.52\textwidth]{c}{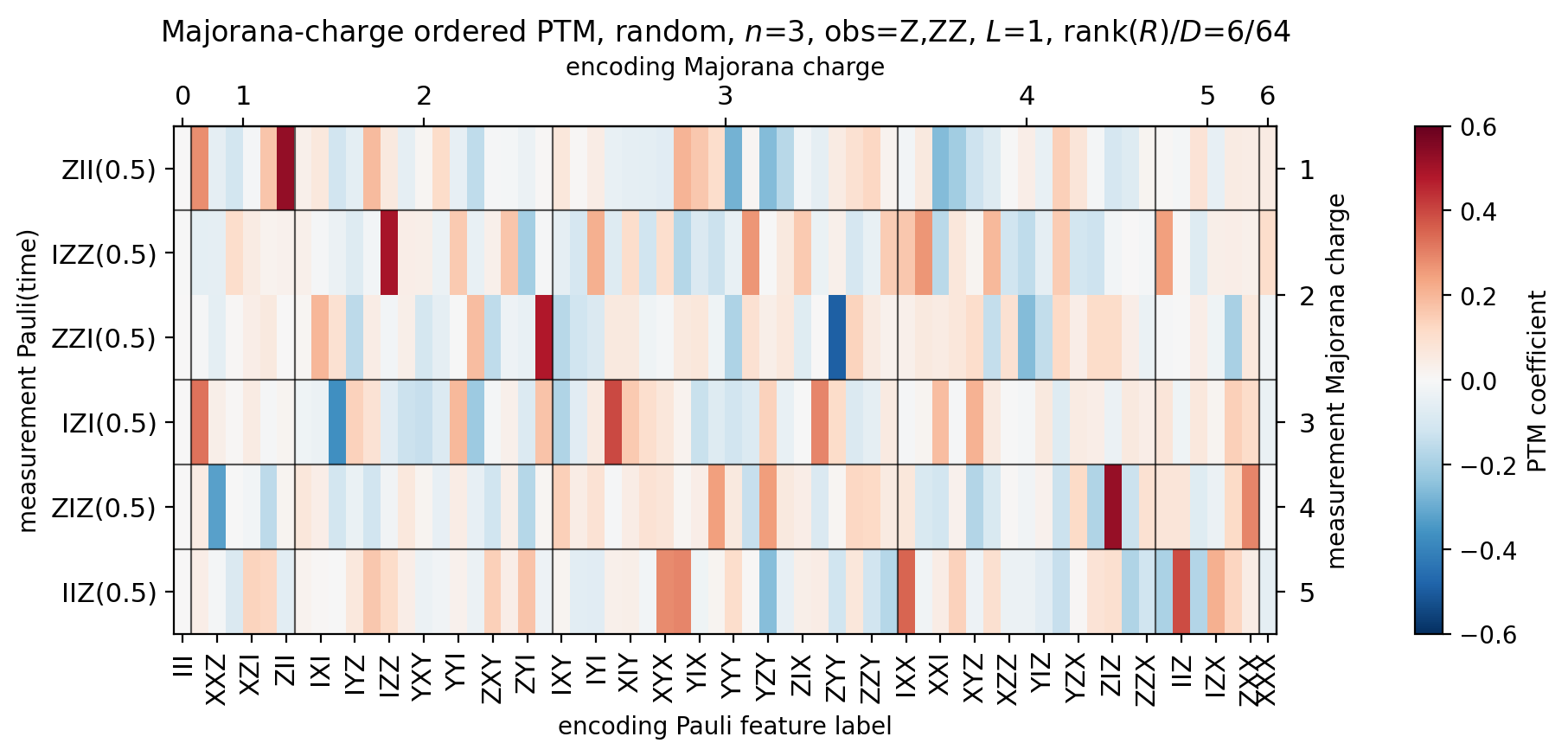}
  \caption{Majorana-charge ordered effective PTM $R$ for $n=3$ qubits, the observable set $\Scal=\{Z_j,Z_iZ_j\}$, and a single evolution time $t=0.5$. Columns and rows denote encoding Pauli and time-evolved measured Pauli observables, respectively, grouped by their Majorana charge. Panels show time evolution under various Hamiltonians: (a) $H\st{TFIM}\ut{xx-z}$ [\cref{eq_H_TFIM_xxz}], (b) $H\st{TFIM}\ut{zz-x}$ [\cref{eq_H_TFIM_zzx}], and (c) a random $H$. In (a,b) we use the Majorana charge convention that makes the corresponding $H$ quadratic, revealing a sparse block structure where $Z$-type observables remain confined to their respective Majorana-charge sectors. By contrast, a random $H$ (c) does not conserve Majorana charge and thus populates the full Pauli basis (for visualization, we use here the same Majorana charge convention as in (b)).}
  \label{fig:ptm_majorana}
\end{figure}

These differences are a consequence of the underlying symmetries of the Hamiltonian: specifically, for the TFIM, the Majorana charge of an observable is conserved under the evolution (see \cref{sec_tfim_sector_rank}). 
\Cref{fig:ptm_majorana} shows the symmetry-resolved sector structure of the PTM, obtained by ordering the observables according to the charge convention that makes the corresponding $H$ quadratic in the Majorana operators.
As seen in (a), time evolution of Pauli-$Z$ observables under $H\st{TFIM}\ut{xx-z}$ leaves encoding Pauli sectors with odd Majorana degree unpopulated, explaining the low decodability scores in \cref{fig:feat_iso}(g).
By contrast, the Hamiltonian $H\st{TFIM}\ut{zz-x}$ touches all nontrivial encoding sectors except the top one with charge $2n$ [\cref{fig:ptm_majorana}(b)]. The spreading pattern still remains highly structured and oscillatory, leading to slightly undersaturated decodability scores [\cref{fig:feat_iso}(c)] compared to a random Hamiltonian (panel d), which does not respect Majorana symmetries.

\begin{figure}[tb]
  \centering
  \panelimg[width=0.38\textwidth]{a}{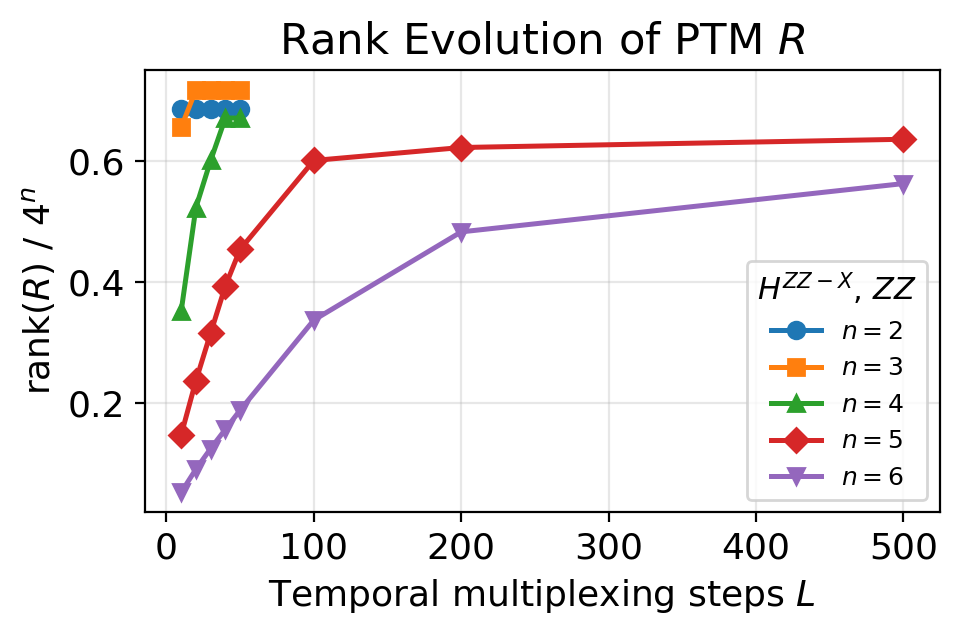}\qquad
  \panelimg[width=0.38\textwidth]{b}{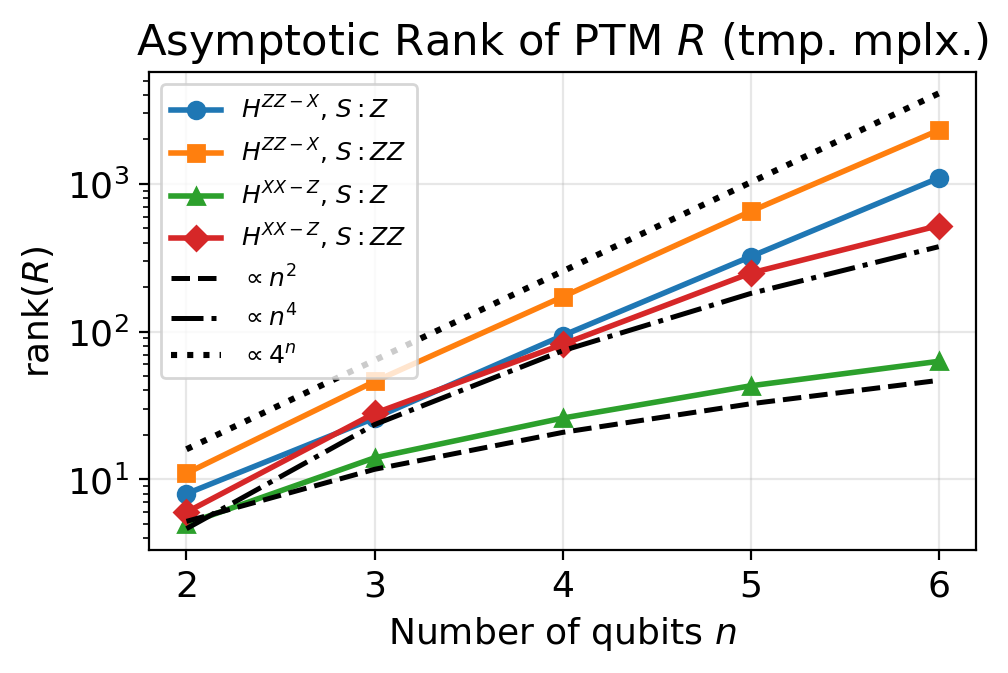}
  \caption{Rank of the effective PTM $R$ [\cref{eq_tmux}] obtained by temporal multiplexing an observable set $\Scal$ (one and two-site Pauli $Z$) under unitary evolution with a TFIM Hamiltonian [\cref{eq_H_TFIM}]. (a) Evolution of $\rank(R)$ with temporal multiplexing steps $\numTM$ for the TFIM \eqref{eq_H_TFIM_zzx} and $\Scal=\{Z_i, Z_i Z_j\}$ observables. (b) Asymptotic PTM rank, $\lim_{\numTM\to\infty} \rank(R)$, obtained numerically (solid curves), compared to the theoretical scaling predictions from \cref{tab:ptm_rank} (broken lines). The dotted curve follows the operator space dimension $4^n$, which upper bounds the exact theoretical result $\mathrm{poly}(n)3^n$.} 
  \label{fig:rank_effptm}
\end{figure}

Furthermore, the analysis based on the Jordan--Wigner/Majorana formalism (see \cref{sec_tfim_sector_rank}) shows that under $H\st{TFIM}\ut{xx-z}$, the Pauli strings generated from an initial $Z$ are mainly consisting of $Z$-strings with $X$/$Y$ endpoints, of which there are $\sim \Ocal(n^2)$. By contrast, under $H\st{TFIM}\ut{zz-x}$ the evolved $Z_j$ spread over exponentially large odd-degree Majorana sectors, producing a Krylov rank $\sim \Ocal(3^n)$ (up to polynomial prefactors).
The implications for the \emph{asymptotic rank} $r_\infty := \lim_{\numTM\to\infty}\rank(R_\numTM)$ are summarized in \cref{tab:ptm_rank} and numerically confirmed in \cref{fig:rank_effptm}(b).
Additionally, \cref{fig:rank_effptm}(a) illustrates the evolution of $\rank(R_\numTM)$ under temporal multiplexing. 
These findings can more generally be understood as a consequence of Hamiltonian symmetries and eigenspace degeneracies, as shown in detail in \cref{app_sym_kry}.

Random Hamiltonians typically populate the full Pauli basis. Under temporal multiplexing they thus consistently lead to a higher rank of $R$ and a more even decodability over Pauli sectors for large $\numTM$.
We find that temporal multiplexing with Pauli $Z$ and $ZZ$ observables and random unitaries requires $\numMtot\sim d^2$ measurements to achieve high decoding scores in all Pauli sectors, along with a saturated rank $\rank(R)\sim |\Scal| \numTM \sim d^2$. 
In practice, however, this exponential measurement budget is rarely needed, since relevant features can be statistically decoded with far fewer measurements (see \cref{fig:cap_nord_tmux,fig:cap_nord_tmux_rel} below).

\begin{table}[tb]
\centering
\begin{tabular}{lll}
\hline
Hamiltonian & Observables & Asymptotic saturated rank $r_\infty$ \\
\hline
$H\st{TFIM}\ut{xx-z}$ & $Z_j$ & $\Ocal(n^2)$ \\
$H\st{TFIM}\ut{xx-z}$ & $Z_j$, $Z_iZ_j$ & $\Ocal(n^4)$ \\
$H\st{TFIM}\ut{xx-z}$ & $X_j,Y_j,Z_j$ & $\Ocal(\mathrm{poly}(n)3^n)$  \\
\hline
$H\st{TFIM}\ut{zz-x}$ & $Z_j$ & $\Ocal(\mathrm{poly}(n)3^n)$ \\
$H\st{TFIM}\ut{zz-x}$ & $Z_j$, $Z_iZ_j$ & $\Ocal(\mathrm{poly}(n)3^n)$ \\
$H\st{TFIM}\ut{zz-x}$ & $X_j,Y_j,Z_j$ & $\Ocal(\mathrm{poly}(n)3^n)$ \\
\hline
\end{tabular}
\caption{Asymptotic saturated rank $r_\infty := \lim_{\numTM\to\infty}\rank(R_\numTM)$ (see \cref{sec_tfim_asymptotic_rank}) of the temporally multiplexed PTM \eqref{eq_tmux} for the two TFIM variants [\cref{eq_H_TFIM}] and different observable families $\Scal$. For temporal multiplexing of pure Pauli $Z$-type observables, the Hamiltonian $H\ut{xx-z}\st{TFIM}$ leads to exponentially lower expressivity, reflected by polynomial-in-$n$ ranks $\ll 3^n$.}
\label{tab:ptm_rank}
\end{table}

\paragraph*{Summary.} 
The decodability framework presented here extends the Krylov observability theory of \cite{cindrak_engineering_2025,cindrak_krylov_2025} by resolving geometry on the level of individual Pauli features.
We remark that the Krylov grade $M$ discussed in \cite{cindrak_engineering_2025} corresponds to the asymptotic rank of a single observable, $r_\infty(\Scal=\{O\}) = M$ [\cref{eq_rank_limit_krylov,eq_symmetry_hamiltonian_krylov_grade}], while the effective dimension score $O_K$ can be regarded as a scalar proxy for our multi-observable $\rank(R_\numTM)$.
The geometric decodability score $\gamma_r^2$ of a feature $r$ [\cref{eq_decod_score}] is a data-independent property of the chosen measurement observables and quantum operations after the encoding. 
For moderate qubit numbers $n$, full Pauli feature decodability is guaranteed when one either performs a full measurement ($\numM=d^2$) on an invertible quantum channel, or one employs a temporal multiplexing scheme leading to a full column rank of $R$, again requiring an exponential measurement budget $\numM \numTM \sim d^2$.
Practically sufficient decodability can already be achieved with modest scores and polynomial-in-$n$ measurement budgets, as typically done in applications \cite{fujii_harnessing_2017,steinegger_predicting_2025}. 
Structured Hamiltonians can be advantageous in the large-$n$ limit to avoid exponential concentration issues \cite{xiong_fundamental_2025,sannia_exponential_2025}. 
However, they can negatively impact expressivity of the model under temporal multiplexing of certain observables, as reflected in polynomial or reduced exponential growth of the PTM rank with qubit number $n$ (see \cref{tab:ptm_rank,fig:rank_effptm}).

\subsection{Statistical decodability of input data}
\label{sec_input_decod}

We focused above on the data-agnostic decodability of Pauli features $\phi_j$, which are generated by the encoding and transform linearly under the action of the reservoir. However, they are themselves nonlinear functions of the classical input $\uv\in\reals^D$. 
For example, with amplitude encoding \eqref{eq_enc_amplsqrt}, a single Pauli-$Z$ generates an affine readout $\bra Z\ket\propto 2u-1$, a single Pauli-$X$ can introduce significant nonlinearities since $\bra X\ket=\sqrt{1-(2u-1)^2}=  1-\frac{1}{2}(2u-1)^2 - \frac{1}{8}(2u-1)^4 + \ldots$ (cf.\ \cref{tab:enc_proj}).
It is therefore important to understand the actual nonlinearities in $u$ present at the readout. 
This is a classical problem in reservoir computing, studied as memory capacity \cite{jaeger_short_2001} or, more generally, information processing capacity \cite{dambre_information_2012,kubota_unifying_2021,martinez-pena_information_2023,cindrak_enhancing_2024}.
The latter notion is often called \emph{nonlinear processing capacity} for models like a QELM, where internal memory is absent.

We show in \cref{sec_cap_factorization} that the classical nonlinear processing capacity can be split into a feature-representation part and a decodability part. The former describes the nonlinear capacity of the encoding (which is discussed in this section), while the latter accounts for transmission of the features by the PTM (discussed in \cref{sec_Pauli_decoding}).

\subsubsection{Nonlinear degree at readout}
\label{sec_nonlin_readout}

In order to estimate the degree of nonlinearity in $u$ available at the readout, one typically calculates the empirical correlation between the readout features $\Fv(\uv)$ [see \cref{eq_qelm_ptm}] and certain nonlinear target functions $y(\uv)$.
Raw monomials like $\{u^j\}$ are strongly correlated under typical distributions of $u$ and thus do not allow for clear identification of the nonlinear degree within this statistical approach. 
Following \cite{dambre_information_2012,cindrak_enhancing_2024}, we therefore use orthogonal polynomials $p_j(u_\alpha)$ (degree $j$) to construct orthogonal degree-$k$ targets as $y_\ev(\uv) = \prod_{\alpha=1}^D p_{e_\alpha}(u_\alpha)$, with $\sum_{\alpha=1}^D e_\alpha = k$.
Specifically, for amplitude encoding, we use (shifted) Legendre polynomials, which are orthonormal w.r.t.\ the uniform distribution on $[-1,1]$ or $[0,1]$; for angle encoding, we use Hermite polynomials, which are orthonormal w.r.t.\ the Gaussian distribution $\Ncal(0,1)$. 
Here the inputs are encoded using $R_Y(u_\alpha)$, rather than the convention of \cref{eq_enc_rot}. We generally assume independent inputs $u_\alpha$.

\begin{figure}[tb!]
  \centering
  \panelimg[width=0.48\textwidth]{a}{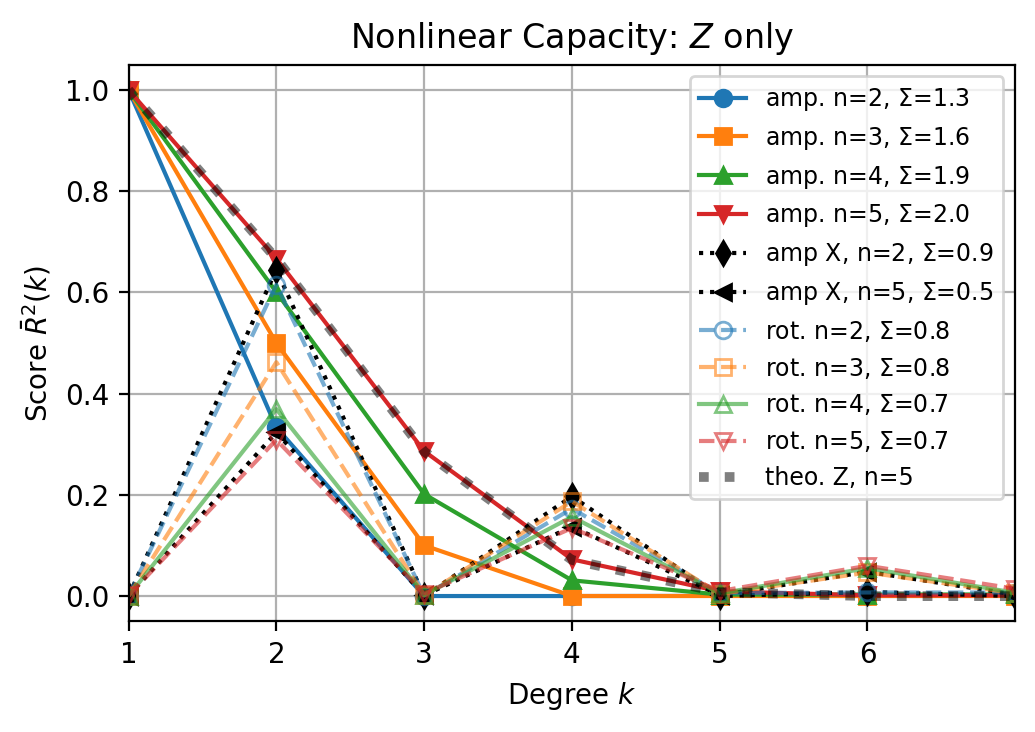}
  \panelimg[width=0.48\textwidth]{b}{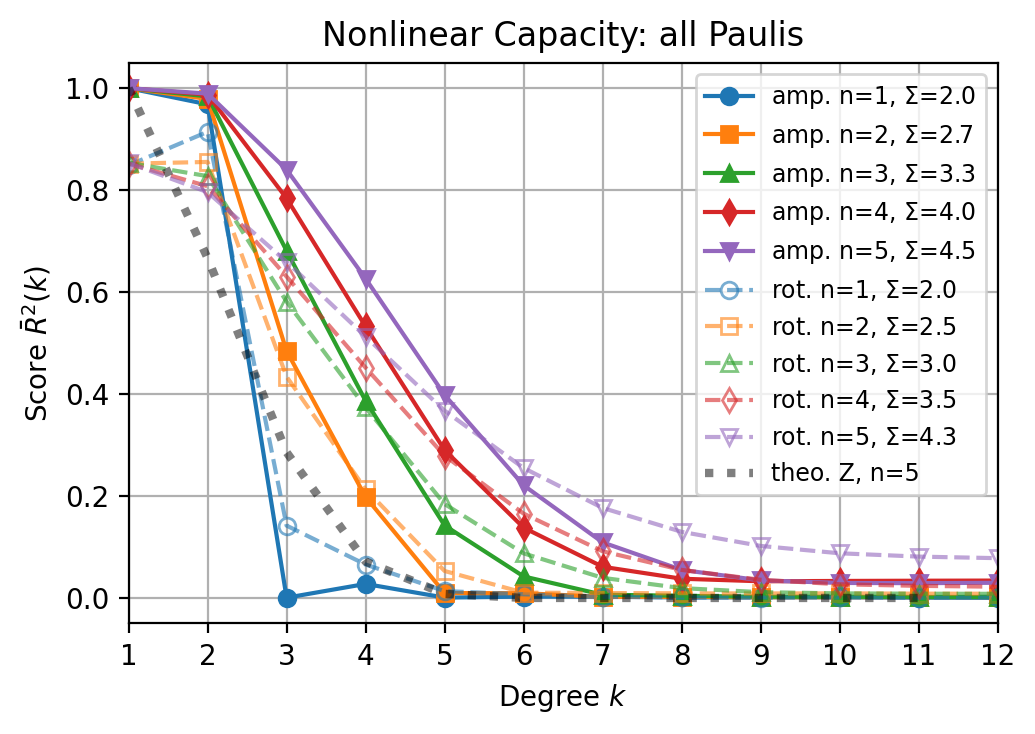}
  \caption{Ideal nonlinear capacity score $\Rcal_\Imat^2(k)$ [\cref{eq_nlcap_R2_score}] for amplitude and rotational encodings [\cref{eq_enc_amplsqrt,eq_enc_rot}]. This score indicates the ability to generate degree-$k$ polynomials of the input $u_\alpha$ ($\alpha=1,\ldots,D$ and $n=D$) for a QELM $f(\uv) = \wv^\top \phiv(\uv)$ without feature mixing [i.e., with $R=\Imat$ in \cref{eq_qelm_classical}]. Unless indicated, only Pauli-$Z$ features are measured in (a).  Nonlinear capacity is significantly enhanced when measuring over all Paulis (b). This is also demonstrated by the score for amplitude encoding with Pauli-$X$ readout included in (a). The thick dotted curve represents the theoretical score for $\Rcal^2_Z(k)$ with amplitude encoding and $n=5$ input qubits [\cref{eq_nlcap_R2_theo_Z}], which agrees exactly with the numerical results. $\Sigma$ denotes the integrated capacity over all $k$. }
  \label{fig:cap_nord_enc}
\end{figure}

We take the coefficient of determination \cite{dambre_information_2012,martinez-pena_information_2023}
\beq \Rcal^2(k) = \frac{1}{N_k} \sum_{|\ev|=k} \Rcal^2[y_\ev],\qquad  \Rcal^2[y_\ev] = 1-\frac{\E_\uv (y_\ev(\uv) - \hat f(\uv))^2}{\mathrm{Var}_\uv y_\ev(\uv)},
\label{eq_nlcap_R2_score}\eeq 
averaged over all $N_k=\binom{D+k-1}{k}$ degree-$k$ orthogonal polynomials, to indicate whether the trained predictor $\hat f(\uv)$ can generate nonlinearities of degree $k$ in the $(D=n)$-dimensional input \footnote{In \cref{eq_nlcap_R2_score}, $\mathrm{Var}_\uv y(\uv)$ denotes the variance of $y(\uv)$ over the distribution of $\uv$.}.
$\Rcal^2$ generalizes the geometric score \eqref{eq_decod_score} by taking statistical correlations into account (see \cref{app_decodability} for further discussion). 
In the special case where the targets are whitened orthonormal features, one has $\Rcal^2[y_r] = \gamma_r^2$ [see \cref{eq_R2_white}].
In the case of amplitude encoding \eqref{eq_enc_amplsqrt} with complete Pauli-$Z$-string library as readout, the number of representable degree-$k$ targets is $\binom{D}{k}$, giving an averaged score 
\beq \Rcal^2_Z(k)=\begin{cases} 
  \binom{D}{k}/\binom{D+k-1}{k} & \text{for } k\leq D,\\ 
  0 & \text{for } k>D
\end{cases}
\label{eq_nlcap_R2_theo_Z}\eeq  
which exponentially decreases with $k$. The capacity calculations use unregularized least squares with a fitted intercept, $\hat f(\mathbf u)=\hat b+\hat{\mathbf w}^{\top}R \phiv(\mathbf u)$.

In \cref{fig:cap_nord_enc}, we first display the nonlinear capacity $\Rcal_\Imat^2$ that can be \emph{ideally} achieved from the encoded state without effects of feature mixing, i.e., we take $R=\Imat$ in \eqref{eq_qelm_classical}, or $R=S$ when restricting to observable subsets ($m\leq q$). 
As shown in panel (a), a Pauli-$X$ readout (black symbols with dotted curves) generates even degree polynomials, behaving similarly to angle encoding with Pauli-$Z$ observables (connected open symbols). 
In \cref{fig:cap_nord_enc}(b) we show the ideal nonlinear capacity score $\Rcal_\Imat^2(k)$ obtained for a complete Pauli observable set. 
Notably, the presence of non-$Z$ Paulis significantly enhances the nonlinearity present in the readout, as becomes clear by comparison to the dotted curve corresponding to $\Rcal^2_Z(k)$ for $D=n=5$ [\cref{eq_nlcap_R2_theo_Z}]. 
Moreover, angle and amplitude encodings show a similar behavior. 
We attribute the decay of $\Rcal_\Imat^2(k)$ with $k$ mainly to the fact that the product state encoding only generates square-free monomials of the Pauli features $\phi_j$, while we test against (the practically relevant case of) all possible degree-$k$ targets \footnote{This argument is not fully rigorous, since features like $\phi_x^{(i)}(u_i)$ have infinite Taylor expansion in $u_i$ and can thus correlate locally with raw monomials $u_i^m$.}.

\begin{figure}[tb!]
  \centering
  \panelimg[width=0.35\textwidth]{a}{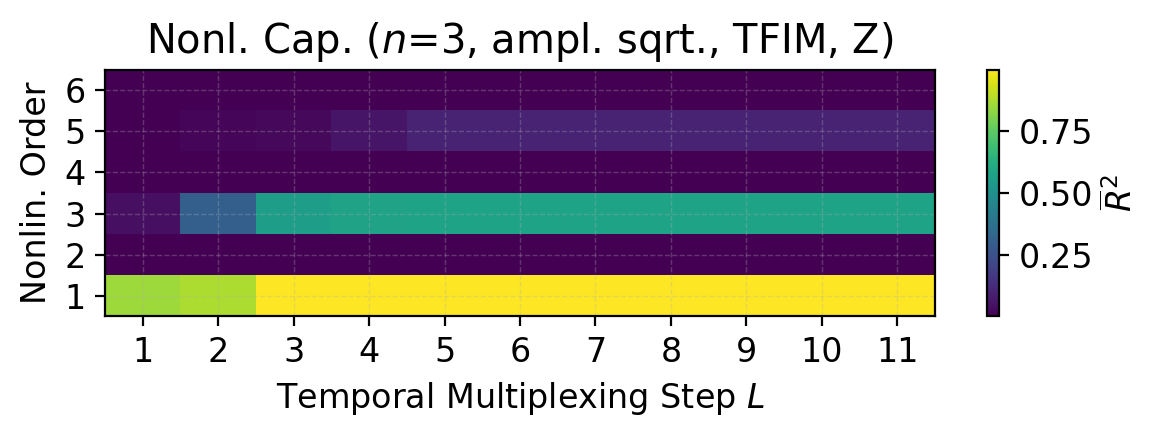}\qquad
  \panelimg[width=0.35\textwidth]{b}{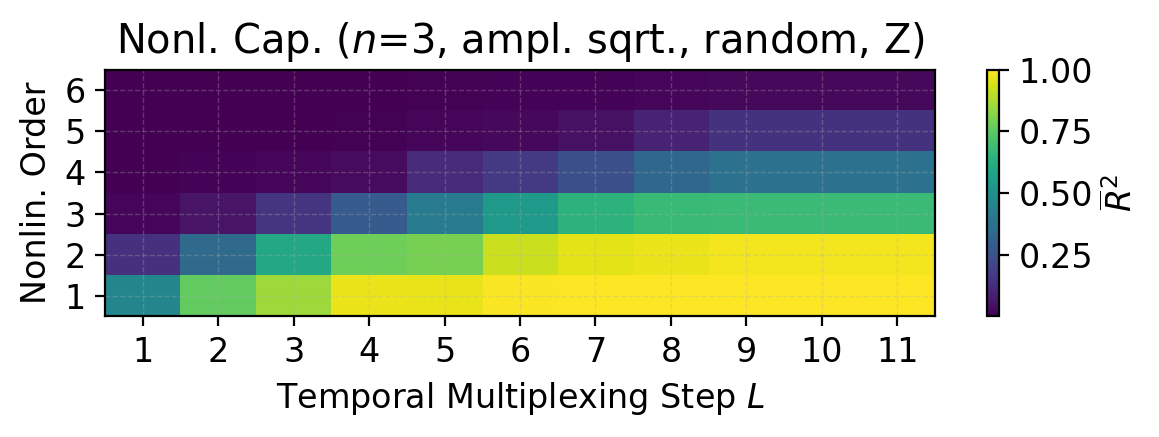}\\
  \panelimg[width=0.35\textwidth]{c}{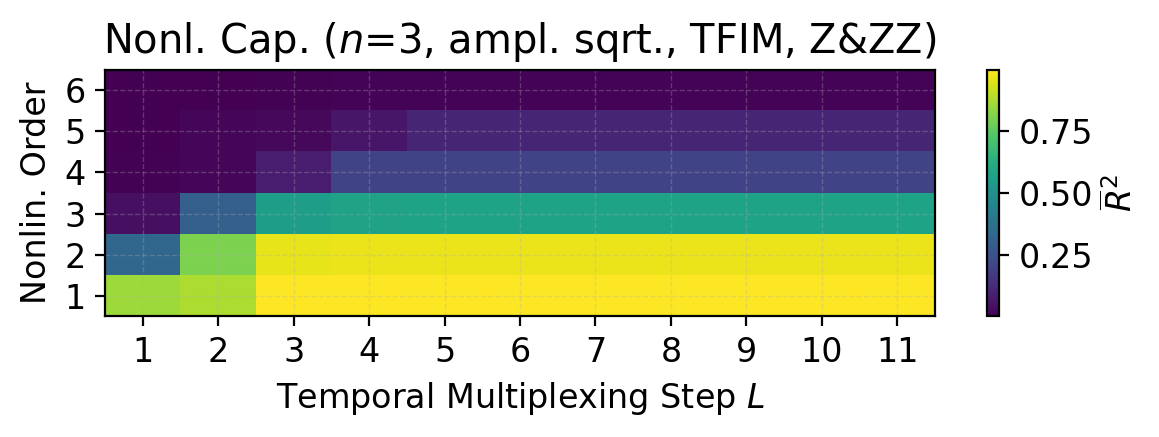}\qquad
  \panelimg[width=0.35\textwidth]{d}{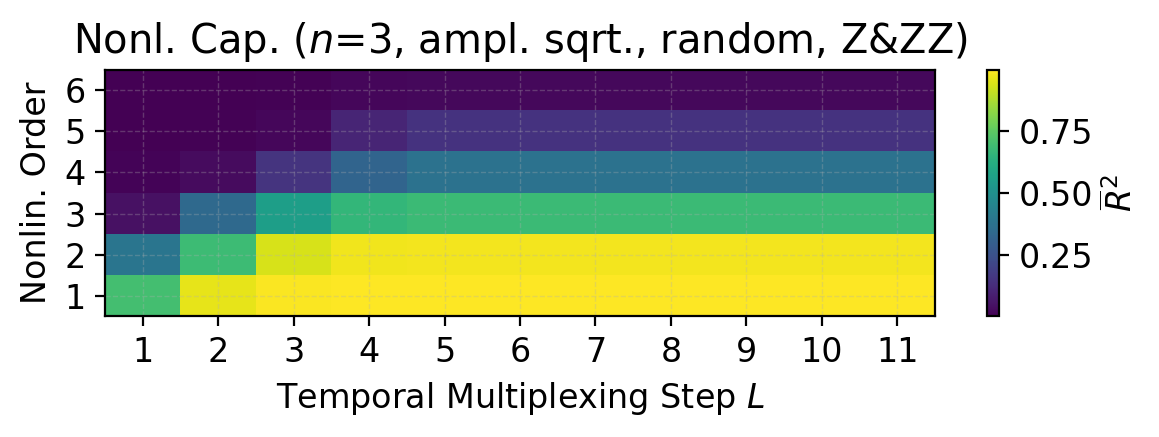}\\
  \caption{Nonlinear capacity score $\Rcal^2(k)$ [\cref{eq_nlcap_R2_score}] as a function of the nonlinear order $k$ and the temporal multiplexing iteration length $\numTM$ [\cref{eq_tmux}]. The unitary evolution time is given by $t_L=\numTM$. We consider an $n=3$ qubit model in the same setup as in \cref{fig:feat_iso}. The unitary is based on the TFIM \eqref{eq_H_TFIM_zzx} (a,c) or a random Hamiltonian (b,d), while the observable set $\Scal$ consists of either only $Z$-Paulis (a,b) or $Z$- and $ZZ$-Paulis (c,d).}
  \label{fig:cap_nord_tmux}
\end{figure}

\begin{figure}[tb]
  \centering
  \panelimg[width=0.35\textwidth]{a}{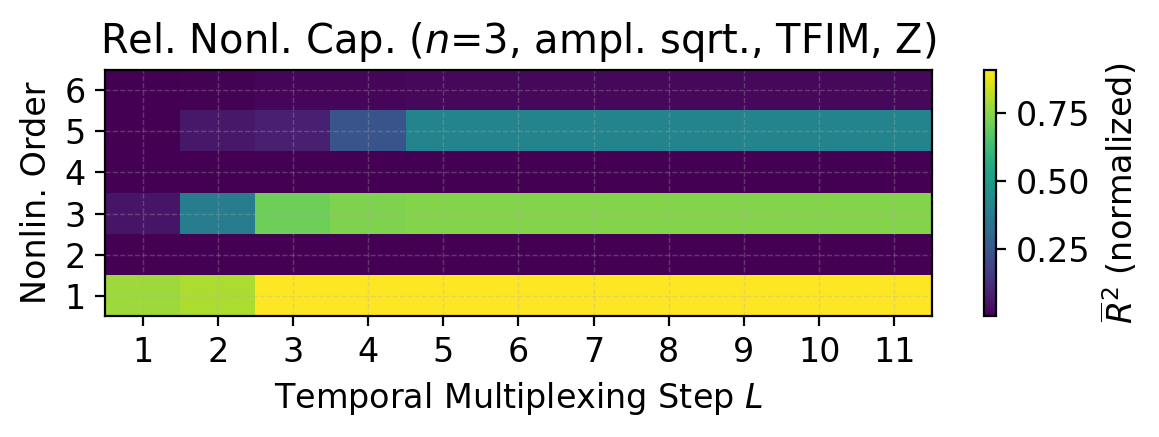}\qquad
  \panelimg[width=0.35\textwidth]{b}{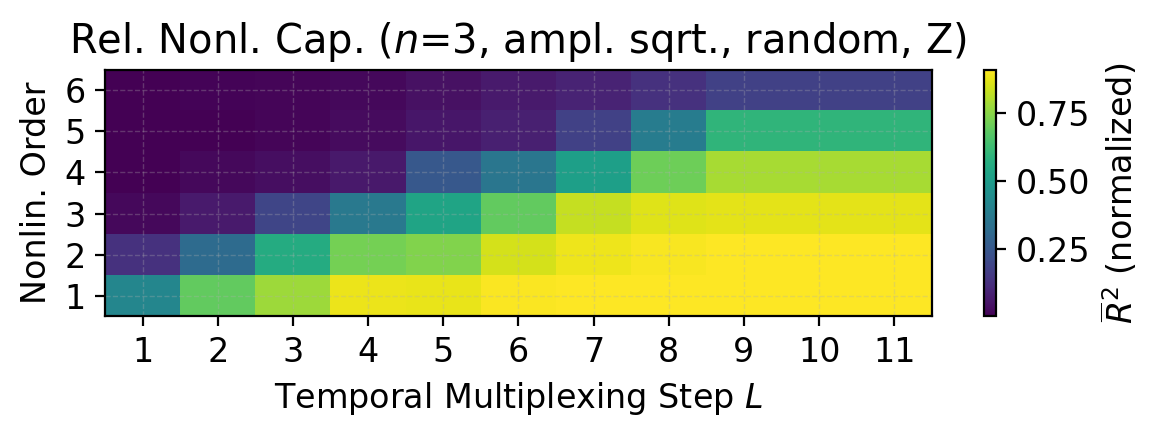}\\
  \panelimg[width=0.35\textwidth]{c}{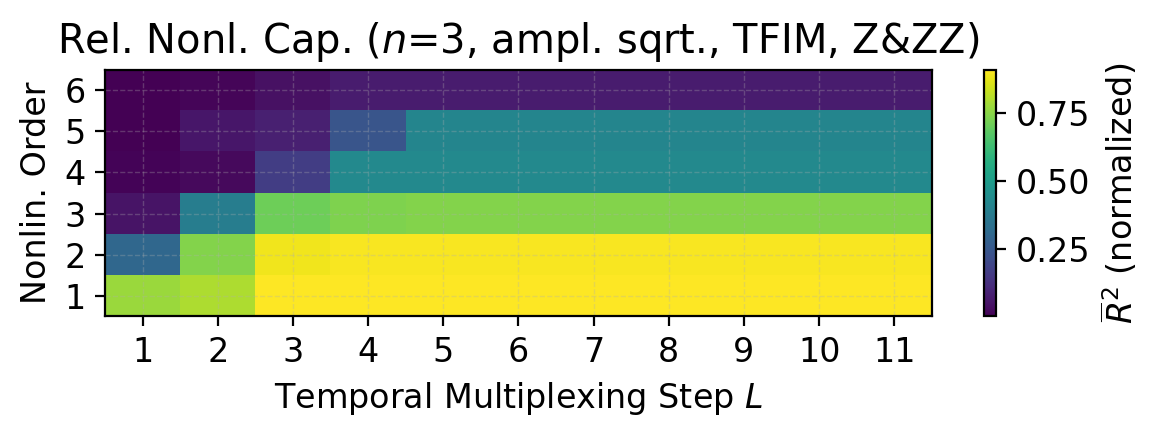}\qquad
  \panelimgshift[width=0.35\textwidth]{-12pt}{0pt}{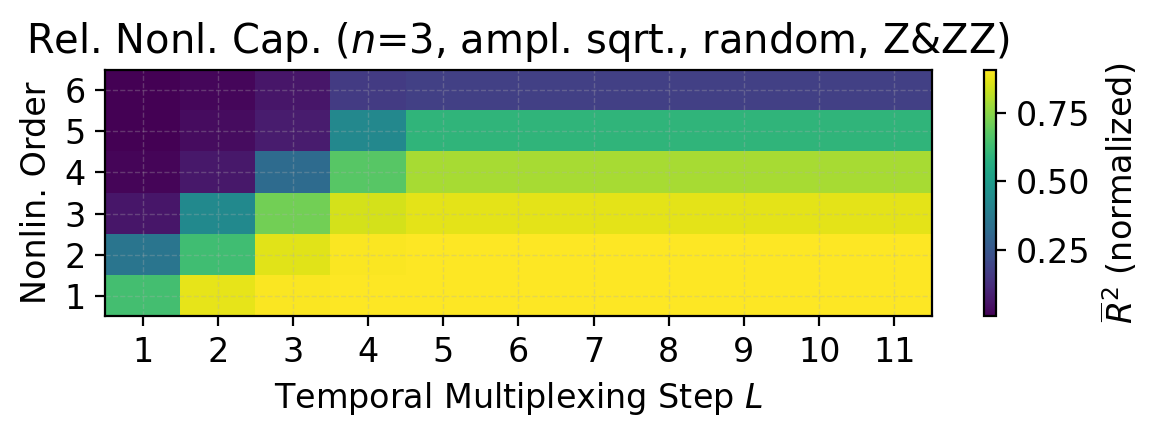}{d}\\
  \caption{Relative nonlinear capacity score $\Rcal^2(k)/(\Rcal^2\st{all}(k)+\delta)$ [\cref{eq_nlcap_R2_score}] as a function of the nonlinear order $k$ and the temporal multiplexing iteration length $\numTM$ [\cref{eq_tmux}]. $\Rcal^2\st{all}$ is the nonlinear capacity score obtained for a complete Pauli observable set, and we choose $\delta=0.1$ as a threshold since scores $\Rcal^2\st{all}\lesssim 0.1$ are not considered meaningful. We use the same setup as in \cref{fig:cap_nord_tmux}. The unitary is based on the TFIM \eqref{eq_H_TFIM_zzx} (a,c) or a random Hamiltonian (b,d), while the observable set $\Scal$ consists of either only $Z$-Paulis (a,b) or $Z$- and $ZZ$-Paulis (c,d). }
  \label{fig:cap_nord_tmux_rel}
\end{figure}

\Cref{fig:cap_nord_tmux} illustrates the nonlinear capacity score $\Rcal^2(k)$ obtained for a QELM, based on the TFIM or a random Hamiltonian, under temporal multiplexing of Pauli-$Z/ZZ$ observables.
The score follows a similar pattern as the data-independent Pauli decodability $\gamma_j^2$ (see \cref{fig:feat_iso}) as they both show a delayed growth with increasing nonlinear order $k$ and multiplexing step $\numTM$.
However, $\Rcal^2(k)$ saturates faster than $\gamma_j^2$ because the considered nonlinear target functions $y_k(\uv)$ can be generated already from relatively few Pauli features.
The decrease of $\Rcal^2(k)$ with $k$ stems from two facts: (i) the encodings used here can generate only a small subset of all possible degree-$k$ monomials (see \cref{eq_nlcap_R2_theo_Z} and \cref{fig:cap_nord_enc}). 
A clearer measure of the effectiveness of temporal multiplexing to generate nonlinearities is thus the score $\Rcal^2(k)$ normalized by the score $\Rcal^2\st{all}(k)$ obtained when measuring over the complete Pauli observable set (see \cref{fig:cap_nord_tmux_rel}).
(ii) Higher-order observables can be missing for highly structured Hamiltonians or insufficient multiplexing steps.
Hamiltonians with a more global and random structure are thus beneficial when measurements are restricted to Pauli-$Z$ observables but higher-order nonlinearities are desired.

\subsubsection{Nonlinear monomials at readout}
\label{sec_mon_decod}

\begin{figure}[tb]
  \centering
  \panelimgshift[width=0.6\textwidth]{-20pt}{0pt}{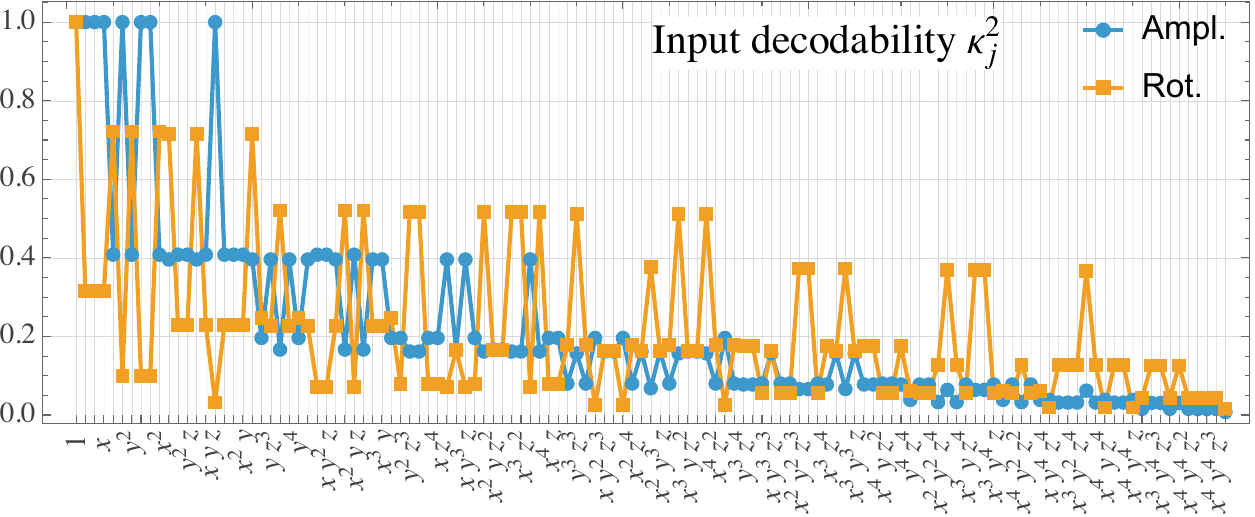}{a}\\
  \panelimgshift[width=0.25\textwidth]{-15pt}{0pt}{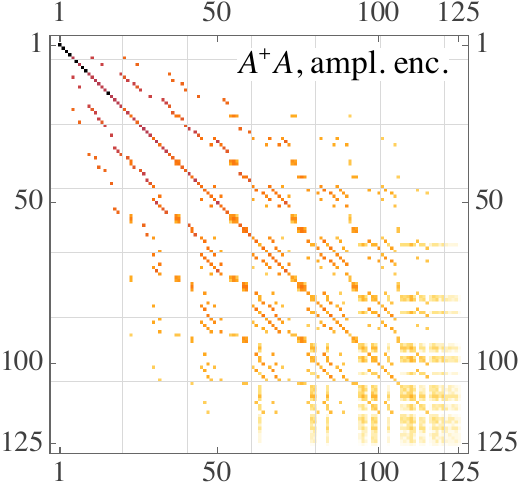}{b}\qquad 
  \panelimgshift[width=0.3\textwidth]{-15pt}{0pt}{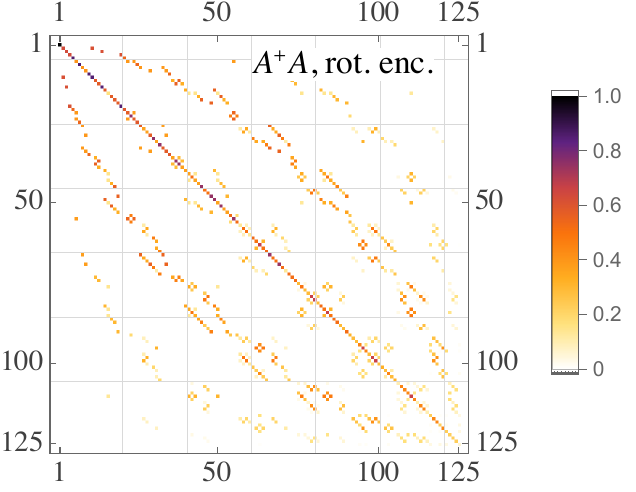}{c}
  \caption{Ability to construct monomials of the input $\{x,y,z\}$ for amplitude encoding \eqref{eq_enc_amplsqrt} and rotational encoding \eqref{eq_enc_rot} with $n=D=3$. (a) Decodability score $\kappa_j^2$ according to \cref{eq_input_dec_score}. (b,c) Projector matrix $A^+ A$, where nonzero off-diagonals correspond to $\kappa_j^2<1$ and indicate decoding is possible only within a certain accuracy. The underlying Taylor expansion is performed up to order $r=4$ in each $u_j\in \{x,y,z\}$ and changes are mild upon further increasing $r$. The scores and matrices are averaged over a uniform grid of expansion points $\uv_0\in [0.1,0.9]^3$ for amplitude encoding (with some margin to the singular boundaries) and $\uv_0\in [0,1]^3$ for rotational encoding. For readability, monomial labels are incompletely displayed in (a) and replaced by column/row indices in (b,c).}
  \label{fig:decod_mon}
\end{figure}

Raw monomials in $u_\alpha$ are often relevant features for practical applications, such as autoregressive learning of dynamical systems.
Let thus $\Bcal(\uv) = \{1, u_1, \ldots,  u_1 u_2, \ldots, \prod_\alpha u_\alpha^r \}$ be the vector of size $b$ of monomials of $\{u_\alpha\}_{1\leq \alpha\leq D}$ up to degree at most $r$ in each variable.
The order-$r$ Taylor expansion of the Pauli features $\{\phi_j(\uv)\}_{1\leq j\leq q}$ within a small region around some operating point $\uv_0$ (e.g., data mean) can then be written as $\phiv(\uv) \simeq A(\uv_0) \Bcal(\uv-\uv_0)$ with a coefficient matrix $A(\uv_0) \in \mathbb{R}^{q\times b}$. 
In order to measure how accurately a certain monomial in $\Bcal$ can be locally constructed from linear combinations of $\phi_j$, we define a geometric decodability score 
\beq \kappa_j^2 \equiv (A^+ A)_{jj},\qquad \sum_{j=1}^{b} \kappa_j^2 = \rank(A)\leq \min(q,b). 
\label{eq_input_dec_score}\eeq  
This score depends on the operating point $\uv_0$ and is thus not a global reconstruction indicator such as $\Rcal_\Imat^2[y]$.
The score values indicate that construction is possible only within an error $\sim (1-\kappa_j^2)^{1/2}$ due to off-diagonal terms in the $j$-th row/column of the projector $A^+ A$.
A somewhat artificial dependence of $\kappa_j^2$ on the Taylor order $r$ (owing to the rank constraint) can be mitigated by considering large $r$ and $b$. 
Indeed, most of the $\kappa_j^2$ decrease upon increasing $r$, except for a few stable ones. These include monomials that arise directly from the tensor product structure of the encoding state. For amplitude encoding \eqref{eq_enc_amplsqrt} including all $Z$-strings, these are all square-free monomials up to order $n$ (e.g., for $n=D=3$: $\Bcal\st{stable}=\{1,x,y,z, xy, xz, yz, xyz\}$). 

The typical behavior of $\kappa_j^2$ and of the projectors $A^+ A$ (averaged over a range of expansion points) is illustrated in \cref{fig:decod_mon}. 
While the score typically decays with increasing nonlinear degree, rotational encoding tends to have a greater expressivity towards higher-order monomials. 
This is associated with less leakage into off-diagonal terms of $A^+ A$ [see \cref{fig:decod_mon}(c)], and contributes to the robust performance often observed for rotational encoding.

\subsection{Factorization of the nonlinear processing capacity}
\label{sec_cap_factorization}

\begin{figure}[tb!]
  \centering
  \panelimg[width=0.48\textwidth]{a}{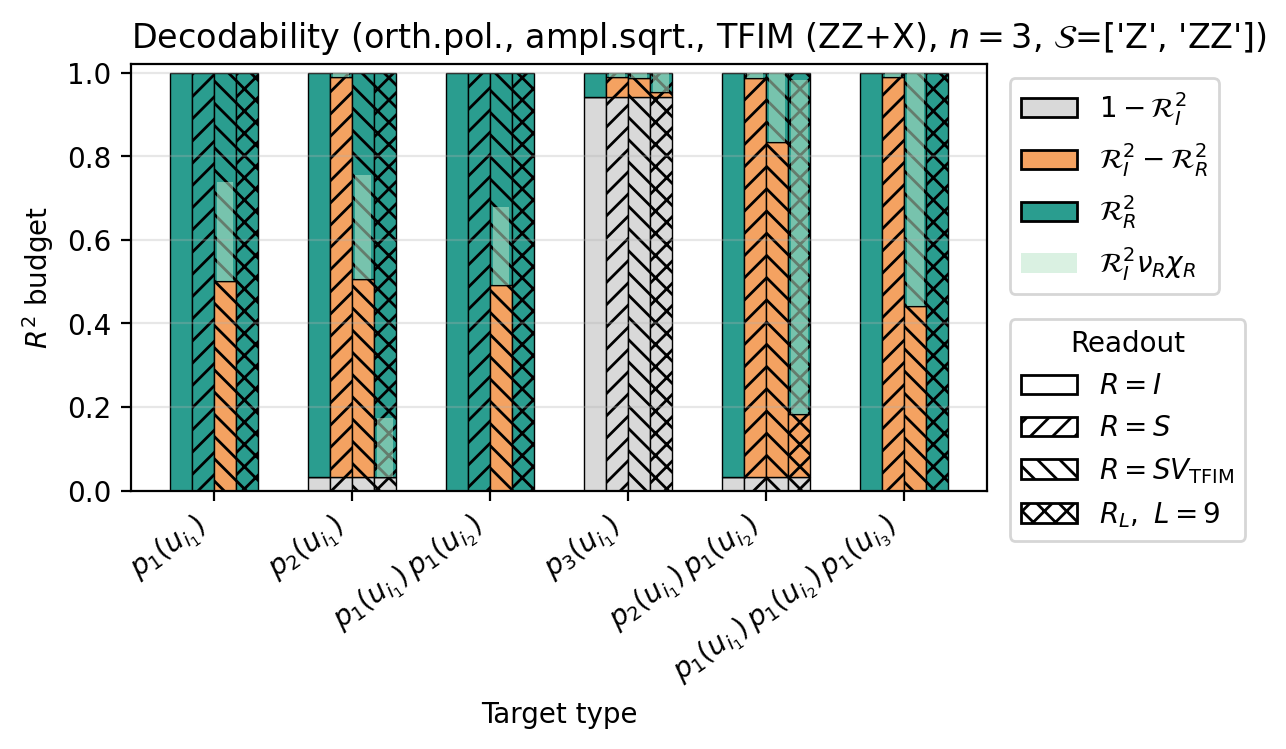}
  \panelimg[width=0.48\textwidth]{b}{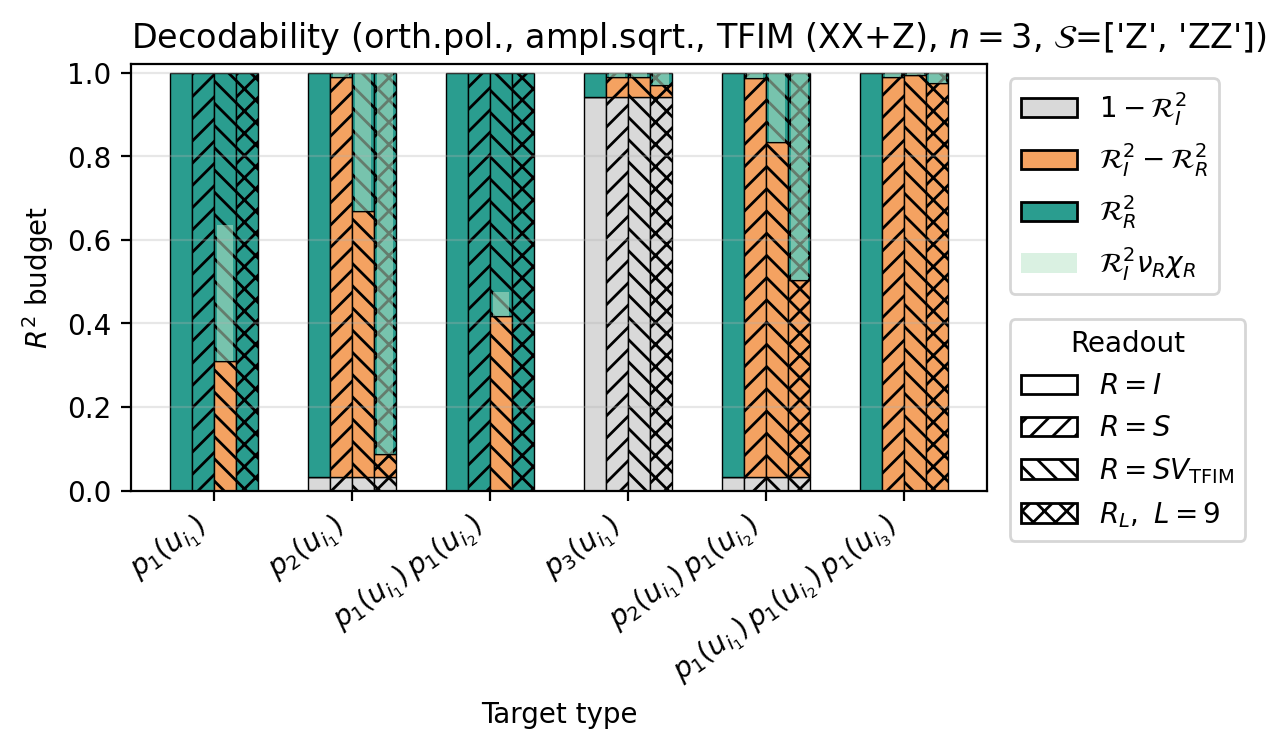}\\
  \panelimg[width=0.48\textwidth]{c}{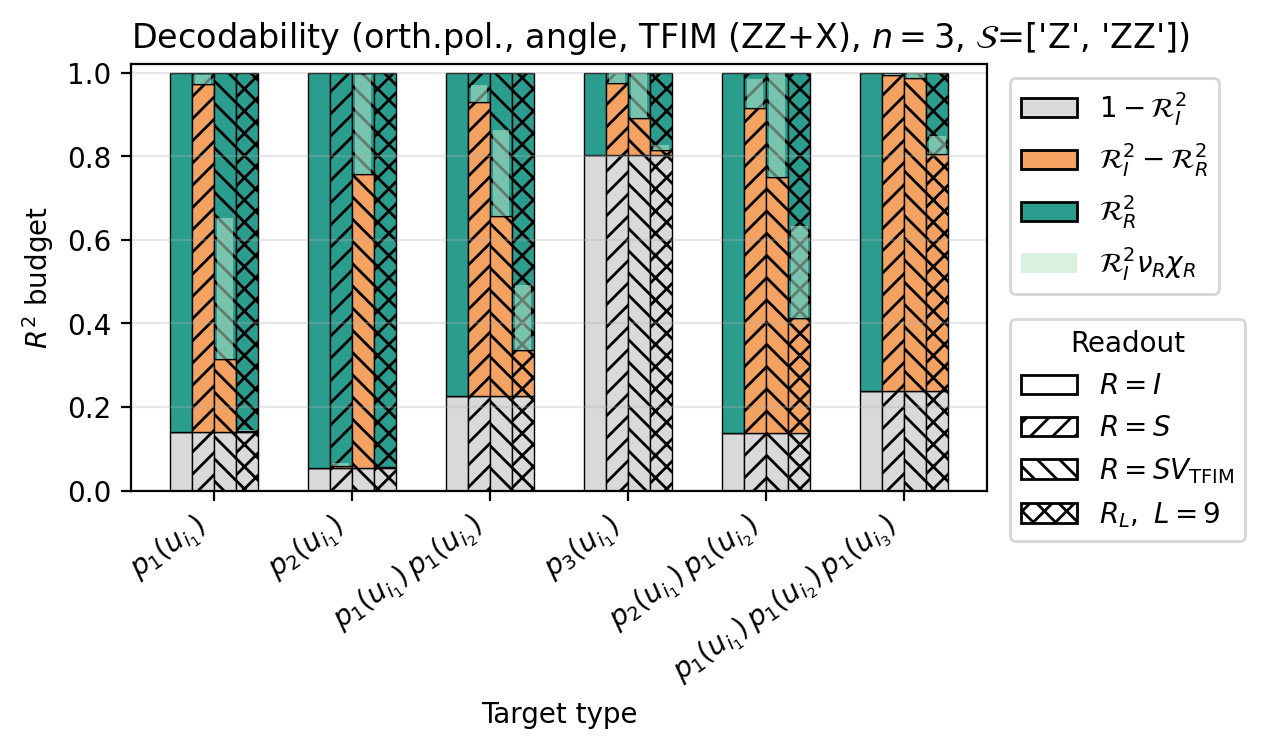}
  \panelimg[width=0.48\textwidth]{d}{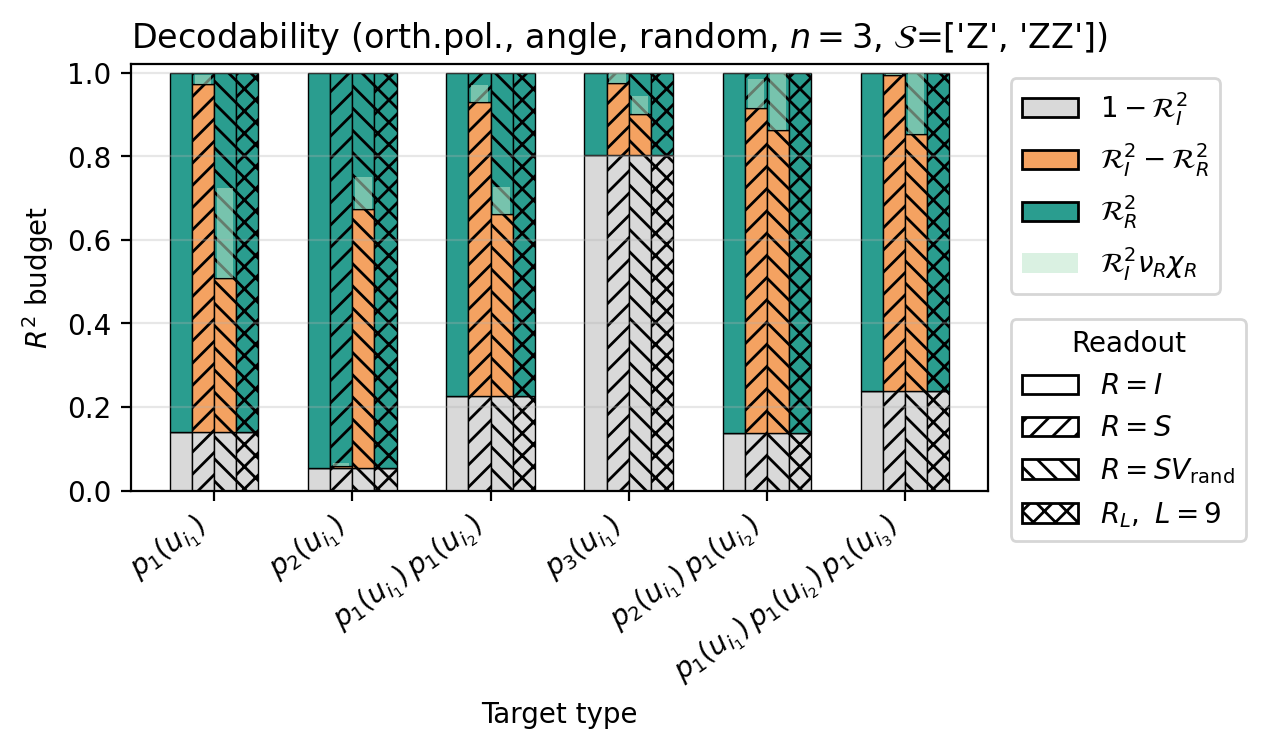}\\
  \caption{Decodability scores $\Rcal_R^2$ for families of orthogonal polynomials $p_j(u_{\alpha_j})$. Following \cref{eq_decod_budget}, the undecodable fraction of the target is decomposed into a part not representable by the encoding ($1-\Rcal_\Imat^2$, gray) and a part not transmitted through the reservoir (``channel loss'', $\Rcal_\Imat^2 - \Rcal_R^2$, orange). Onto the decodable fraction $\Rcal_R^2$ (green) the gain from geometrically hidden but statistically rescued features $\Rcal_\Imat^2\nu_R[y]\chi_R[y]$ is overlayed (after clipping and scaling). The readout is constructed by either: measuring all Pauli observables directly after encoding ($R=\Imat$); just the selected set directly after encoding ($R=S$, here $Z$ and $ZZ$); the selected set after Hamiltonian evolution ($R=S V$); the selected set after temporal multiplexing with the considered Hamiltonian for $\numTM$ steps ($R_\numTM$). We consider amplitude or angle encoding [\cref{eq_enc_amplsqrt,eq_enc_rot} in the form $R_Y(u)$] and various Hamiltonians, as indicated.}
  \label{fig:decod_budget}
\end{figure}

The diagnostics introduced above probe different but interrelated aspects of the information processing capacity of a QELM: a target $y(\uv)$ must be approximated by functions constructible from the encoding-generated Pauli features $\phiv(\uv)$ (\cref{sec_nonlin_readout,sec_mon_decod}), and these features must in turn be decoded from the readout $\Fv=R\phiv$ (\cref{sec_Pauli_decoding}). 
As is shown in \cref{app_decodability}, the nonlinear processing capacity factorizes along this chain [see \cref{eq_cap_factorization}]:
\beq 
\Rcal_R^2[y]=\Rcal_\Imat^2[y]\tau_R[y].
\label{eq_cap_fact}
\eeq 
The ideal factor $\Rcal_\Imat^2[y]$ measures the fraction of the target representable by the encoding-generated Pauli-feature library (\cref{sec_nonlin_readout}), whereas $\tau_R[y]$ measures the fraction of that optimal feature-space representative transmitted through the effective PTM and the selected measurements. 
Both $\Rcal_\Imat^2$ and $\tau_R$ take statistical correlations in the features into account.
For a target aligned with a single isotropic or whitened feature coordinate, the channel factor reduces to the corresponding geometric score, $\gamma_r^2=(R^+R)_{rr}$ or $\widetilde\gamma_r^2=(\Pi_{\widetilde R})_{rr}$ [see \cref{eq_R2_white,eq_decod_score_white}]. 
Analogously, the geometric specialization of the encoding factor for the local Taylor map is the monomial score $\kappa_j^2=(A^+A)_{jj}$ [see \cref{eq_input_dec_score,eq_decod_monomial_geom}]. 
Averaging $\Rcal_R^2[y]$ over the degree-$k$ target family gives $\Rcal_R^2(k)$ [see \cref{eq_nlcap_R2_score}].

The channel transmission $\tau_R$ can be refined further by separating geometric visibility from statistical recoverability (see \cref{app_decodability}):
\beq
\tau_R[y]=1-\nu_R[y]\bigl(1-\chi_R[y]\bigr),
\label{eq_channel_trans_score}\eeq
where $\nu_R[y]$ quantifies how much of the target lies in the nullspace of the PTM, while $\chi_R[y]$ quantifies how much of the target is predicted from statistical correlations of accessible features.
Thus a large channel transmission score can arise in two distinct ways: the task-relevant representative may be geometrically visible to the PTM/readout, corresponding to small $\nu_R[y]$, or it may be geometrically hidden but statistically recoverable from correlated accessible features, corresponding to large $\chi_R[y]$.

From \cref{eq_cap_fact} and the refined channel identity, the normalized target variance can be decomposed as [see \cref{eq_refined_decod_budget}]
\beq 
1=\underbrace{1-\Rcal_\Imat^2[y]}_{\text{not represented by encoding}}
+\underbrace{\Rcal_\Imat^2[y]\nu_R[y]\bigl(1-\chi_R[y]\bigr)}_{\text{hidden and not statistically rescued}}
+\underbrace{\Rcal_R^2[y]}_{\text{decoded}} ,
\label{eq_decod_budget}
\eeq 
where the middle term is equivalent to $\Rcal_\Imat^2[y]-\Rcal_R^2[y]$.
The score $\Rcal_\Imat^2[y]\nu_R[y]\chi_R[y]$ describes the reduction in channel mean-squared error obtained by predicting the geometrically hidden component from the accessible coordinates, relative to the predictor that uses only the directly visible component.

The decodability budget is visualized in \cref{fig:decod_budget} and the stacked bars indicate what the encoding failed to generate, what the encoding generated but the PTM/readout failed to transmit even after statistical rescue, and what is ultimately decodable, either statistically or geometrically.
In general, the reservoir cannot enhance decodability in cases where the target cannot already be represented by the encoding alone.
As expected, if the reservoir hinders decodability of a representable feature, this effect can be mitigated by temporal multiplexing.
Notably, when measuring with Pauli-$Z$ and $ZZ$ observables, decodability can grow or decrease when the reservoir unitary is removed from the model. 
Moreover, comparing \cref{fig:decod_budget}(a) and (b), the reduced geometric decodability of features for certain structured Hamiltonians observed in \cref{fig:feat_iso} is also reflected by the statistical counterpart (see also \cref{fig:decod_budget_rawZ} for the monomial case).

\begin{table}[tb]
  \centering
  \renewcommand\arraystretch{1.45}
  \begin{tabular}{l|c|c|c}
      \hline
       & \textbf{encoding} & \textbf{channel} & \textbf{composite} \\
      \hline
      geometric, per direction
      &
      $\kappa_j^2=(A^+A)_{jj}$
      &
      $\gamma_r^2=(R^+R)_{rr}$
      &
      --
      \\
      \hline
      statistical, per task $y$
      &
      $\Rcal_\Imat^2[y]$
      &
      $\tau_R[y]$
      &
      $\Rcal_R^2[y]=\Rcal_\Imat^2[y]\tau_R[y]$
      \\
      \hline
      refined channel split
      &
      --
      &
      $\tau_R[y]=1-\nu_R[y]\bigl(1-\chi_R[y]\bigr)$
      &
      $\Rcal_R^2[y]
      =
      \Rcal_\Imat^2[y]
      \left[1-\nu_R[y]\bigl(1-\chi_R[y]\bigr)\right]$
      \\
      \hline
      statistical rescue
      &
      --
      &
      $\nu_R[y]\chi_R[y]$
      &
      $\Rcal_\Imat^2[y]\nu_R[y]\chi_R[y]$
      \\
      \hline
  \end{tabular}
  \caption{Summary of the decodability diagnostics. The encoding factor $\Rcal_\Imat^2[y]$ measures how much of a target is represented by the encoding-induced Pauli-feature library, while the channel factor $\tau_R[y]$ measures how much of this representative is transmitted by the effective PTM and readout. The refined decomposition [see \cref{eq_decod_budget}] separates the channel loss into the nullspace load $\nu_R[y]$ [\cref{eq_nullspace_load}] and the statistically rescued fraction $\chi_R[y]$ [\cref{eq_statistical_rescue_score}].}
  \label{tab:cap_factorization}
\end{table}

\subsection{Summary and Discussion}
\label{sec_decod_summary}

Decodability of Pauli features can be understood as the underlying concept that determines the expressivity of a QELM with initial-state encoding and data-independent reservoir dynamics. 
The decodability score $\Rcal^2$ corresponds to the (nonlinear) information processing capacity \cite{dambre_information_2012} of the QELM, and we have shown here that it separates into two parts: (i) an \emph{encoding part} $\Rcal_\Imat^2[y]$ (geometric proxies: $\kappa_j^2$, $A$), measuring whether the encoding-generated dictionary of Pauli features can build the monomials and orthogonal polynomials relevant for the task, and (ii) a \emph{channel part} $\tau_R[y]$ (with geometric proxies $\gamma_r^2$, $\tilde\gamma_r^2$, $\rank\, R$), measuring the invertibility of the effective PTM on the task-relevant feature directions, conditioned on the encoding distribution. 
The channel score receives contributions both from features that are decodable purely algebraically and from features that are algebraically invisible but still decodable via statistical correlations.
The different notions of decodability are summarized in \cref{tab:cap_factorization}. 

Nonlinear processing capacity has been found to track Krylov observability very closely \cite{cindrak_engineering_2025}, which is explained by the present framework in terms of the PTM-induced mixing of the fundamental Pauli features.
High decodability scores are associated with high-rank effective PTMs $R$, which are achieved for random unitaries.
Strongly mixing quantum channels can help expressivity and nonlinear processing in QELMs \cite{gotting_exploring_2023,vetrano_state_2025,lorenzis_entanglement_2025}, and are thus often beneficial for training performance apart from concentration effects \cite{xiong_fundamental_2025}.
By contrast, QRCs with internal memory states often work best at a point where random matrix behavior just begins to dominate (``edge of chaos'') (see, e.g., \cite{martinez-pena_dynamical_2021,xia_reservoir_2022,ivaki_quantum_2025}).
An alternative to increasing randomness is to devise task-aligned reservoir structures and measurement schemes \cite{sakurai_quantum_2022,hayashi_impact_2023,palacios_role_2024,lorenzis_entanglement_2025,gross_kernelbased_2026}.
Knowing the connection between the readout and the encoding-generated features via the PTM framework can thus inform feature engineering strategies to enhance expressivity.

Before proceeding, it is useful to also briefly discuss the \emph{injectivity} property of the feature map \cite{martinez-pena_inputdependence_2025}, which ensures that distinct inputs map to distinct readouts.
A sufficient condition for local injectivity of $\Fv(\uv)=R\phiv(\uv)$ [\cref{eq_qelm_classical}] is a full-rank Jacobian:
\beq \rank\, R J_{\phiv}(\uv)=D, \qquad  J_{\phiv}(\uv) = \frac{\pd \phiv(\uv)}{\pd\uv} \in \reals^{\numF\times D}.
\label{eq_inject_cond}\eeq 
For the considered product state encodings it is straightforward to check that $\phiv(\uv)$ is injective on the respective domains in $\reals^D$ (possibly after removing one endpoint).
Hence, local injectivity of $\Fv(\uv)$ is determined by whether $R$ is injective on the tangent space $\operatorname{range}\,J_{\phiv}(\uv)$, i.e. whether 
\beq \nullsp\, R\cap \operatorname{im}J_{\phiv}(\uv)=\{0\}.
\label{eq_inject_crit}
\eeq  
In a generic setting, this condition is satisfied (at fixed $\uv$) if $\rank\, R\ge D$, which is the case for adequate measurement schemes and sufficiently scrambling dynamics.


\section{Interpretable learning of dynamical systems}
\label{sec_numerics}

\subsection{Time series modeling with QELMs}
The classical representation \eqref{eq_qelm_classical} allows us to identify the surrogate model a QELM learns during the training process.  
Specifically, for the product state encodings considered in \cref{sec_encodings}, the function library $\phiv:\mathbb{R}^D\to\mathbb{R}^{\numF}$ consists of $\numF = 3^n$ monomials of the single-Pauli features $\phi_k$ (recall that observables containing Pauli-$Y$ factors have vanishing expectation values, see \cref{tab:enc_proj}).
We consider the standard problem of learning to forecast a chaotic dynamical system by training on its sampled trajectory $\uv_t=\uv(t_0+t\Delta t)\in\mathbb{R}^D$. 
The continuous and discrete evolutions are given by
\beq \text{dynamical system:}\qquad  \dot \uv = \gv(\uv),\qquad \uv_{t+1} = \psiv_\gv(\uv_t),
\label{eq_flow_map}\eeq
where $\psiv_\gv(\uv) = \uv + \Delta t \gv(\uv) + \mathcal{O}(\Delta t^2)$ is the flow map (propagator) that can be computed from the vector field $\gv(\uv)\in \mathbb{R}^D$ via a Taylor/Lie series expansion:
\beq 
\psiv_\gv(\uv)= \exp(\Delta t\,\mathcal{L}_\gv)\,\uv=\sum_{k=0}^\infty \frac{\Delta t^k}{k!}\,(\mathcal{L}_\gv^k\uv),\qquad (\mathcal{L}_\gv f)(\xv)=\gv(\xv)\cdot\nabla f(\xv).
\label{eq_flow_map_exp}\eeq 
The non-integrable nature and fractal attractors of chaotic systems make them natural benchmarks for system identification, which addresses the question to which extent the underlying data-generating process can be uniquely discovered from the data \cite{ljung_system_1999,shumaylov_when_2025}.

If the dynamics is Markovian, it can be modeled solely with an NG-RC or QELM based on an instantaneous feature map (see \cite{gross_kernelbased_2026} for examples).
Accordingly, we formulate the training problem of the QELM \eqref{eq_qelm_classical} as 
\beq \text{QELM:}\qquad \uv_{t+1} = \fv(\uv_t) = W^\top R \phiv(\uv_t)
\label{eq_qelm_timeser}\eeq
and seek to learn an approximation to the flow map $\psiv_\gv$ via minimizing the least-squares loss \eqref{eq_QELM_opt} on the dataset $\{\xv=\uv_t,\yv=\uv_{t+1}\}_{t=1}^P$. The generalization of \eqref{eq_QELM} to multiple outputs is straightforward and involves training a separate predictor $f_r$ for each output dimension $r$ via the weight matrix $W = [\wv_1,\ldots,\wv_D]\in \reals^{\numMtot\times D}$.

We use the Lorenz-63 and Halvorsen models as benchmarks ($D=3$ dimensions) \cite{sprott_elegant_2010}:
\beq
\label{eq_lorenz_63_halvorsen}
  \text{Lorenz-63:} \; \begin{cases} \dot x = \sigma(y-x) \\ \dot y = x(\rho-z)-y \\ \dot z = xy - \beta z \end{cases}
  \qquad
  \text{Halvorsen:} \; \begin{cases} \dot x = -ax - 4y - 4z - y^2 \\ \dot y = -ay - 4z - 4x - z^2 \\ \dot z = -az - 4x - 4y - x^2 \end{cases}
\eeq
with parameters $\sigma=10$, $\beta=8/3$, $\rho=28$ for Lorenz-63, and $a=1.4$ for Halvorsen. This setup leads to chaotic behavior with largest Lyapunov exponents $\lambda_{\text{max}}\ut{Lor63}\approx 0.89$ and $\lambda_{\text{max}}\ut{Halv}\approx 0.76$, respectively.
The polynomial form of the vector field in \eqref{eq_lorenz_63_halvorsen} then suggests to use amplitude encoding \eqref{eq_enc_amplsqrt} with $n=3$ qubits, leading to a library of the following 27 terms:
{\scriptsize
\begin{align}
  \phiv\st{amp}^{n=3}(\uv) = \Bigl\{& 1,2 \sqrt{(1-x) x},2 x-1,2 \sqrt{(1-y) y},2 (2 x-1) \sqrt{-((y-1) y)},4 \sqrt{(x-1) x (y-1) y},2 y-1, \nonumber \\
  & 2 \sqrt{-((x-1) x)} (2 y-1),(2 x-1) (2   y-1),2 \sqrt{(1-z) z},2 (2 x-1) \sqrt{-((z-1) z)},\nonumber \\
  & 2 (2 y-1) \sqrt{-((z-1) z)}, 2 (2 x-1) (2 y-1) \sqrt{-((z-1) z)},4 \sqrt{(x-1) x (z-1) z},\nonumber \\
  &4 (2 y-1)   \sqrt{(x-1) x (z-1) z}, 4 \sqrt{(y-1) y (z-1) z},4 (2 x-1) \sqrt{(y-1) y (z-1) z},\label{eq_lib_amplsqrt} \\ 
  & 8 \sqrt{-((x-1) x (y-1) y (z-1) z)},2 z-1,2 \sqrt{-((x-1) x)} (2
   z-1),(2 x-1) (2 z-1), \nonumber \\
   & 2 \sqrt{-((y-1) y)} (2 z-1),2 (2 x-1) \sqrt{-((y-1) y)} (2 z-1),4 (2 z-1) \sqrt{(x-1) x (y-1) y}, \nonumber \\ 
   & (2 y-1) (2 z-1),2 \sqrt{-((x-1) x)} (2 y-1) (2 z-1),(2 x-1) (2 y-1) (2 z-1) \Bigr\} \nonumber 
\end{align}}
We will also consider the amplitude encoding scheme \eqref{eq_enc_amplsq}, which has a similar polynomial structure.

In order to confine the trajectory to the interval $\uv\in [0,1]^3$ or $[-1,1]^3$, as required by the encoding, we apply the transformation $\uv\to \uv'=\boldsymbol{\alpha}\odot (\uv-\mv)$ with $\boldsymbol{\alpha},\mv\in \reals^3$ to the state (we typically use the same $\alpha$ for all coordinates).
This leads to the rescaled Lorenz-63 model:
{\scriptsize
\begin{align}
    \dot u &= -\sigma u + \sigma \frac{\alpha_x}{\alpha_y} v + \sigma \alpha_x (m_y - m_x), \nonumber \\
    \dot v &= -v + \frac{\alpha_y}{\alpha_x}(\rho - m_z) u - \frac{\alpha_y}{\alpha_z} m_x w - \frac{\alpha_y}{\alpha_x \alpha_z} u w + \alpha_y m_x (\rho - m_z) - \alpha_y m_y, \label{eq_lor63_rescaled} \\
    \dot w &= -\beta w + \frac{\alpha_z}{\alpha_x \alpha_y} u v + \frac{\alpha_z}{\alpha_x} m_y u + \frac{\alpha_z}{\alpha_y} m_x v + \alpha_z (m_x m_y - \beta m_z), \nonumber 
\end{align}}
and the rescaled Halvorsen model:
{\scriptsize
\begin{align}
    \dot u &= -a u - 4 \frac{\alpha_x}{\alpha_y} v - 4 \frac{\alpha_x}{\alpha_z} w - 2 \frac{\alpha_x m_y}{\alpha_y} v - \frac{\alpha_x}{\alpha_y^2} v^2 + \alpha_x(-a m_x - 4 m_y - 4 m_z - m_y^2), \nonumber \\
    \dot v &= -a v - 4 \frac{\alpha_y}{\alpha_z} w - 4 \frac{\alpha_y}{\alpha_x} u - 2 \frac{\alpha_y m_z}{\alpha_z} w - \frac{\alpha_y}{\alpha_z^2} w^2 + \alpha_y(-a m_y - 4 m_z - 4 m_x - m_z^2), \label{eq_halvorsen_rescaled} \\
    \dot w &= -a w - 4 \frac{\alpha_z}{\alpha_x} u - 4 \frac{\alpha_z}{\alpha_y} v - 2 \frac{\alpha_z m_x}{\alpha_x} u - \frac{\alpha_z}{\alpha_x^2} u^2 + \alpha_z(-a m_z - 4 m_x - 4 m_y - m_x^2). \nonumber 
\end{align}
}
This vector field-level transformation simplifies comparison between the learned coefficients and the true ones.

\subsection{Numerical results}
We use here a complete measurement operator set of $4^n$ Pauli observables. In this setup, the reservoir unitary becomes immaterial [see \cref{eq_ols_weights_unitary}].
We generate trajectories using a 5th-order Runge--Kutta integrator with a time step of $\Delta t=0.01$. Reported values of the training error and forecast horizon are averaged over multiple initial conditions.
After training the QELM, we obtain a classical surrogate model $\hat \fv(\uv)$ by inserting into \cref{eq_qelm_classical} the learned weights $\hat \wv$.

The RMS training error
\beq \epsilon\st{train}=\Big(\frac{1}{P}\sum_{t=1}^P \|\uv_{t+1} - \hat\fv(\uv_t)\|^2\Big)^{1/2}
\label{eq_train_err}\eeq 
measures one-step accuracy of the trained predictor $\hat \fv$. 
As a measure of multi-step prediction accuracy, the autonomously generated trajectory $\hat\uv_{t+1} = \hat \fv(\hat \uv_t)$ is compared to the true trajectory via the forecast horizon
\beq T\st{fch}/\Delta t = \inf \left\{ t : \| \uv_t - \hat\uv_t \| > \varsigma \right\},
\label{eq_fc_hor}\eeq 
where $\bar \uv = \frac{1}{P}\sum_{t=1}^P \uv_{t}$ and $\varsigma^2= \frac{1}{P}\sum_{t=1}^P \| \bar \uv - \uv_{t}\|^2$ are the mean and variance of the trajectory, respectively. 
We average $T\st{fch}$ over multiple starting points $\uv_0\in \mathbb{R}^D$ and express it in units of the Lyapunov time $T_L=1/\lambda\st{max}$.

\begin{figure}[tb]
  \centering
  \includegraphics[width=0.65\textwidth]{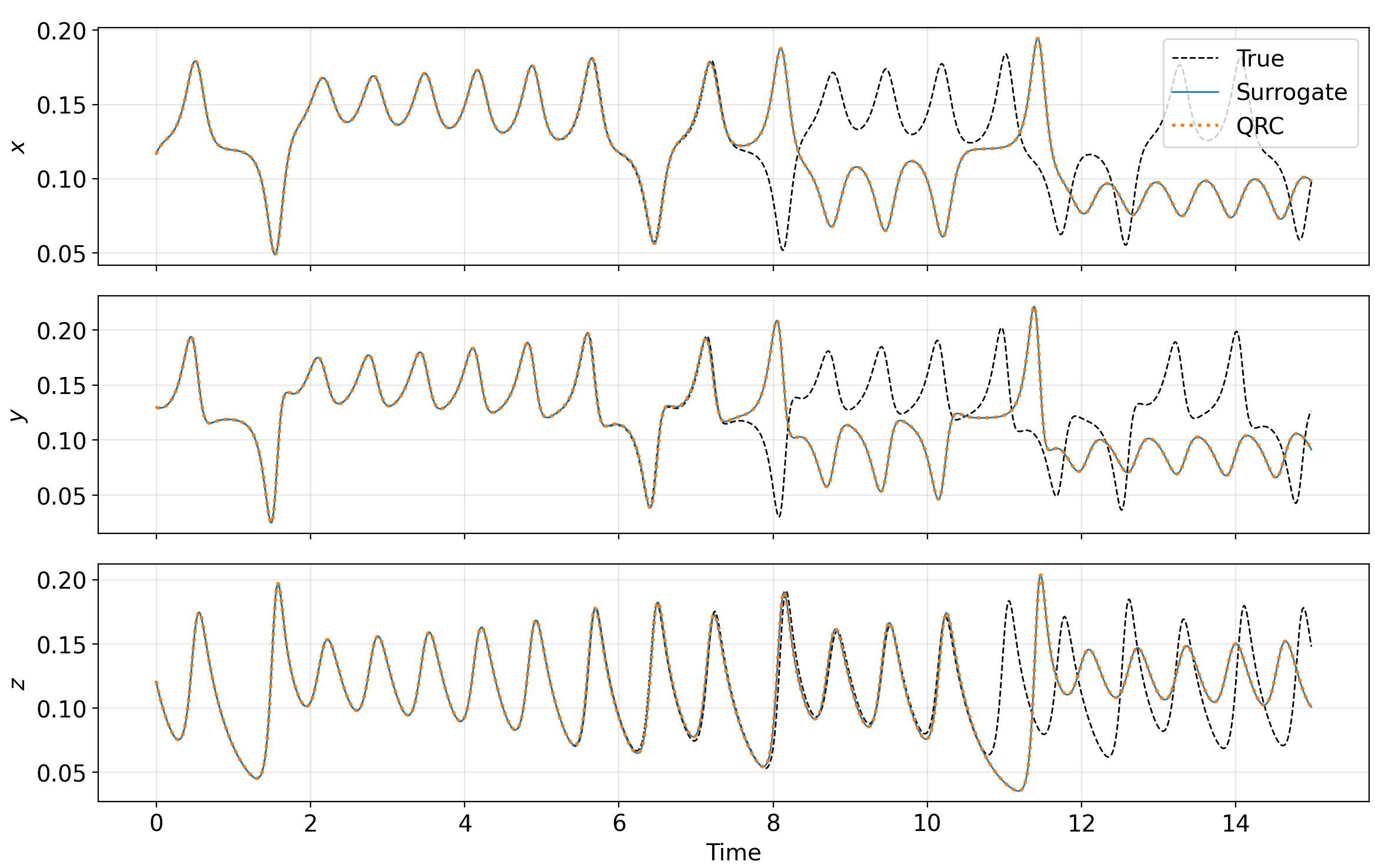}
  \caption{QELM vs.\ its classical representation: The forecast obtained from a QELM [\cref{eq_QELM}] trained on the rescaled Lorenz-63 model is matched exactly by the forecast of its iterated classical (surrogate) representation [\cref{eq_qelm_classical}]. Both eventually diverge from the true trajectory due to the Lyapunov instability. The QELM employs amplitude encoding \eqref{eq_enc_amplsqrt} with $n=3$ qubits. Ground-truth trajectories are generated with a resolution of $\Delta t=0.01$. Time is given in model time units. For the scaling transformation in \cref{eq_lor63_rescaled}, we use parameters $\alpha_{x,y,z}=0.004$ and $\mv=[-30,-30,-5]$.}
  \label{fig:surro_forec_lor63}
\end{figure}

\begin{table}[b]
  \centering
  \begin{tabular}{l| l| l| l | l | c | c}
    \hline
    System & type & QELM setup & $\alpha$ & $\lambda$ & $\epsilon\st{train}$ & $T\st{fch}/T_L$ \\
    \hline
    Lorenz-63 & std. & ampl. enc. $[0,0.3]$, \cref{eq_enc_amplsqrt} & - & $10^{-6}$ & $10^{-5}$ & $4.3\pm 1.7$ \\
    Lorenz-63 & resc. & ampl. enc. $[0,1]$, \cref{eq_enc_amplsqrt} & 0.005 & $10^{-8}$ & $10^{-6}$ & $6.2\pm 1.4$ \\
    Halvorsen & std. & ampl. enc. $[0.1,0.8]$, \cref{eq_enc_amplsqrt} & - & $10^{-8}$ & $10^{-4}$ & $2.3\pm 0.8$ \\
    Halvorsen & std. & \ldots + add squares to readout vector & - & $10^{-8}$ & $10^{-5}$ & $5.8\pm 1.7$ \\
    Halvorsen & resc. & ampl. enc. $[0,1]$, \cref{eq_enc_amplsqrt} & 0.027 & $10^{-7}$ & $10^{-4}$ & $3.4\pm 1.0$ \\
    Halvorsen & resc. & \ldots + add squares to readout vector & 0.027 & $10^{-7}$ & $10^{-6}$ & $6.8\pm 1.7$ \\
    Halvorsen & std. & ampl. enc. $[-0.01,0.01]$, \cref{eq_enc_amplsq} & - & $10^{-15}$ & $10^{-7}$ & $5.9\pm 1.2$ \\
    Halvorsen & resc. & ampl. enc. $[-1,1]$, \cref{eq_enc_amplsq} & 0.001 & $10^{-15}$ & $10^{-7}$ & $5.9\pm 1.5$ \\
    \hline
  \end{tabular}
  \caption{Training RMS errors $\epsilon\st{train}$ [\cref{eq_train_err}] and forecast horizons $T\st{fch}$ [\cref{eq_fc_hor}] for the benchmark systems and (unitary-free) QELM setups discussed in the text. We generally use $n=3$ input qubits (no delay states) and a time resolution of $\Delta t=0.01$. $T\st{fch}$ with its standard deviation (not error of the mean) is given in units of the Lyapunov time $T_L$. $\lambda$ denotes the Ridge regularization parameter. The column `type' indicates whether we use the standard [\cref{eq_lorenz_63_halvorsen}] or rescaled dynamical system. In the former case, we uniformly scale the trajectories to the interval reported in the 3rd column. In the latter case, we solve the dynamical system in \cref{eq_lor63_rescaled,eq_halvorsen_rescaled} with the rescaling factor $\alpha$ (same for all state coordinates).}
  \label{tab:qelm_trainerr_fch}
\end{table}

Results of training a QELM in various setups are summarized in \cref{tab:qelm_trainerr_fch}. 
We emphasize that our aim here is not to beat previous benchmarks, but to perform controlled interpretability experiments. We thus keep the setup deliberately simple. Significant improvements in training error and forecast horizon can be achieved by further optimizing the encoding and feature-engineering of the readout.
\Cref{fig:surro_forec_lor63} demonstrates consistency between the autonomously generated prediction of the trained QELM and the iterated classical model, exemplified for the rescaled \emph{Lorenz-63} model.
Both remain close to the true trajectory for a time $T\st{fch}/T_L \approx 6$, when the intrinsic Lyapunov instability of the chaotic system leads to a divergence, triggered by errors in the initial condition, discretization, and extra terms in the surrogate model (see below) \footnote{We find that training on increments via $\uv_{t+1}-\uv_{t} = \fv(\uv_t)$ does not significantly improve the forecast horizon, unless one uses small integrator time steps $\Delta t\lesssim 0.001$.}.

When repeating the experiments with the \emph{Halvorsen} model \eqref{eq_halvorsen_rescaled}, we obtain slightly lower forecast horizons of $T\st{fch}/T_L\approx 3$ and RMS training errors of $\epsilon\st{train}\approx 10^{-4}$. 
One reason for this reduced performance compared to the Lorenz-63 model lies in the absence of squares of the state variables $\{x,y,z\}$ in the library \eqref{eq_lib_amplsqrt}. 
These are crucial for the Halvorsen equations \eqref{eq_halvorsen_rescaled}, whereas the Lorenz-63 model \eqref{eq_lor63_rescaled} depends only on cross terms of the state variables.
Indeed, when enhancing the readout vector by its squares, prediction accuracy for the Halvorsen model improves significantly (see \cref{tab:qelm_trainerr_fch}).
Alternatively, a similar performance can also be achieved when using the amplitude encoding variant in \eqref{eq_enc_amplsq}, which already involves squares of the state variables.
In the context of reservoir computing it is, of course, well known that prediction performance can be enhanced by feature engineering \cite{yang_design_2018,gauthier_next_2021,steinegger_predicting_2025}.
The present analysis exemplifies that, within the PTM framework, this optimization can be systematically phrased as an inverse problem, where the structure of the ground-truth model dictates the required terms in the encoding.

\begin{figure}[bp]
  \centering
  \panelimg[width=0.42\textwidth]{a}{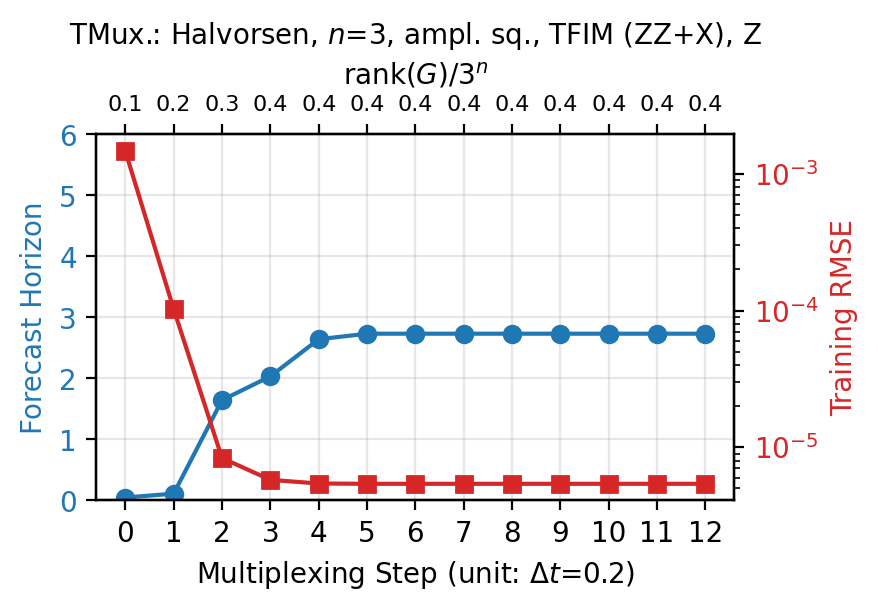}\qquad
  \panelimg[width=0.42\textwidth]{b}{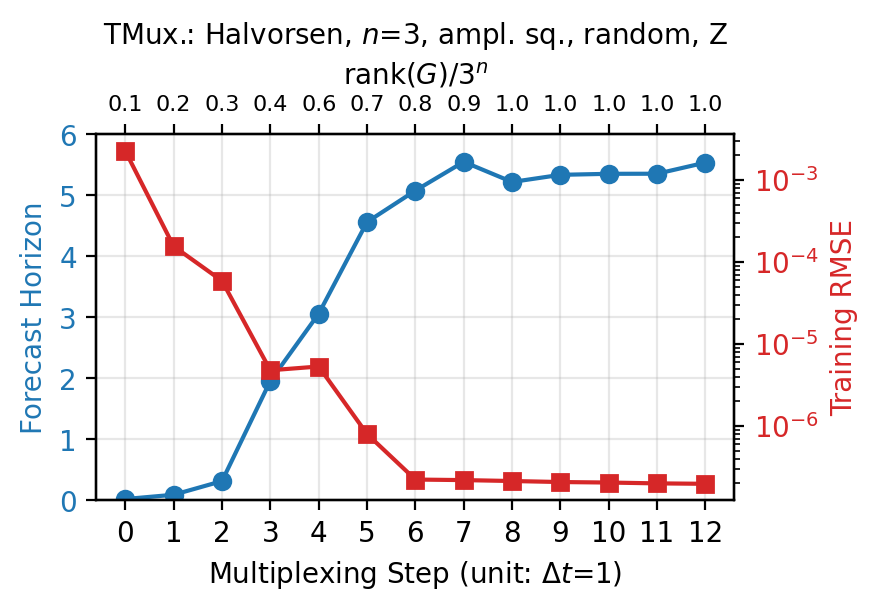}
  \caption{Prediction performance (forecast horizon in units of the Lyapunov time $T_L$, and RMS training error) under temporal multiplexing of a QELM trained on the Halvorsen model \eqref{eq_halvorsen_rescaled}. We use amplitude encoding \eqref{eq_enc_amplsq} with $n=3$ qubits, single-qubit Pauli-$Z$ observables, and (a) the TFIM Hamiltonian $H\st{TFIM}\ut{zz-x}$ [\cref{eq_H_TFIM_zzx}] or (b) a random Hamiltonian. The top horizontal axis shows the numerically obtained rank of the readout matrix $G$ [\cref{eq_gram_matrix_readout}] relative to the maximum achievable rank $q=3^n$ [\cref{eq_gram_rank}]. \emph{No additional features} are added to the readout. For the TFIM variant $H\st{TFIM}\ut{xx-z}$ [\cref{eq_H_TFIM_xxz}], we obtain a maximum rank of $\rank(G)/3^n\simeq 0.22$, RMSE $\epsilon\st{train}\simeq 7\times 10^{-4}$, and forecast horizon $T\st{fch}\simeq 0$.}
  \label{fig:tmux_training}
\end{figure}

When employing \emph{temporal multiplexing}, the rank of the readout matrix $G$ [\cref{eq_gram_matrix_readout}] can serve as an indicator of the training accuracy, as illustrated in \cref{fig:tmux_training}. 
This rank is upper bounded by the number of nonzero Pauli features $\numF$, which, for the encoding \eqref{eq_enc_amplsq} used here, is $\numF=3^n$. 
When conditioning is controlled and exponential concentration effects are avoided \cite{xiong_fundamental_2025,schutte_expressivity_2025,sannia_exponential_2025}, achieving $\rank(G) \simeq \numF$ makes essentially all Pauli features available to the readout and can thus improve performance. 
Indeed, the QELM in \cref{fig:tmux_training}(b) achieves training error and forecast horizon essentially identical to the ones reported in \cref{tab:qelm_trainerr_fch} obtained with a complete Pauli observable set. 
By contrast, \cref{fig:tmux_training}(a) shows that performance remains poor when the readout utilizes only a few effective observables.
In the present setting, this is a consequence of using a structured Hamiltonian like the TFIM [\cref{eq_H_TFIM_zzx}], which spreads observables less effectively over the full operator space, so that $\rank(G)\simeq\rank(R)\ll q$ [see \cref{eq_gram_ptm_rank,fig:feat_iso}]. 
This effect is most pronounced when using the specific TFIM variant $H\st{TFIM}\ut{xx-z}$ [\cref{eq_H_TFIM_xxz}] with $Z$-type observables. Numerical experiments confirm the close relation \eqref{eq_gram_ptm_rank} between the readout and PTM rank.

\subsection{Interpretability}
\begin{figure}[tb!]
  \centering
  \panelimg[width=0.7\textwidth]{a}{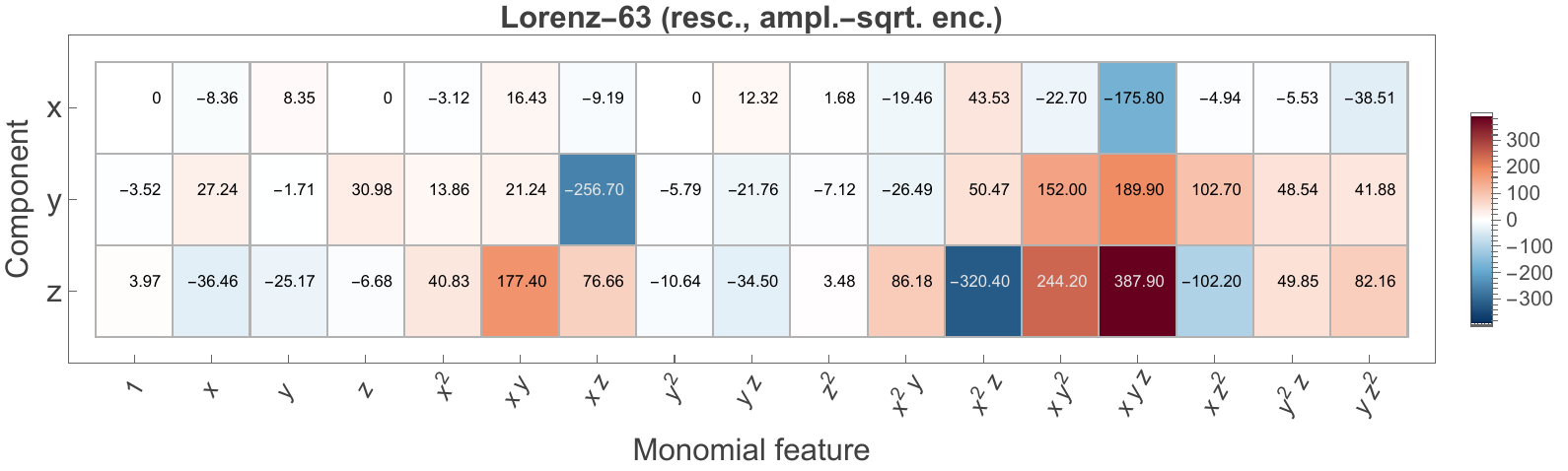}
  \panelimg[width=0.7\textwidth]{b}{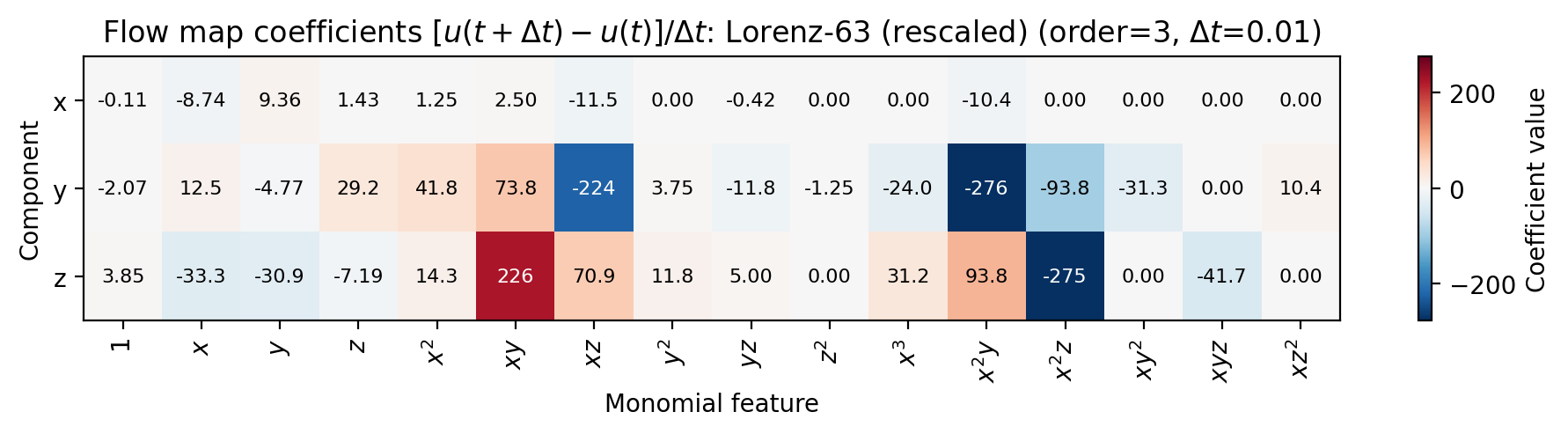}
  \panelimg[width=0.7\textwidth]{c}{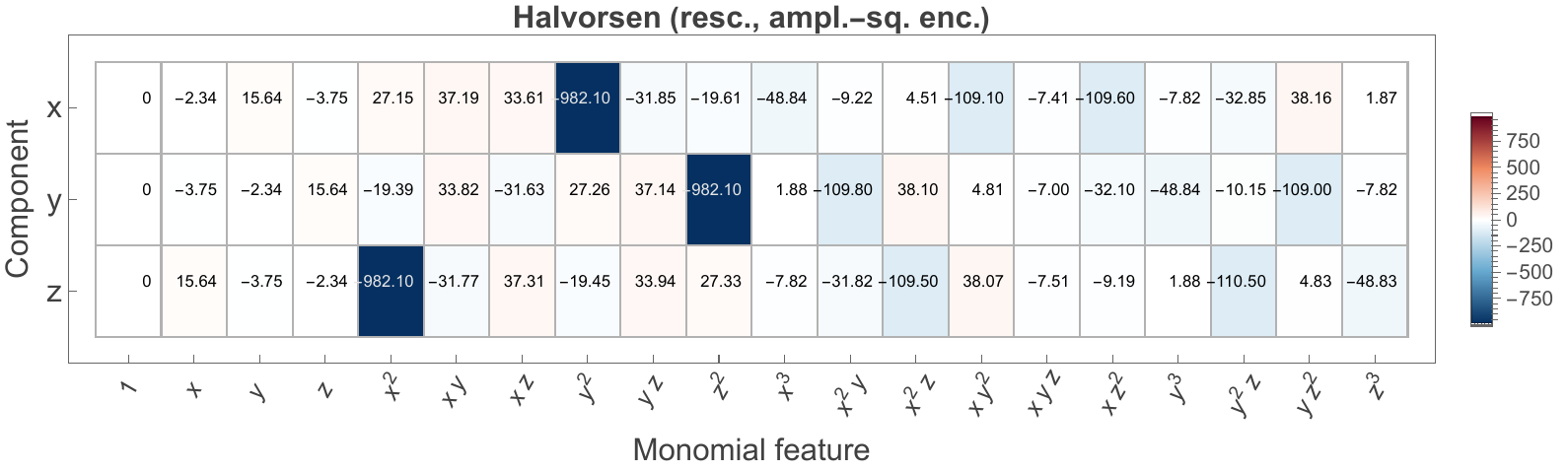}
  \panelimg[width=0.73\textwidth]{d}{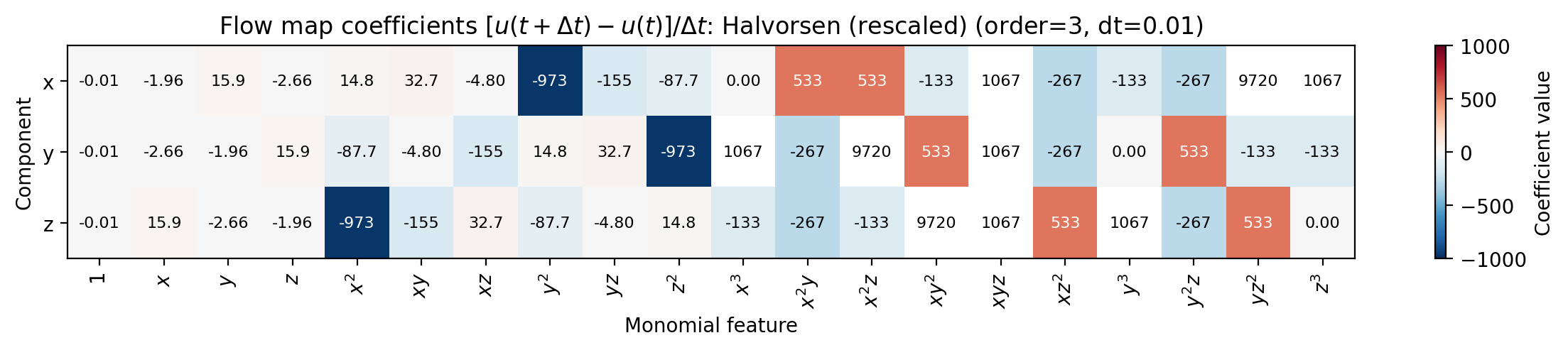}
  \caption{Illustration of the flow map learned by a QELM, obtained from the Taylor-approximation approach \cref{eq_qelm_flowmap}. The QELM is trained on (a) the Lorenz-63 model \eqref{eq_lor63_rescaled} and (c) the Halvorsen model \eqref{eq_halvorsen_rescaled}, using the \emph{amplitude} encoding variants \eqref{eq_enc_amplsqrt} and \eqref{eq_enc_amplsq}, respectively. The rescaling parameters are here $\mv=[-30,-30,-5], \alpha_{x,y,z}=0.004$ (Lorenz-63) and $\mv=[-10,-10,-10],\alpha_{x,y,z}=0.001$ (Halvorsen). The corresponding true flow map increments $(\psiv_\gv(\uv)-\uv)/\Delta t$ [\cref{eq_flow_map}], expanded to third order in the state variables $u_j$, are shown in (b,d) for comparison (terms identically zero are omitted). }
  \label{fig:flowmap}
\end{figure}

\begin{figure}[tb!]
  \centering
  \panelimg[width=0.7\textwidth]{a}{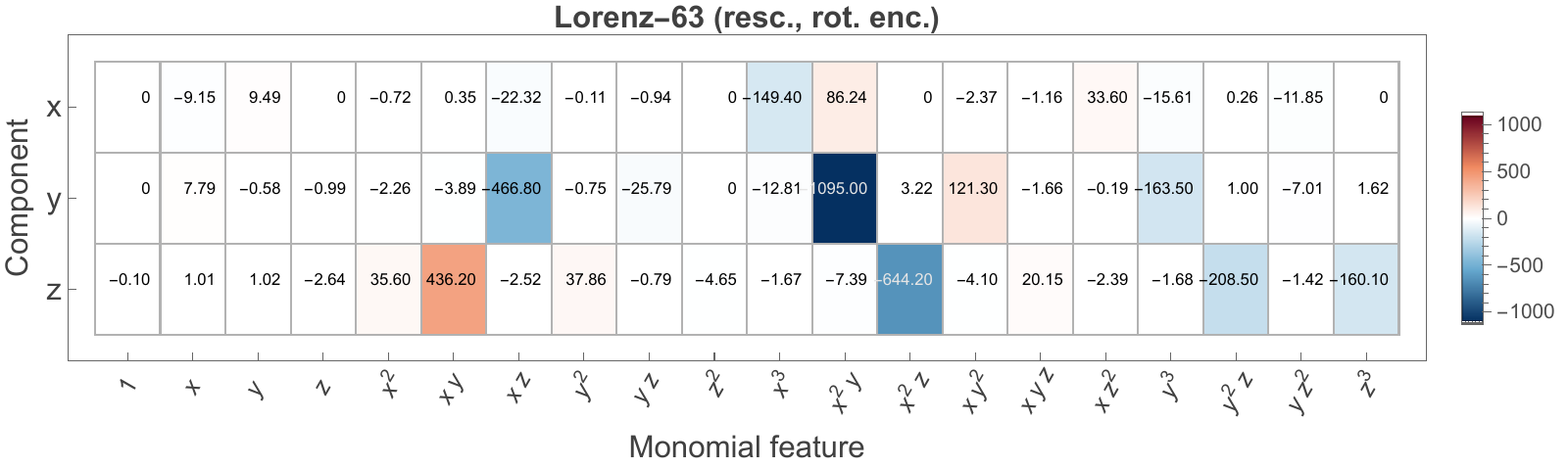}
  \panelimg[width=0.7\textwidth]{b}{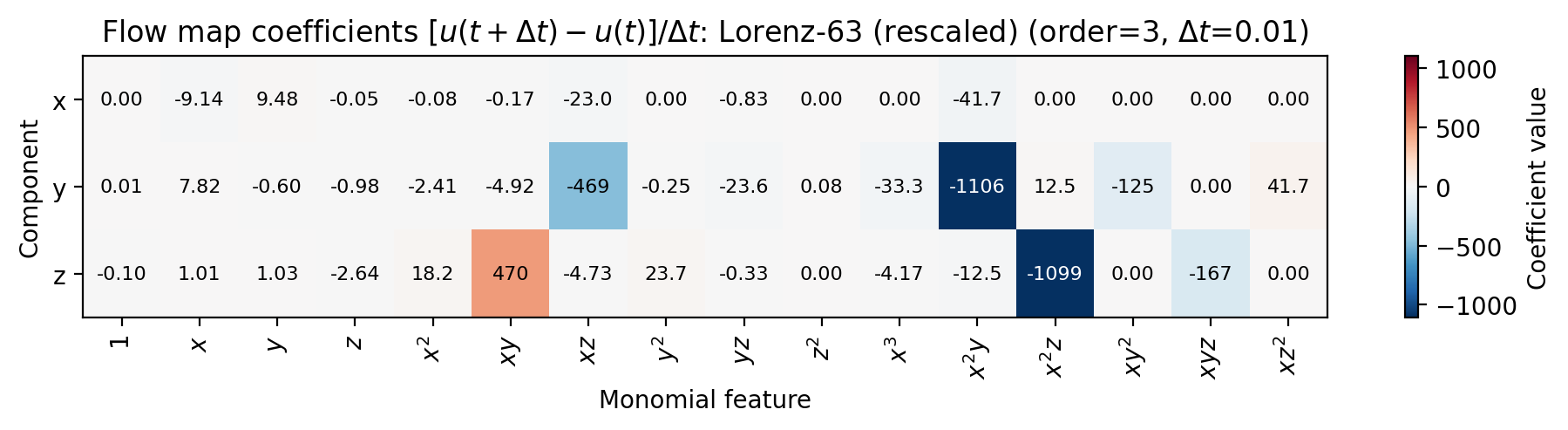}
  \panelimg[width=0.7\textwidth]{c}{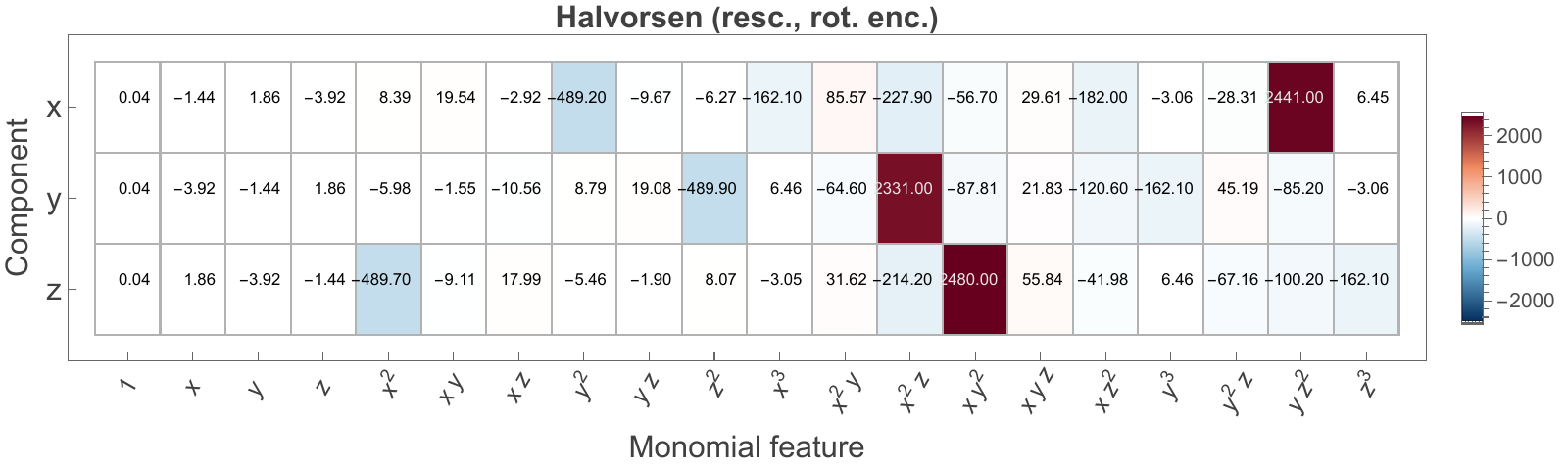}
  \panelimg[width=0.73\textwidth]{d}{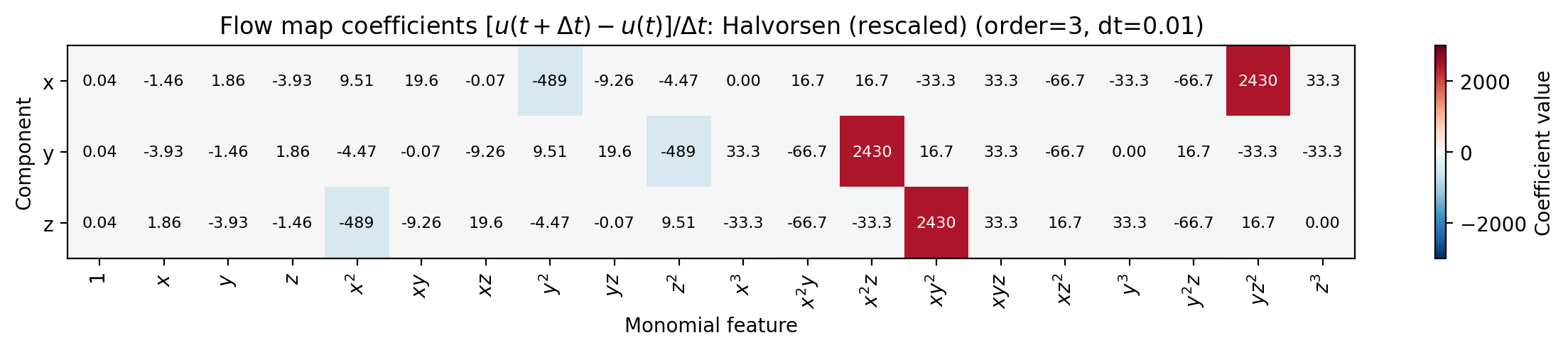}
  \caption{Illustration of the flow map learned by a QELM, obtained from the Taylor-approximation approach \cref{eq_qelm_flowmap}. The QELM uses \emph{rotational} encoding [\cref{eq_enc_rot}] and is trained on (a) the Lorenz-63 model \eqref{eq_lor63_rescaled} and (c) the Halvorsen model \eqref{eq_halvorsen_rescaled}. The rescaling parameters are here $\mv=[1,1,20], \alpha_{x,y,z}=0.002$ (Lorenz-63) and $\mv=[-3,-3,-3],\alpha_{x,y,z}=0.002$ (Halvorsen). The corresponding true flow map increments $(\psiv_\gv(\uv)-\uv)/\Delta t$ [\cref{eq_flow_map}], expanded to third order in the state variables $u_j$, are shown in (b,d) for comparison. The coefficients differ from \cref{fig:flowmap} due to the different expansion points. }
  \label{fig:flowmap_rot}
\end{figure}

When training an NG-RC with an instantaneous feature library on a Markovian dynamical system, the model learns an approximation of the flow map \cite{brunton_discovering_2016,gauthier_next_2021,chen_next_2022,zhang_how_2025,gross_flow_2026}.
Libraries $\phiv$ of the form \eqref{eq_lib_amplsqrt}, however, do not render directly interpretable models due to the structural mismatch with the ground truth model \footnote{Moreover, non-orthogonal function libraries are typically highly redundant, reflected in large condition numbers of the corresponding design matrices [\cref{eq_gram_matrix,eq_gram_matrix_readout}].}.
In the present teacher-forcing setup, a way to render the model interpretable is thus to express the feature map $\phiv$ in terms of an $N$-dimensional library $\Bcal$ of functions that make up the flow map $\psiv_\gv$ [\cref{eq_flow_map}],
\beq \phiv(\uv) \simeq K \Bcal(\uv) ,
\label{eq_fmap_trafo}\eeq 
and insert this into the learned model \eqref{eq_qelm_classical},
\beq \hat\fv(\uv) \simeq \hat W^\top R K \Bcal(\uv) = \psiv\st{approx}(\uv).
\label{eq_qelm_flowmap}\eeq 
This renders an approximation $\psiv\st{approx}$ to the flow map via \cref{eq_qelm_timeser}, with a transformation matrix $K\in \reals^{q\times N}$.

While there are several methods to determine $K$ (such as collocation), we perform here a Taylor expansion of $\phiv(\uv)$ around some operating point $\bar\uv$ (data mean).
This is natural here, since, for the considered polynomial dynamical systems, we can take $\Bcal = \Bcal_r$ to be the set of \emph{monomials} in $u_j$ up to order $r$, of which there are $N  = {D+r \choose r}$. 
Recall that low-order monomials are well constructible from the library $\phiv$ (see \cref{sec_mon_decod}).
However, due to the tensor product structure of $\phiv$ [see \cref{eq_phi_tensor}] and the fact that the derivatives of the single-qubit feature vector lie in a 2-dimensional space (vanishing Pauli-$Y$), $K$ typically does not have full column rank here when $N<q$. 
A consequence of this is that only a $\rank(K)$-dimensional subspace of the library $\Bcal$ is accessible through the encoding features $\phiv$ and that hence some monomials in $\Bcal$ are not independently identifiable.
Nevertheless, deviations are small for the settings considered here \footnote{Deviations are small for $N<q$: in practice, for $n=D=3$ (thus $q=27$) and monomials of order $r=3$ (thus $N=19$ excluding the constant term), we get $\rank(K)=16$.}.

\Cref{fig:flowmap} shows the learned flow map increments $\psiv(\uv)-\uv$, resulting from \cref{eq_qelm_flowmap} after  training a QELM with amplitude encoding [\cref{eq_enc_amplsqrt,eq_enc_amplsq}] on the (rescaled) Lorenz-63 and Halvorsen models (no feature engineering in the readout). 
Comparison to the theoretical expectation [\cref{eq_flow_map_exp}] reveals good agreement especially for low-order coefficients. 
While deviations can be large for higher-order coefficients, their influence on the overall training error is still small. 

As illustrated in \cref{fig:flowmap_rot}, also in the case of rotational encoding [\cref{eq_enc_rot}] the QELM forms linear combinations of the feature library $\phiv(\uv)$ so that its Taylor expansion approximately matches the true flow map $\psiv_\gv$ [\cref{eq_flow_map_exp}].
Compared to amplitude encoding, rotational encoding renders a slightly more accurate flow map approximation and lower training error.
We find that the lowest training error is achieved when scaling the training trajectories $\uv(t)$ to an interval $\sim 10^{-2}$ around their mean. 


\section{Summary}

We have employed the Pauli transfer matrix (PTM) formalism to describe training of a QELM (a memoryless QRC) as a decoding problem of features mixed by quantum channels, and discussed its implications for system identification and time-series prediction.
Writing states and channels in the Pauli basis (or any other orthogonal operator basis) turns the QELM into a sequence of real linear maps $T_{\Ecal_j}$ acting on the vector of Pauli features $\phiv(\uv)$ representing the encoded state $\rho\enc(\uv)$ \cite{fujii_quantum_2020,martinez-pena_quantum_2023}. 
The QELM thus takes the form of a classical regression model (or ``next-generation'' RC \cite{gauthier_next_2021}) over random features $T\phiv(\uv)$, which carry the full dependence of the QELM on the encoded input $\uv$ and determine the nonlinearities that the model can possibly express.
We remark that classical representations of QML models have been explored in various previous works \cite{sweke_potential_2023,schreiber_classical_2023,mingard_exploiting_2024,masot-llima_prospects_2025,lorenzis_entanglement_2025}. 

While the PTM-based regression representation of a QELM is essentially a known consequence of the linear nature of a data-independent quantum channel \cite{fujii_quantum_2020,martinez-pena_quantum_2023,kobayashi_coherence_2024,martinez-pena_inputdependence_2025}, our work provides several new insights into QELMs with initial-state encoding. 
We have shown that nonlinear information processing capacity of a QELM can be expressed in terms of two factors: (i) how well a target can be approximated in terms of the Pauli features $\phi_k$, and (ii) how well these features are transmitted through the reservoir (see \cref{sec_cap_factorization}).
Notably, even features obstructed by the reservoir channel can be reconstructed from statistical correlations with visible features.
This provides a more fine-grained perspective on the different contributions to the information processing capacity \cite{dambre_information_2012} and the training error beyond universal approximation theorems \cite{chen_learning_2019,chen_temporal_2020,goto_universal_2021,martinez-pena_quantum_2023,monzani_universality_2024,gonon_universal_2025}.
In fact, learning of the flow map of a dynamical system benefits from a feature library matching the structure of the vector field \cite{gross_flow_2026}. 
For the encodings used here, the terms in the flow map can be approximated on a restricted domain from the encoded features, which are Fourier-like or of other forms (see \cref{sec_encodings}). 
Applying the present insights to construct reduced-order models in cases where the exact data-generating process is not known (black-box system identification) is an interesting avenue for future work \cite{ljung_system_1999,brunton_datadriven_2019,gil-fuster_understanding_2024}.

The present work emphasizes Hamiltonian structure as an indicator for QRC performance (see also, e.g., \cite{schuld_effect_2021,hayashi_impact_2023,palacios_role_2024,lorenzis_entanglement_2025,han_multiple_2025}): specifically, for temporal multiplexing, greater reservoir randomness is advantageous under restricted measurement schemes, while highly structured Hamiltonians can lead to low expressivity.
We have explained these findings in terms of the underlying Hamiltonian symmetries, which partition the reachable Krylov space (see \cref{app_sym_kry}), and provided a theoretical analysis of the effective PTM. As a specific application, we provided asymptotic rank estimates for the transverse field Ising chain based on the Majorana formalism (see \cref{sec_tfim_sector_rank}).

\begin{figure}[tbp]
  \centering
  \panelimgshift[width=0.28\textwidth]{-12pt}{8pt}{Figures/pauli_wt_ZZX_t6.png}{a}\quad
  \panelimg[width=0.3\textwidth]{b}{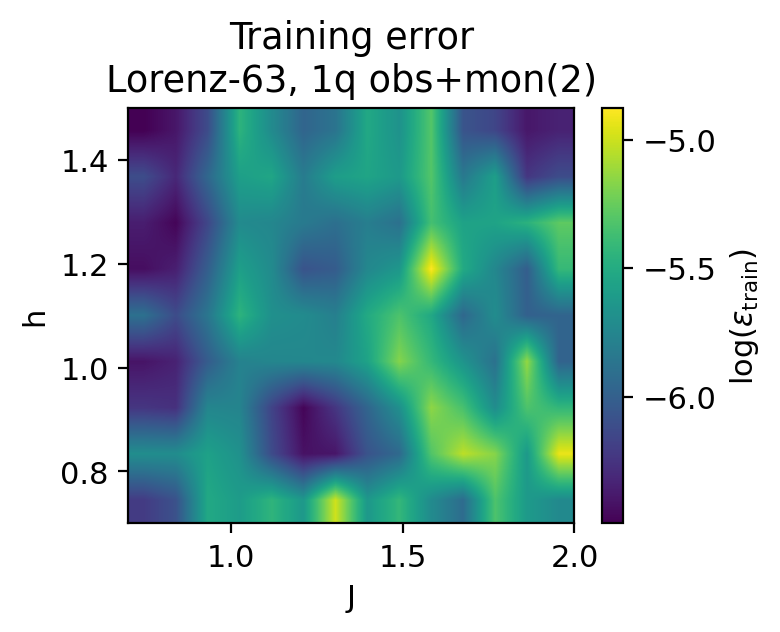}\quad
  \panelimg[width=0.28\textwidth]{c}{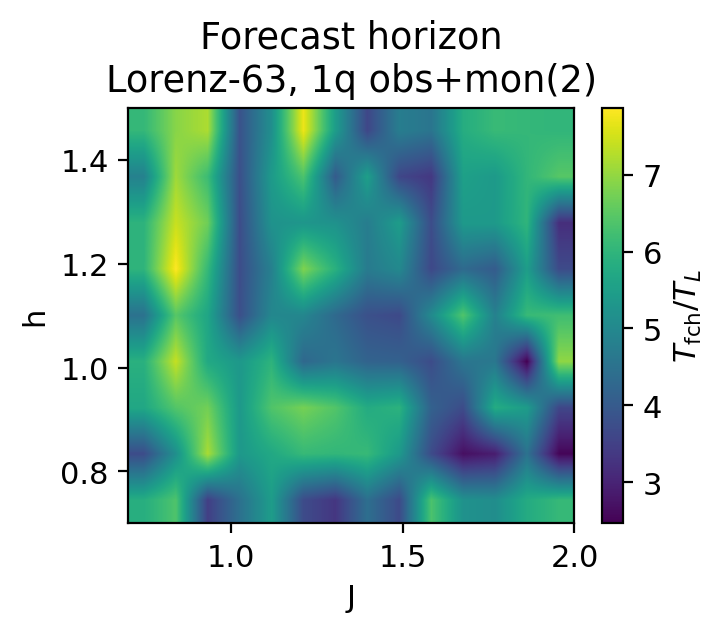}
  \caption{The Pauli weight average (a) of Heisenberg-evolved measurement observables can predict the training error (b) and forecast horizon (c) of a QELM without temporal multiplexing. We use a TFIM Hamiltonian [\cref{eq_H_TFIM_zzx}, varying coupling parameters $J,h$, fixed $t=6$] with $n=3$ qubits and amplitude encoding \eqref{eq_enc_amplsqrt}. The QELM in (b,c) is trained on the Lorenz-63 model and employs 1-qubit Pauli-$Z$ observables in the readout, enhanced by monomials up to degree 2 to increase sensitivity to nonlinear terms. The Pauli weight average is data-independent.}
  \label{fig:pauliwt_metric}
\end{figure}

Numerous previous studies noted that simply enhancing the quantum nature of a QML model often does not improve its performance on practical datasets \cite{bowles_better_2024,schnabel_quantum_2025,basilewitsch_quantum_2025,rohe_questionable_2025,gundlach_quantum_2025,fellner_quantum_2026,freinberger_role_2026}. 
The present study adds further perspectives on this finding: 
first, as discussed above, the specific mathematical structure of the features generated by a quantum encoding is not necessarily optimal for approximating a classical target function. For instance, the target function may not be exactly representable from the linear span of the features generated by the encoding even before any transformation by a quantum channel (ideal decodability score $\Rcal_\Imat^2<1$).
This is the classic representability problem in QML \cite{vidal_input_2020,perez-salinas_data_2020,schuld_effect_2021,kubler_inductive_2021}.
Second, a reservoir unitary mixes features and reversing this potentially detrimental effect requires complex measurement schemes (e.g., high-order Pauli observables, optimized pre-measurement unitaries, or temporal multiplexing schemes). 
This insight is supported here by comparing the Pauli weight average of single-qubit observables to its training accuracy: as illustrated in \cref{fig:pauliwt_metric}, for a minimal QELM without temporal multiplexing trained on a time series prediction task, large Pauli weight average strongly correlates with high training error and short forecast horizon.
Consistent with \cite{cindrak_engineering_2025,cindrak_krylov_2025}, what matters for task performance is therefore not just how a quantum state evolves, but what part of the Heisenberg/Krylov operator space is actually accessible under the chosen observables and temporal multiplexing.
The decodability framework presented here allows one to separate the impact of the reservoir from that of the encoding on the nonlinear processing capacity of the model (see \cref{app_decodability}).

The $4^n$-dimensional space of Pauli features \footnote{For the real product encodings considered here, any Pauli string containing a $Y$ factor has zero pre-reservoir expectation, leaving at most $\sim 3^n$ features.} is analogous to the exponential frequency content of QML models that employ gate-based rotational encoding \cite{schuld_effect_2021,peters_generalization_2022,shin_exponential_2023}. These and similar properties stemming from the exponential Hilbert space dimensionality point to possible quantum advantages, but require specific measurement strategies.
Moreover, simply increasing qubit number $n$ does not generally improve QRC performance, as  ill-conditioning and exponential concentration effects can become prevalent \cite{xiong_fundamental_2025,pfeffer_reducedorder_2023,schutte_expressivity_2025,lorenzis_entanglement_2025}. 
In that regime, more sophisticated reservoir design strategies will be required to steer feature propagation through the quantum layers.
Finally, using the Pauli basis in the transfer matrix definition is justified by the measurement process, but not the only possibility.
For future work, it could thus be interesting to study other orthogonal operator sets and assess the generality of the present findings.

\acknowledgments
This project was made possible by the DLR Quantum Computing Initiative and the Federal Ministry for Research, Technology and Space; \texttt{https://qci.dlr.de/NeMoQC}


\appendix 
\crefalias{section}{appendix}

\begin{table}[tb!]
    \centering
    \begin{tabular}{l l l}
        \hline
        Symbol & Meaning & Defined in \\
        \hline
        $\Hcal$ & Hilbert space & \cref{sec_model} \\
        $\Herm(\Hcal)$ & Space of Hermitian operators on $\Hcal$ & \cref{sec_model} \\
        $n$ & Number of qubits & \cref{sec_model} \\
        $d$ & Hilbert space dimension ($d=2^n$) & \cref{sec_model} \\
        $D$ & Input data dimension & \cref{sec_model} \\
        $P$ & Number of training samples & \cref{eq_QELM_opt} \\
        $\numM$ & Number of measurement operators & \cref{eq_QELM} \\
        $\numTM$ & Number of temporal multiplexing steps & \cref{eq_tmux} \\
        $\numMtot$ & Total number of measurements (e.g., $\numMtot=\numM \numTM$) & \cref{eq_qelm_classical} \\
        $\numF$ & Number of nonzero Pauli features ($\numF\leq d^2$) & \cref{eq_qelm_classical} \\
        \hline
        $\xv,\uv$ & Input data ($\xv,\uv \in \reals^D$) & \cref{eq_QELM,eq_qelm_classical} \\
        $y$ & Target value ($y \in \reals$) & \cref{eq_QELM_opt} \\
        $\rho(\xv)$ & Density matrix of the quantum state (feature map) & \cref{eq_QELM} \\
        $M_k$ & Measurement operators & \cref{eq_QELM} \\
        $P_k$ & $n$-qubit Pauli operators ($k=0,\ldots,d^2-1$) & \cref{eq_pauli_feature} \\
        $\sigma^a$ & single-site Pauli matrices ($\sigma^0\equiv \Imat$, $X$, $Y$, $Z$) & \cref{eq_single_qubit_phi} \\
        $U_R,U\st{enc}$ & Reservoir and encoding unitary operators & \cref{eq_q_feature_map} \\
        $T_{\Ecal}$ & Pauli transfer matrix (channel $\Ecal$, $T_{\Ecal} \in \reals^{d^2\times d^2}$) & \cref{eq_PTM} \\
        $V$ & Unitary PTM ($V \in \mathrm{SO}(d^2)$) & \cref{eq_PTM_uni} \\
        $R$ & Effective PTM (temporal multiplexing) ($R \in \reals^{\numMtot\times q}$) & \cref{eq_tmux} \\
        $\Scal$ & Selected observable set ($|\Scal|=\numM$) & \cref{eq_qelm_ptm} \\
        $S$ & Selector matrix corresponding to $\Scal$ & \cref{eq_qelm_ptm} \\
        $\phiv(\xv)$ & Pauli feature vector ($\phiv(\xv)\in \reals^q$) & \cref{eq_pauli_feature} \\
        $\Fv(\xv)$ & Readout feature vector ($\Fv(\xv) = R \phiv(\xv)$) & \cref{eq_qelm_classical} \\
        $\wv$ & Readout weights ($\wv\in \reals^\numMtot$) & \cref{eq_qelm_classical} \\
        $\lambda$ & Regularization parameter & \cref{eq_QELM_opt} \\
        $f(\xv)$ & QELM output function ($f(\xv) = \wv^\top \Fv(\xv)$) & \cref{eq_QELM} \\
        $\nu$ & Pauli weight & \cref{eq_pauli_wt_avg} \\
        $\gamma_r$ & Geometric channel decodability score & \cref{eq_decod_score} \\
        $\kappa_j$ & Geometric encoding decodability score & \cref{eq_input_dec_score} \\
        $\Rcal_R^2$ & Information processing capacity & \cref{eq_nlcap_R2_score,eq_cap_fact} \\
        $\Rcal_\Imat^2$ & Information processing capacity for an identity reservoir & \cref{eq_cap_fact} \\
        $\tau_R$ & Channel transmission score & \cref{eq_channel_trans_score} \\
        $\nu_R$ & Nullspace load & \cref{eq_channel_trans_score} \\
        $\chi_R$ & Statistical rescue score & \cref{eq_channel_trans_score} \\
        $\boldsymbol{\alpha}$ & Dynamical system rescaling parameter & \cref{eq_lor63_rescaled,eq_halvorsen_rescaled} \\
        \hline
    \end{tabular}
    \caption{Summary of notation used in this work. A trained quantity is indicated by a hat (e.g., $\hat{\wv}$).}
    \label{tab:notation}
\end{table}

\section{Kernel representation}
\label{app_kernel}
Training over a Hilbert--Schmidt orthonormal basis of measurement operators $\{M_k\}_{k=1}^{d^2}$ results in a quantum-kernel form of the QELM \cite{gross_kernelbased_2026}:
\begin{equation}
 f(\mathbf{x}) =\sum_{j=1}^P \alpha_j^* K(\mathbf{x},\mathbf{x}_j) =\tr\left(M^*\rho(\mathbf{x})\right), \qquad K(\mathbf{x},\mathbf{x}') =\tr\left(\rho(\mathbf{x})\rho(\mathbf{x}')\right).
\label{eq_QELM_kernel}
\end{equation}
For the least-squares loss, the dual parameters are
$\boldsymbol{\alpha}^*=(\mathbf{K}+\lambda\mathbb{I}_P)^{-1}\mathbf{y}$,
where $K_{ij}=K(\mathbf{x}_i,\mathbf{x}_j)$, and the optimal measurement operator is
\begin{equation}
 M^*=\sum_{j=1}^P \alpha_j^*\rho(\mathbf{x}_j).
\label{eq_Mopt}
\end{equation}
Compared with the equivalent primal formulation in \cref{eq_QELM}, this representation is computationally more efficient when $P\ll d^2$.

If the reservoir is a data-independent unitary $U_R$, the resulting state can be written as [see \cref{eq_encoding_Uenc}]
$\rho_U(\mathbf{x})=U_R\rho_{\mathrm{enc}}(\mathbf{x})U_R^\dagger$.
For a complete measurement operator basis, unitary invariance of the Hilbert--Schmidt inner product gives
\begin{equation}
K(\mathbf{x},\mathbf{x}') =\tr\left(\rho_U(\mathbf{x})\rho_U(\mathbf{x}')\right) =\tr\left( \rho_{\mathrm{enc}}(\mathbf{x})\rho_{\mathrm{enc}}(\mathbf{x}') \right),
\label{eq_kernel}
\end{equation}
which is the usual fidelity kernel. Thus the Gram matrix, dual coefficients, and predictions are independent of $U_R$, whereas the optimal observable transforms covariantly as $M_U^*=U_R M_{\mathrm{enc}}^*U_R^\dagger$, with $M_{\mathrm{enc}}^*$ obtained from \cref{eq_Mopt} using $\rho_{\mathrm{enc}}$.

Consider now a restricted Pauli observable set $\mathcal{S}$. 
In this case, the kernel induced by the primal model is generally not the full fidelity kernel. To see this, define the Hilbert--Schmidt projector onto the measured operator subspace by
\begin{equation}
 \Pi_{\mathcal{S}}(\rho ) =\frac{1}{d}\sum_{k\in\mathcal{S}}\tr(P_k \rho)P_k.
\end{equation}
Using the normalized Pauli basis $P_k/\sqrt{d}$, the corresponding kernel is
\begin{equation}
 K_{\mathcal{S},U}(\mathbf{x},\mathbf{x}') =\frac{1}{d}\sum_{k\in\mathcal{S}}
 \tr\left(P_k\rho_U(\mathbf{x})\right) \tr\left(P_k\rho_U(\mathbf{x}')\right) \\
 =\tr\left[ \Pi_{\mathcal{S}}\left(\rho_U(\mathbf{x})\right) \Pi_{\mathcal{S}}\left(\rho_U(\mathbf{x}')\right) \right].
\label{eq_kernel_restricted}
\end{equation}
The reservoir unitary drops out of this restricted kernel if conjugation preserves the measured subspace,
\beq
 U_R^\dagger \Vcal_\Scal U_R = \Vcal_\Scal,\qquad \Vcal_\Scal = \operatorname{span}\{P_l : l\in\Scal\},
\label{eq_kernel_subspace_invariance}
\eeq
which follows from cyclicity of the trace, $\tr(P_k U_R \rho_\mathrm{enc} U_R^\dagger) = \tr(Q \rho_\mathrm{enc})$ with some $Q\in \Vcal_\Scal$.
In this case, the restricted feature vector changes only by an orthogonal transformation, so ridge regression results in the same Gram matrix and dual solution. 
This condition holds, for instance, for certain Clifford unitaries that map $\Scal$ onto itself up to signs (such as CNOT circuits when $\Scal$ comprises all $Z$-strings).
However, a generic Pauli subset $\Scal$ does not satisfy \cref{eq_kernel_subspace_invariance} and the corresponding restricted primal QELM is then represented by the kernel \eqref{eq_kernel_restricted}, which still depends on the reservoir unitary $U_R$.

\section{Decoding reformulation of linear regression}
\label{app_ols_decoding}

Let $\phiv\in\mathbb{R}^{q}$ be a vector and $R\in\mathbb{R}^{B\times q}$ be a fixed mixing matrix, and $\mathbf{e}_r\in\mathbb{R}^{q}$ the $r$-th standard basis vector. 
We ask for the reconstruction of a single component $\phi_r=\mathbf{e}_r^\top \phiv$ by finding weights $\mathbf{w}_r\in\mathbb{R}^{B}$ such that
\beq
\phi_r \simeq \mathbf{w}_r^\top R\phiv
\label{eq_decod_cond}\eeq
to within a certain error tolerance. 
We discuss below situations where equality holds and where reconstruction can only be approximate. We remark that related ideas are pursued in compressed sensing \cite{brunton_datadriven_2019}.

In general, this problem is solved by ordinary least squares (OLS) regression.
Collecting $P$ centered training samples as $\Phi=[\phiv_1, \cdots , \phiv_P]\in\mathbb{R}^{q\times P}$, the corresponding target vector is $\mathbf{e}_r^\top \Phi$. 
The OLS problem 
\beq
\widehat{\mathbf{w}}_r = \underset{\mathbf{w}\in\mathbb{R}^{B}}{\mathrm{argmin}}
\frac{1}{P}
\left\| \mathbf{e}_r^\top \Phi-\mathbf{w}^\top R\Phi \right\|_2^2
\label{eq_decod_ols_problem}\eeq
then has the well-known minimum-Euclidean-norm solution 
\beq
\widehat{\mathbf{w}}_r = \left(R\Sigma R^\top \right)^{+} R\Sigma\mathbf{e}_r,\qquad \Sigma=\frac{1}{P}\Phi\Phi^\top ,
\label{eq_decod_ols}\eeq
with $\Sigma$ the empirical covariance matrix of the training data.
If the data are not centered, an intercept can be included, which takes the form $b_r=\overline{\phi}_r-\mathbf{w}_r^\top R\overline{\phiv}$.
We now discuss how the properties of $R$ and the data covariance $\Sigma$ determine the exactness of the reconstruction of $\phi_r$.

\subsection{Geometric decodability}
\label{app_geom_decod}

Exact geometric reconstruction of the single component $\phi_r$ is possible if $\mathbf{e}_r\in\mathrm{im}(R^\top )$ or, equivalently, if there is a weight vector $R^\top \mathbf{w}_r=\mathbf{e}_r$. This, in turn, is equivalent to
\beq
\mathbf{e}_r^\top \mathbf{v}=0 \quad\text{for every } \mathbf{v}\in\nullsp(R),
\label{eq_decod_crit_geom}\eeq
i.e., every vector of the nullspace of $R$ must have zero $r$-th component. 
If this condition is met, the minimum-norm solution for the weights is provided by
\beq \mathbf{w}_r^* = (R^+)^\top \mathbf{e}_r.
\label{eq_decod_weights_geom}\eeq 
However, the solution need not be unique, since any term in $\nullsp(R^\top )$ can be added, so that $\mathbf{w}_r=\left(R^{+}\right)^\top \mathbf{e}_r + \left(\mathbb{I}_B-RR^{+}\right)\mathbf{u}$,
with $\mathbf{u}\in\mathbb{R}^{B}$. The pseudoinverse singles out the solution with minimum Euclidean norm.
Geometric decodability is thus a property of the mixing matrix $R$ alone.

In certain special cases, the test \eqref{eq_decod_crit_geom} can be circumvented:

\paragraph*{Square and invertible mixing matrix.}
The first case is an invertible $R$ with $B=q$, and positive definite $\Sigma$. Then $\left(R\Sigma R^\top \right)^{-1}R\Sigma = R^{-\top}$ and the minimum-norm solution is
\beq
\mathbf{w}_r^{*}=R^{-\top}\mathbf{e}_r,
\label{eq_decod_weights_square}\eeq
so that $\phi_r$ can be exactly reconstructed, $\widehat{\phi}_r = \mathbf{e}_r^\top R^{-1}R\phiv =\phi_r$.

\paragraph*{Tall mixing matrix with full column rank.}
In the case $B\geq q$ and $\rank(R)=q$, the mapping $\phiv\mapsto R\phiv$ is injective, and every component can be reconstructed exactly. The minimum-norm weights are
\beq
\mathbf{w}_r^{*} = \left(R^{+}\right)^\top \mathbf{e}_r = R\left(R^\top R\right)^{-1}\mathbf{e}_r,\label{eq_decod_weights_tall}\eeq
which fulfill $\mathbf{w}_r^{*\top}R\phiv=\phi_r$.

\subsection{Statistical decodability}

When $R$ is rank deficient ($\rank(R)<q$) and \cref{eq_decod_crit_geom} is not satisfied, a feature $\phi_r$ may still be decodable using statistical correlations.
To this end, recall that, from \cref{eq_decod_ols_problem}, the minimum mean-squared error is $\mathcal{E}_r^{\min} =\min_{\mathbf{w}} (\mathbf{e}_r-R^\top \mathbf{w})^\top  \Sigma (\mathbf{e}_r-R^\top \mathbf{w})$.
Since $\Sigma$ is positive semidefinite, this error is zero only if the residual vector $\mathbf{e}_r-R^\top \mathbf{w}$ lies in $\nullsp(\Sigma)$ for some $\mathbf{w}$, or equivalently if $\mathbf{e}_r \in \mathrm{im}(R^\top ) + \nullsp(\Sigma)$.
This, in turn, is equivalent to
\beq
\mathbf{e}_r^\top \mathbf{v}=0 \quad \text{for every } \mathbf{v} \in \nullsp(R)\cap\mathrm{im}(\Sigma),
\label{eq_decod_crit_stat}\eeq
which says that two statistically admissible inputs producing the same mixed vector may differ in unobserved directions, but such differences must not alter the component $\phi_r$.
Exact recovery through correlations requires an exact, lower-dimensional relation in the data, while strong but imperfect correlations reduce the error but do not make it vanish.
We analyze now a number of specific cases.

\subsubsection{Isotropic data}
For isotropic data, $\Sigma=\sigma_{\phi}^{2}\mathbb{I}_q$, the minimum-norm solution \eqref{eq_decod_ols} reduces to
\beq
\mathbf{w}_r^{*} = \left(RR^\top \right)^{+}R\mathbf{e}_r = \left(R^{+}\right)^\top \mathbf{e}_r.
\label{eq_decod_weights_uncorrel}\eeq
The resulting predictor is $\widehat{\phi}_r=\mathbf{e}_r^\top R^{+}R\phiv$, wherein the component of $\phiv$ in $\nullsp(R)$ is set to zero.

\subsubsection{Positive definite covariance matrix}

If $\Sigma$ is positive definite, then $\nullsp(\Sigma)=\{\mathbf{0}\}$, and the statistical-support criterion reduces to \cref{eq_decod_crit_geom}, i.e., $\mathbf{e}_r\in\mathrm{im}(R^\top )$, with corresponding minimum-norm weights \cref{eq_decod_weights_geom}. 

\subsubsection{Correlated data}

If $\mathbf{e}_r\notin\mathrm{im}(R^\top )$, then part of $\phi_r$ depends on directions that are not directly observable through $R$, but may still be linearly predictable from second-order correlations.
In order to quantify the effect, consider the compact singular value decomposition ($s=\rank(R)$),
\beq
R=U_sD_sV_s^\top ,\qquad D_s=\mathrm{diag}(d_1,\ldots,d_s),
\label{eq_decod_R_SVD}\eeq
and complete $V_s$ to an orthogonal basis, $V=[V_s\ V_0]$, where the columns of $V_0$ span $\nullsp(R)$. 
This allows us to decompose the input into observable and unobservable coordinates,
\beq
\mathbf{a}=V_s^\top \phiv,
\qquad
\mathbf{b}=V_0^\top \phiv,
\eeq
so that
\beq
\phiv=V_s\mathbf{a}+V_0\mathbf{b},\qquad R\phiv=U_sD_s\mathbf{a}.
\eeq
Thus, the mixed vector $R\phiv$ determines only $\mathbf{a}$ exactly.
In this basis, the target component can be written as
\beq
\phi_r=\vv_a^\top \mathbf{a}+\vv_b^\top \mathbf{b},\qquad  \vv_a \equiv V_s^\top \mathbf{e}_r,\qquad \vv_b \equiv V_0^\top \mathbf{e}_r
\eeq
while the covariance matrix takes the form
\beq
V^\top \Sigma V
=
\begin{pmatrix}
\Sigma_{aa} & \Sigma_{ab} \\
\Sigma_{ba} & \Sigma_{bb}
\end{pmatrix}.
\eeq
Using the SVD \eqref{eq_decod_R_SVD} and assuming that $\Sigma_{aa}$ is invertible, the first term in the weight solution \eqref{eq_decod_ols} results as
\beq
\left(R\Sigma R^\top \right)^+ = \left(U_sD_sV_s^\top \Sigma V_sD_sU_s^\top \right)^+ = \left(U_sD_s\Sigma_{aa}D_sU_s^\top \right)^+ = U_sD_s^{-1}\Sigma_{aa}^{-1}D_s^{-1}U_s^\top .
\eeq
Since $\mathbf{e}_r=V_s\vv_a+V_0\vv_b$, the second term in \eqref{eq_decod_ols} reads
\beq
R\Sigma\mathbf{e}_r = U_sD_sV_s^\top \Sigma \left(V_s\vv_a+V_0\vv_b\right) =
U_sD_s \left(\Sigma_{aa}\vv_a+\Sigma_{ab}\vv_b\right).
\eeq
Altogether, one obtains
\beq
\mathbf{w}_r^{*} = U_sD_s^{-1}\left(\vv_a+\Sigma_{aa}^{-1}\Sigma_{ab}\vv_b \right),
\label{eq_decod_correl_weights}\eeq
which results in the predictor 
\beq
\widehat{\phi}_r= \vv_a^\top \mathbf{a}+ \vv_b^\top  \Sigma_{ba}\Sigma_{aa}^{-1}\mathbf{a}.
\label{eq_decod_correl_pred}\eeq
The first term directly reconstructs the observable contribution to $\phi_r$, while the second term predicts its unobservable contribution using correlations between $\mathbf{a}$ and $\mathbf{b}$. 
The residual error variance is given by
\beq
\begin{aligned}
\mathcal{E}_r &= \E[(\phi_r- \hat\phi_r)^2] = \E\!\left[\left(\vv_b^\top \left(\mathbf{b}-\Sigma_{ba}\Sigma_{aa}^{-1}\mathbf{a}\right)\right)^2\right] \\
&= \vv_b^\top  \left( \Sigma_{bb} - \Sigma_{ba}\Sigma_{aa}^{-1}\Sigma_{ab} \right) \vv_b ,
\end{aligned}
\label{eq_decod_res_err}\eeq 
where the matrix in parentheses is the residual covariance of the unobserved coordinates given the observed ones. If $\vv_b=\mathbf{0}$, the target is directly observable and no statistical prediction is needed. If $\vv_b\neq\mathbf{0}$ but $\Sigma_{ab}\vv_b=\mathbf{0}$, correlations provide no improvement. Otherwise, correlations reduce the reconstruction error by predicting the target-relevant part of the unobserved coordinates.

\subsubsection{Simple example: how correlations help}

We give a simple example that shows how correlations help reconstruct targets not directly decodable from $R$.
Consider $q=2$, $B=1$, and
\beq
R=
\begin{pmatrix}
1 & 0
\end{pmatrix},
\eeq
so that the observed variable is $R\phiv=\phi_1$. Suppose the goal is to reconstruct $\phi_2$, and let the data covariance be
\beq
\Sigma
=
\begin{pmatrix}
\sigma_1^2 & \rho\sigma_1\sigma_2 \\
\rho\sigma_1\sigma_2 & \sigma_2^2
\end{pmatrix}.
\eeq
The optimal weight and predictor follow as
\beq
w_2^{*}=\rho\frac{\sigma_2}{\sigma_1},\qquad \widehat{\phi}_2=\rho\frac{\sigma_2}{\sigma_1}\phi_1.
\eeq
Thus, without correlation ($\rho=0$), $\phi_1$ provides no linear information about $\phi_2$. As $\lvert\rho\rvert$ approaches one, the reconstruction error $\Ecal_2= \E[(\phi_2-\hat\phi_2)^2]=\sigma_2^2\left(1-\rho^2\right)$ decreases. For $\lvert\rho\rvert=1$, the two components satisfy an exact linear relation and $\phi_2$ can be reconstructed perfectly, even though $R$ is rank deficient.

\section{General formalism of decodability scores}
\label{app_decodability}

We now develop a formalism of decodability scores based on the least-squares approach of \cref{app_ols_decoding}. 
\Cref{app_decod_gen,app_decod_feat} closely follow the information processing capacity formulation of \cite{dambre_information_2012}. Going beyond this, the subsequent sections show how the decodability scores factorize into quantum encoding and channel parts, and how to distinguish statistical from geometric contributions.
We retain the centered finite-sample setting of \cref{app_ols_decoding} and the notation of \cref{sec_classical}, including the following definitions:
\beq
\begin{aligned}
&\Phi=[\phiv_1,\ldots,\phiv_P]\in\reals^{\numF\times P},\quad
R\in\reals^{\numMtot\times\numF},\quad
G=R\Phi\in\reals^{\numMtot\times P},\quad
\Sigma=\frac{1}{P}\Phi\Phi^\top, \\
&\yv=\bigl(y(\uv_1),\ldots,y(\uv_P)\bigr)^\top,\quad
\mathbf b_y=\frac{1}{P}\Phi\yv,\quad
s_y^2=\frac{1}{P}\yv^\top\yv>0
\end{aligned}
\label{eq_decod_design_matrices}
\eeq
where $y$ is a centered scalar target.
An intercept can be included before centering, as described in \cref{app_ols_decoding}.
Unless stated otherwise, the capacities below refer to the unregularized least-squares method.

\subsection{General decodability score}
\label{app_decod_gen}

The minimum-norm OLS weights and fitted target are
\beq
\widehat{\wv}_y=(R\Sigma R^\top)^+R\mathbf b_y,
\qquad
\widehat{\yv}_R=G^\top\widehat{\wv}_y=\Pi_G\yv,
\qquad
\Pi_G\equiv G^\top(GG^\top)^+G,
\label{eq_decod_general_fit}
\eeq
where $\Pi_G$ projects targets onto the subspace of functions representable by the measured QELM features $G=R\Phi$ (i.e., the row-space $\rowsp(G)\subseteq\reals^P$). 
We define the coefficient of determination $\Rcal^2$ as the \emph{general decodability score}
\beq
\begin{aligned}
\Dcal_R[y]
&\equiv 1-\frac{\min_{\wv}\,P^{-1}\|\yv-G^\top\wv\|_2^2}{s_y^2} =\frac{\yv^\top\Pi_G\yv}{\yv^\top\yv} \\
&=\frac{\mathbf b_y^\top R^\top(R\Sigma R^\top)^+R\mathbf b_y}{s_y^2},
\end{aligned}
\label{eq_R2_general}
\eeq
which corresponds to the information processing capacity introduced in \cite{dambre_information_2012}.
The second equality follows because the OLS residual is orthogonal to the fitted vector. The last equality follows from $GG^\top=P R\Sigma R^\top$ and $G\yv=P R\mathbf b_y$. 
Consequently, we have $0\leq\Dcal_R[y]\leq1$ and
\beq
\Dcal_R[y]=1 \quad\Longleftrightarrow\quad \yv\in\rowsp(G).
\label{eq_decod_general_exact}
\eeq

\subsection{Feature-direction, statistical, and geometric scores}
\label{app_decod_feat}

For a feature-space direction $\av\in\reals^{\numF}$, consider the target
$y_{\av}(\uv_p)=\av^\top\phiv_p$, with variance $s_{\av}^2=\av^\top\Sigma\av$. If $\av\notin\nullsp(\Sigma)$, \cref{eq_R2_general} gives
\beq
\Dcal_{R,\Sigma}(\av) \equiv \Dcal_R[y_{\av}]
= \frac{\av^\top\Sigma R^\top(R\Sigma R^\top)^+R\Sigma\av} {\av^\top\Sigma\av}.
\label{eq_decod_direction}
\eeq
It is convenient to introduce the whitened quantities
\beq
\widetilde R=R\Sigma^{1/2}, \qquad \widetilde{\av}=\Sigma^{1/2}\av, \qquad \Pi_{\widetilde R}=\widetilde R^+\widetilde R,
\eeq
in terms of which \cref{eq_decod_direction} takes the form
\beq
\Dcal_{R,\Sigma}(\av)=
\frac{\widetilde{\av}^\top\Pi_{\widetilde R}\widetilde{\av}} {\widetilde{\av}^\top\widetilde{\av}}=
1-\frac{\|(\Imat-\Pi_{\widetilde R})\widetilde{\av}\|_2^2} {\|\widetilde{\av}\|_2^2},
\label{eq_decod_direction_projector}
\eeq
where $\Sigma^{1/2}$ is the symmetric square root.
Note that the formulas remain valid for singular $\Sigma$, with whitening understood on $\mathrm{im}(\Sigma)$.

Since $\Pi_{\widetilde R}$ is an orthogonal projector, exact and vanishing decodability are characterized by
\beq
\begin{aligned}
\Dcal_{R,\Sigma}(\av)=1
&\quad\Longleftrightarrow\quad
\av\in\rowsp(R)+\nullsp(\Sigma),\\
\Dcal_{R,\Sigma}(\av)=0
&\quad\Longleftrightarrow\quad
R\Sigma\av=0.
\end{aligned}
\label{eq_decod_direction_criteria}
\eeq
For $\av=\ev_r$, the first relation reproduces the statistical recovery criterion of \cref{eq_decod_crit_stat}. The corresponding single-feature statistical score is
\beq
\Dcal_{R,\Sigma}(\ev_r)=
\frac{\ev_r^\top\Sigma R^\top(R\Sigma R^\top)^+R\Sigma\ev_r}{\Sigma_{rr}} =\Rcal_R^2[\phi_r].
\label{eq_decod_stat_feature}
\eeq
This quantity can be nonzero even when the direction $\ev_r$ is absent from $\rowsp(R)$, because correlations encoded in $\Sigma$ can predict the unresolved component.

For isotropic data $\Sigma=\sigma_\phi^2\Imat$, we obtain the geometric decodability score:
\beq
\Dcal_{R,\sigma_\phi^2\Imat}(\ev_r) =\ev_r^\top R^+R\ev_r\equiv\gamma_r^2,
\label{eq_R2_white}
\eeq
consistent with the OLS analysis in \cref{app_geom_decod}. 
Moreover, $1-\gamma_r^2=\|(\Imat-R^+R)\ev_r\|_2^2$ is the squared coefficient-space projection error, and
\beq
\sum_{r=1}^{\numF}\gamma_r^2 =\tr(R^+R) =\rank(R).
\label{eq_decod_geom_sumrule}
\eeq
If $\Sigma$ is nonsingular (or after restriction to its support), the single-coordinate scores in whitened feature coordinates are 
\beq
\widetilde\gamma_r^2 \equiv(\Pi_{\widetilde R})_{rr}
=\Dcal_{\widetilde R,\Imat}(\ev_r),
\qquad \sum_r\widetilde\gamma_r^2=\rank(R\Sigma^{1/2}),
\label{eq_decod_score_white}
\eeq
which generalizes \cref{eq_R2_white}.

The tensor-product statement used in the main text follows directly from
$(\bigotimes_i R_i)^+=\bigotimes_i R_i^+$. If
$R=\bigotimes_{i=1}^nR_i$ and
$\ev_r=\bigotimes_{i=1}^n\ev_{r_i}$, then
\beq
R^+R=\bigotimes_{i=1}^n(R_i^+R_i),
\qquad
\gamma_r^2=\prod_{i=1}^n\gamma_{r_i,i}^2,
\qquad
\wv_r^*=(R^+)^\top\ev_r
=\bigotimes_{i=1}^n (R_i^+)^\top\ev_{r_i}.
\label{eq_decod_tensor_factorization}
\eeq
More generally, if also $\Sigma=\bigotimes_i\Sigma_i$ and
$\av=\bigotimes_i\av_i$, then the statistical score factorizes as
\beq
\Dcal_{\otimes_i R_i,\,\otimes_i\Sigma_i} \!\left(\bigotimes_i\av_i\right)
= \prod_i\Dcal_{R_i,\Sigma_i}(\av_i).
\label{eq_decod_stat_tensor_factorization}
\eeq

\subsection{Representation--transmission factorization}
\label{app_decod_factorization}

Consider the orthogonal projector $\Pi_\Phi$ from the target space onto the subspace representable by the feature library $\Phi$ (unrestricted by measurement or reservoir mixing):
\beq
\Pi_\Phi=\Phi^\top(\Phi\Phi^\top)^+\Phi,
\qquad
\widehat{\yv}_\phi=\Pi_\Phi\yv,
\qquad
\av_y=\Sigma^+\mathbf b_y,
\label{eq_decod_ideal_projection}
\eeq
so that $\widehat{\yv}_\phi=\Phi^\top\av_y$ and $\Sigma\av_y=\mathbf b_y$. 
Here $\av_y$ is the minimum-norm feature-space representative of the target projection. 
The ideal feature-representation score, corresponding to a QELM with $R=\Imat$ (see \cref{sec_nonlin_readout}), is
\beq
\Rcal_\Imat^2[y]\equiv \Dcal_\Imat[y]
=
\frac{\|\widehat{\yv}_\phi\|_2^2}{\|\yv\|_2^2}
=
\frac{\mathbf b_y^\top\Sigma^+\mathbf b_y}{s_y^2}
=
\frac{\av_y^\top\Sigma\av_y}{s_y^2}.
\label{eq_decod_ideal_score}
\eeq
This quantity measures the fraction of the target variance that can be represented by the encoding-generated feature library, independently of subsequent reservoir or measurement action.

For $\widehat{\yv}_\phi\neq \mathbf 0$, we define the \emph{channel-transmission} score
\beq
\begin{aligned}
\tau_R[y] 
&\equiv
\frac{\|\Pi_G\widehat{\yv}_\phi\|_2^2}{\|\widehat{\yv}_\phi\|_2^2} =\Dcal_{R,\Sigma}(\av_y) \\
&=
\frac{\mathbf b_y^\top R^\top(R\Sigma R^\top)^+R\mathbf b_y}{\mathbf b_y^\top\Sigma^+\mathbf b_y}
=
\frac{\widetilde{\mathbf b}_y^\top\Pi_{\widetilde R}\widetilde{\mathbf b}_y}{\widetilde{\mathbf b}_y^\top\widetilde{\mathbf b}_y},
\end{aligned}
\label{eq_channel_transmission}
\eeq
where $\widetilde{\mathbf b}_y\equiv(\Sigma^+)^{1/2}\mathbf b_y$, $\widetilde R=R\Sigma^{1/2}$, and $\Pi_{\widetilde R}=\widetilde R^+\widetilde R$.
If $\widehat{\yv}_\phi=\mathbf 0$, then $\Rcal_R^2[y]=0$ for every $R$, and the channel factor is immaterial.
The general score factorizes as
\beq
\Dcal_R[y]=\Rcal_R^2[y]=\Rcal_\Imat^2[y]\,\tau_R[y]= \Dcal_\Imat[y]\,\Dcal_{R,\Sigma}(\av_y).
\label{eq_cap_factorization}
\eeq
This follows from $\|\Pi_G\yv\|_2^2=\|\Pi_G\Pi_\Phi\yv\|_2^2=\tau_R[y]\|\Pi_\Phi\yv\|_2^2$, since $\rowsp(G)\subseteq\rowsp(\Phi)$ and hence the fitted component can first be projected onto the feature library and then onto the readout subspace.
Equivalently, the result is obtained by substituting $\Sigma\av_y=\mathbf b_y$ into \cref{eq_decod_direction}.

\subsection{Geometric and statistical contributions to the channel-transmission}
\label{app_decod_geometric_statistical}

We now split $\tau_R[y]$ [\cref{eq_channel_transmission}] into a geometric and a statistical contribution. Let
\beq
R=U_sD_sV_s^\top
\eeq
be the compact singular value decomposition of $R$, with $s=\rank(R)$. The columns of $V_s\in\reals^{\numF\times s}$ span $\rowsp(R)$, while the columns of $V_0$ span $\nullsp(R)$.
Writing $V=[V_s\,\,V_0]$, we decompose the feature vector and the ideal target direction as
\beq
\zv=V_s^\top\phiv, \qquad
\boldsymbol{\eta}=V_0^\top\phiv, \qquad
\av_\parallel=V_s^\top\av_y, \qquad
\av_\perp=V_0^\top\av_y .
\label{eq_row_null_coordinates}
\eeq
Since $R\phiv=U_sD_s\zv$, the coordinates $\zv$ are geometrically accessible from the readout.
By contrast, the coordinates $\boldsymbol{\eta}$ lie in $\nullsp(R)$ and are not geometrically accessible.

The ideal feature representative decomposes as
\beq
\widehat y_\phi(\uv) = \av_\parallel^\top \zv(\uv) + \underbrace{\av_\perp^\top\boldsymbol{\eta}(\uv)}_{h_y(\uv)} .
\label{eq_target_row_null_decomp}
\eeq
The first term is directly transmitted by $R$.
The second term,
\beq
h_y(\uv)=\av_\perp^\top V_0^\top\phiv(\uv),
\eeq 
is geometrically hidden from the readout.
It may nevertheless be statistically predictable from the accessible coordinates $\zv$.
To quantify this effect, let
\beq
V^\top\Sigma V =
\begin{pmatrix}
\Sigma_{\parallel\parallel} & \Sigma_{\parallel\perp}\\
\Sigma_{\perp\parallel} & \Sigma_{\perp\perp}
\end{pmatrix}.
\label{eq_row_null_covariance}
\eeq
The variance of the hidden component is then
\beq
\mathrm{var}(h_y)= \av_\perp^\top\Sigma_{\perp\perp}\av_\perp ,
\eeq
while the part of $h_y$ linearly predictable from the accessible coordinates $\zv$ has variance
\beq
\mathrm{var}\!\left(\Pi_{\zv}h_y\right) =
\av_\perp^\top \Sigma_{\perp\parallel} \Sigma_{\parallel\parallel}^{+} \Sigma_{\parallel\perp} \av_\perp ,
\eeq
where $\Pi_{\zv}$ denotes the projection onto the span of the rowspace coordinates $\zv$.
This motivates the \emph{statistical rescue} score
\beq
\chi_R[y] \equiv
\begin{cases}
\displaystyle
\frac{\av_\perp^\top\Sigma_{\perp\parallel}\Sigma_{\parallel\parallel}^{+}\Sigma_{\parallel\perp}\av_\perp}
{\av_\perp^\top\Sigma_{\perp\perp}\av_\perp},
&
\av_\perp^\top\Sigma_{\perp\perp}\av_\perp>0,
\\[2.2ex]
0,
&
\av_\perp^\top\Sigma_{\perp\perp}\av_\perp=0,
\end{cases}
\label{eq_statistical_rescue_score}
\eeq
which fulfills $0\leq\chi_R[y]\leq 1$.
A value $\chi_R[y]\simeq 0$ means that the geometrically hidden part of the target-relevant feature representative is also not statistically predictable from the measured rowspace.
By contrast, $\chi_R[y]\simeq 1$ indicates that the hidden component is almost fully recoverable from correlations in the data.

However, $\chi_R[y]\simeq 1$ does not exclude that the decodability character of a task is actually geometric, which is the case if almost no task-relevant feature lies in $\nullsp(R)$.
To quantify how much of a feature lies in geometrically hidden directions, we define the \emph{nullspace load}
\beq
\nu_R[y]\equiv \frac{\av_\perp^\top\Sigma_{\perp\perp}\av_\perp}{\av_y^\top\Sigma\av_y}.
\label{eq_nullspace_load}
\eeq
In general, since $\Sigma_{\parallel\perp}$ need not vanish, $\nu_R[y]$ can exceed unity and should be understood as a nullspace-load diagnostic. By contrast, the product $\nu_R[y](1-\chi_R[y])\in[0,1]$ entering \cref{eq_refined_channel_identity} is a proper fraction.
We now collect the coordinates defined in \cref{eq_row_null_coordinates} over the
training set and obtain coordinate matrices $Z=V_s^\top\Phi$ and $N=V_0^\top\Phi$, so that
\beq
\widehat{\yv}_\phi=Z^\top\av_\parallel+N^\top\av_\perp .
\eeq 
Since $G=R\Phi=U_sD_sZ$, we have $\Pi_G=\Pi_Z$, where $\Pi_Z=Z^\top(ZZ^\top)^+Z$.
The visible component $Z^\top\av_\parallel$ already lies in $\rowsp(Z)$, and therefore the only channel error comes from the part of the hidden component not predictable from $Z$:
\beq
\frac{1}{P}\left\|\widehat{\yv}_\phi-\Pi_G\widehat{\yv}_\phi\right\|_2^2
= \av_\perp^\top
\left(
\Sigma_{\perp\perp}-\Sigma_{\perp\parallel}\Sigma_{\parallel\parallel}^{+}\Sigma_{\parallel\perp}
\right)\av_\perp .
\label{eq_hidden_residual_variance}
\eeq
Dividing by $\|\widehat{\yv}_\phi\|_2^2/P=\av_y^\top\Sigma\av_y$ yields the refined transmission score
\beq
\tau_R[y]=1-\nu_R[y]\bigl(1-\chi_R[y]\bigr).
\label{eq_refined_channel_identity}
\eeq
For a fixed target, full channel transmission thus occurs if the geometrically hidden component is either absent or exactly predictable from the accessible rowspace:
\beq
\tau_R[y]=1
\quad\Longleftrightarrow\quad
(\Imat-\Pi_Z)N^\top\av_\perp=\mathbf 0
\quad\Longleftrightarrow\quad
\nu_R[y]\bigl(1-\chi_R[y]\bigr)=0 .
\eeq
Full transmission for every target is equivalent to
\beq
\nullsp(R)\cap\mathrm{im}(\Sigma)=\{\mathbf{0}\}
\quad\Longleftrightarrow\quad
\rank(R\Sigma^{1/2})=\rank(\Sigma).
\label{eq_decod_full_transmission}
\eeq
Thus full column rank of $R$ is sufficient but not necessary, since it is enough that $R$ be injective on the populated feature subspace.

Upon inserting \cref{eq_refined_channel_identity} into \cref{eq_cap_factorization}, the decodability score takes the form
\beq
\Rcal_R^2[y]=\Rcal_\Imat^2[y]\left[1-\nu_R[y]\bigl(1-\chi_R[y]\bigr)\right].
\label{eq_refined_cap_factorization}
\eeq
Equivalently, the normalized target variance can be decomposed as
\beq
1=\underbrace{1-\Rcal_\Imat^2[y]}_{\text{not represented by encoding}}
+\underbrace{\Rcal_\Imat^2[y]\,\nu_R[y]\bigl(1-\chi_R[y]\bigr)}_{\text{geometrically hidden and not statistically rescued}}
+\underbrace{\Rcal_R^2[y]}_{\text{decoded}} .
\label{eq_refined_decod_budget}
\eeq
The channel-transmission deficit $1-\tau_R[y]$ is the amount of target-relevant feature weight lying outside $\rowsp(R)$ that is \emph{not} statistically rescued by correlations.
The quantity
\beq
\Rcal_\Imat^2[y]\,\nu_R[y]\chi_R[y] \qquad \text{(geometrically hidden but statistically rescued)}
\label{eq_statistically_rescued_load}
\eeq
measures the reduction in normalized mean-squared error relative to predicting with only the directly visible component.

For a nonsingular $\Sigma$, or in an orthonormal basis of its support, write
$\widetilde{\mathbf b}_y/\|\widetilde{\mathbf b}_y\|_2
=\sum_{r\in\Jcal}c_r\ev_r$, with $\sum_r c_r^2=1$. 
The whitened single-direction scores of \cref{eq_decod_score_white} bound the transmission loss according to
\beq
1-\tau_R[y] 
\leq
\left(\sum_{r\in\Jcal}|c_r|\sqrt{1-\widetilde\gamma_r^2}\right)^2 
\leq
\sum_{r\in\Jcal}\bigl(1-\widetilde\gamma_r^2\bigr),
\label{eq_tau_bound}
\eeq
which follows from the triangle inequality applied to $(\Imat-\Pi_{\widetilde R})\sum_r c_r\ev_r$ and Cauchy--Schwarz.

\subsection{Specializations to nonlinear-capacity and monomial scores}

For the degree-$k$ target family $\{y_{\ev}:|\ev|=k\}$ used in \cref{eq_nlcap_R2_score}, the nonlinear processing capacity is simply the average of the general score,
\beq
\Rcal_R^2(k)
=\frac{1}{N_k}\sum_{|\ev|=k}\Dcal_R[y_{\ev}].
\label{eq_decod_degree_average}
\eeq
Applying \cref{eq_cap_factorization} target by target gives the exact degree-resolved factorization
\beq
\Rcal_R^2(k)
=\Rcal_\Imat^2(k)\,\overline\tau_R(k),
\qquad
\overline\tau_R(k)
\equiv
\frac{\sum_{|\ev|=k}\Rcal_\Imat^2[y_{\ev}]\tau_R[y_{\ev}]}
     {\sum_{|\ev|=k}\Rcal_\Imat^2[y_{\ev}]},
\label{eq_cap_profile_fact}
\eeq
whenever the denominator is nonzero. Hence the relative nonlinear-capacity score is an average of channel transmission weighted by the ideal representation capacities.

The local monomial score of \cref{eq_input_dec_score} is another specialization. 
Let $\mathsf B=[\Bcal(\uv_1-\uv_0)-\overline{\mathcal B},\ldots,\Bcal(\uv_P-\uv_0)-\overline{\mathcal B}]\in\reals^{b\times P}$, $\overline{\mathcal B}=\frac{1}{P}\sum_{p=1}^{P}\mathcal B(\mathbf u_p-\mathbf u_0)$, 
$\Sigma_{\Bcal}=P^{-1}\mathsf B\mathsf B^\top$, and use the local approximation
$\Phi\simeq A\mathsf B$. The ideal and post-channel empirical decodability of the $j$-th monomial are respectively
\beq
\Dcal_{A,\Sigma_{\Bcal}}(\ev_j),
\qquad
\Dcal_{RA,\Sigma_{\Bcal}}(\ev_j),
\qquad
\Dcal_{RA,\Sigma_{\Bcal}}(\ev_j)
\leq\Dcal_{A,\Sigma_{\Bcal}}(\ev_j),
\label{eq_decod_monomial_stat}
\eeq
where the inequality follows from
$\rowsp(RA\Sigma_{\Bcal}^{1/2})\subseteq\rowsp(A\Sigma_{\Bcal}^{1/2})$. For isotropic or whitened monomial coordinates, this reduces to the geometric scores
\beq
\kappa_j^2(A)
=(A^+A)_{jj}
=\Dcal_{A,\Imat}(\ev_j),
\qquad
\kappa_j^2(RA)
=((RA)^+RA)_{jj}
=\Dcal_{RA,\Imat}(\ev_j).
\label{eq_decod_monomial_geom}
\eeq
Consequently $\kappa_j^2(RA)\leq\kappa_j^2(A)$, with equality for all monomial directions if
$\nullsp(R)\cap\mathrm{im}(A)=\{\mathbf{0}\}$.

Finally, let $\{\yv_a\}$ be a complete empirically orthonormal basis of the centered sample space,
$P^{-1}\yv_a^\top\yv_b=\delta_{ab}$. Parseval's identity for the projector $\Pi_G$ yields the empirical capacity sum rule
\beq
\sum_a\Dcal_R[y_a]
=\tr(\Pi_G)
=\rank(G)
=\rank(R\Sigma^{1/2}).
\label{eq_cap_sumrule}
\eeq
For whitened feature-coordinate targets this becomes the sum rule in \cref{eq_decod_score_white}; for $\Sigma=\Imat$, it reduces further to $\sum_r\gamma_r^2=\rank(R)$.

\subsection{Summary}

In summary, the row space of the mixing matrix $R$ determines which input components are directly reconstructible, its nullspace determines the unresolved ambiguities, and its singular values determine the stability of the reconstruction. 
Geometric decodability is a property of the mixing matrix alone, while statistical decodability depends additionally on the correlations of the data.
Correlations can reduce the error by predicting target-relevant nullspace components, but exact statistical recovery requires a deterministic relation among the data.
We discuss geometric decodability for some examples of PTMs in \cref{app_exact_feat_iso}.

\begin{figure}[tb!]
  \centering
  \panelimg[width=0.96\textwidth]{a}{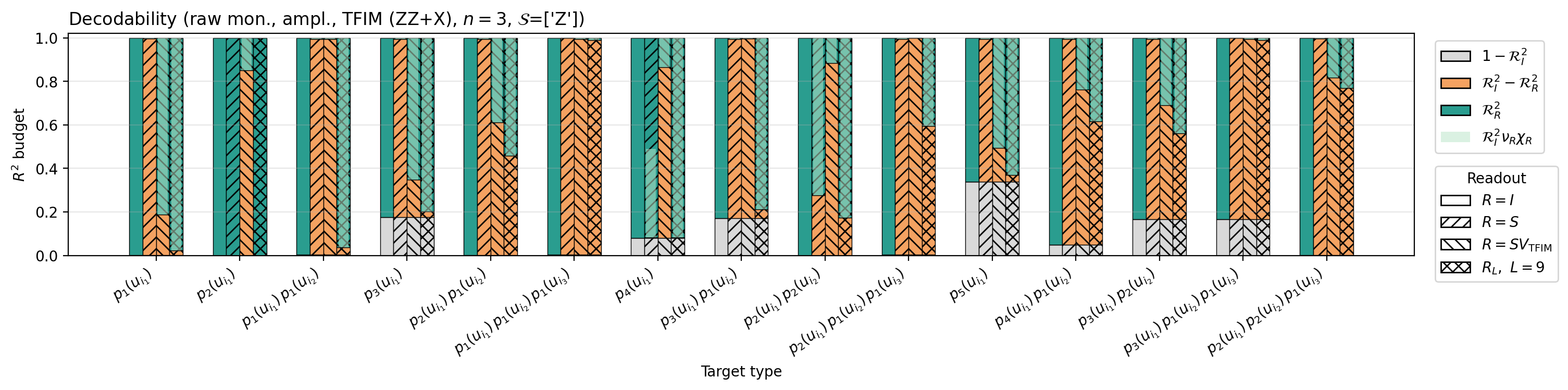}\\
  \panelimg[width=0.96\textwidth]{b}{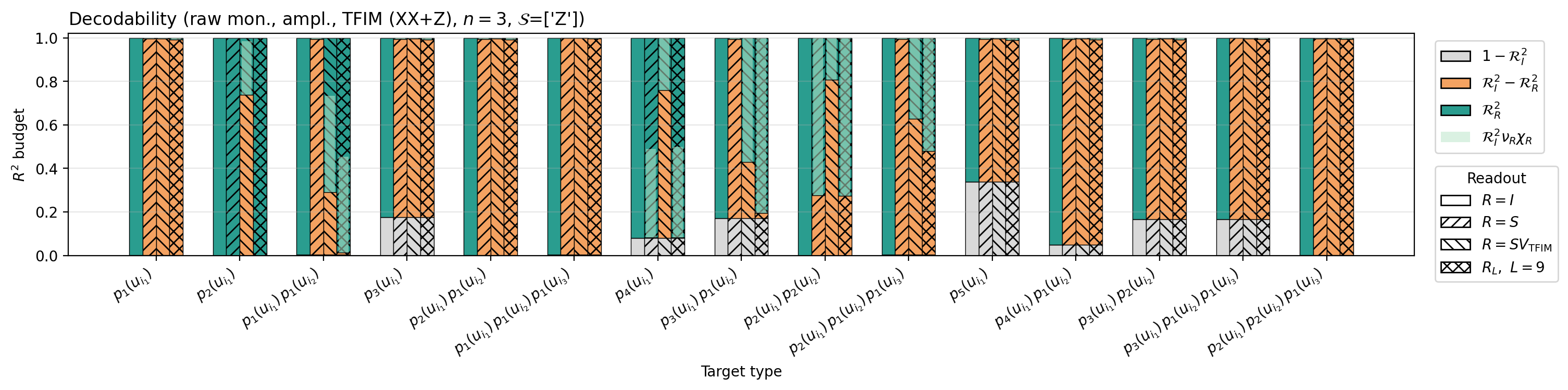}
  \caption{Decodability budget for degree-$j$ raw monomial targets $p_j(u_{\alpha_j})=u_{\alpha_j}^j$ when the selected measurements are restricted to Pauli-$Z$ observables. The bars use the decomposition in \cref{eq_refined_decod_budget} (from bottom to top): gray marks the component absent from the encoding, $1-\Rcal_\Imat^2$; orange marks the encoded component that the reservoir/readout does not recover, $\Rcal_\Imat^2-\Rcal_R^2$; green marks the decoded component, $\Rcal_R^2$, including the part (overlayed) that is geometrically hidden but statistically rescued, $\Rcal_\Imat^2\nu_R\chi_R$ [\cref{eq_statistically_rescued_load}, clipped and scaled]. The four readout choices compare access to the full Pauli feature vector immediately after encoding ($R=\Imat$), direct measurement of the selected $Z$ features only ($R=S$), the same selected features after Hamiltonian evolution ($R=S V$), and their $\numTM$-step temporally multiplexed version ($R_\numTM$). Both panels use amplitude encoding \eqref{eq_enc_amplsq}, but differ by the TFIM Hamiltonian [\cref{eq_H_TFIM}].}
  \label{fig:decod_budget_rawZ}
\end{figure}

\Cref{fig:decod_budget_rawZ} illustrates the decodability score decomposition of \cref{eq_refined_decod_budget} for monomial targets and a TFIM-based PTM with Pauli-$Z$ observables (corresponding to the setup in \cref{fig:tmux_training}).
The reduced geometric decodability of the TFIM variant \eqref{eq_H_TFIM_xxz} compared to \eqref{eq_H_TFIM_zzx} (already visualized in \cref{fig:feat_iso}) is directly reflected by the scores. In fact, for the amplitude encoding \eqref{eq_enc_amplsq}, the temporally multiplexed $Z$-readout satisfies $\Fv(-\uv)=\Fv(\uv)$, because $[P_Z, H\st{TFIM}\ut{xx-z}]=0$ and $\rho\st{enc}(-\uv)=P_Z \rho\st{enc}(\uv)P_Z$ with $P_Z=Z_1\cdots Z_n$. Hence, under a symmetric input distribution on $[-a,a]$ ($a<1$), every odd-degree monomial is orthogonal to the readout space and thus has vanishing capacity, although they are available under full Pauli readout ($R=\Imat$).
This is not necessarily the case for other encodings, such as the variant \eqref{eq_enc_amplsqrt}, for which higher-order monomials have a high statistical decodability.

Note that, while $0\leq \chi_R[y]\leq 1$, the nullspace load $\nu_R[y]$ can exceed 1. Hence, the quantity \eqref{eq_statistically_rescued_load} can in principle exceed $\Rcal_R^2$: it is the error reduction when using correlations relative to using only directly visible components for predictions. We therefore show in the plots (also in \cref{fig:decod_budget}) a clipped and scaled version $\mathrm{min}(\Rcal_\Imat^2 \nu_R \chi_R, \Rcal_R^2)/\Rcal_R^2$.

\section{Geometric decodability for common PTMs} 
\label{app_exact_feat_iso}

\begin{figure}[tbp]
  \centering
  \panelimg[width=0.32\textwidth]{a}{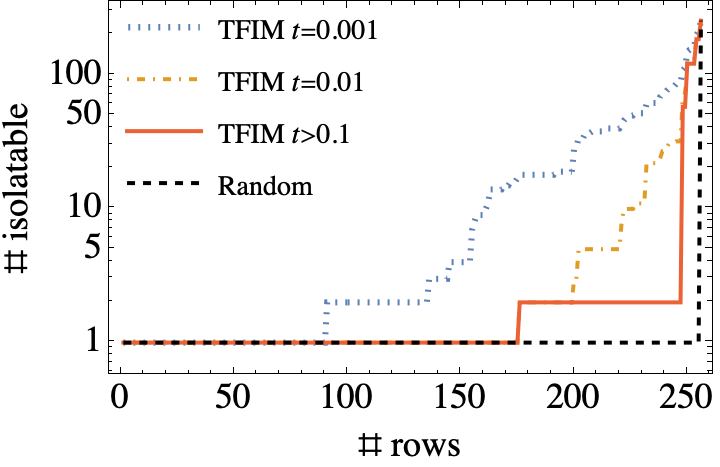}\qquad
  \panelimg[width=0.32\textwidth]{b}{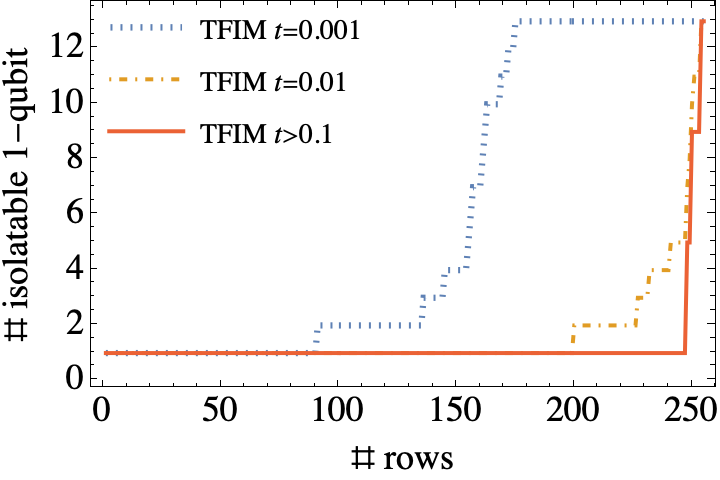}
  \caption{Geometric decodability according to the null-space criterion \eqref{eq_decod_crit_geom}: Number of exactly reconstructible features as a function of the number of observables (=rows of the PTM) $\numM$, for the TFIM with $n=4$ qubits ($d^2=256$) at various evolution times $t$. Panel (a) shows the total number of reconstructible features, including a random-unitary reference, while panel (b) shows the number of reconstructible 1-qubit features. In contrast to statistical decodability, geometric decodability is an exact criterion independent of the data (see \cref{app_ols_decoding}).}
  \label{fig_isocond_TFIM}
\end{figure}

We analyze here geometric decodability criterion \eqref{eq_decod_crit_geom} for common PTMs.
Consider the case where $R\in \reals^{\numM\times \numF}$ is dense and \emph{quasi-random}, arising, e.g., for a Haar-random unitary $U$. 
For $\numM<\numF$, the null space is necessarily nonzero and in random orientation. Accordingly, the probability that the same component vanishes for all vectors of $\nullsp(R)$ is essentially zero, such that no features are geometrically decodable.
Once $\numM\geq \numF$, $R$ has full column rank, hence $\nullsp(R)=\{0\}$ and the null-space condition in \cref{eq_decod_crit_geom} is trivially fulfilled, implying that all $\numF$ features become instantly reconstructible for a random $R$.
In this case one may define $W=R^+ = (R^\top R)^{-1} R^\top$ and assign $\wv^\top = \ev_r^\top W$.

For a \emph{Clifford} unitary, which maps between elements of the Pauli group, $\mathrm{row}(R)$ is spanned by a subset of standard basis vectors, hence $\rank(R)$ features are exactly reconstructible even if $R$ is wide.
For unitaries based on \emph{structured Hamiltonians} or at small evolution times, one may expect \cref{eq_decod_crit_geom} to be fulfilled for a certain set of features, provided $\numM$ is sufficiently large (but can still be $\numM\ll \numF$).
This is confirmed numerically in \cref{fig_isocond_TFIM}, where the behavior of the feature isolation criterion \eqref{eq_decod_crit_geom} is illustrated for various unitaries as a function of the number of observables $\numM$ (rows of $R$).

\section{Pauli transfer matrix formalism: further details}
\label{app_ptm}

We provide here some further useful information on the Pauli transfer matrix (PTM) formalism for QELMs.
When ordering the basis with identity first, we have $\phiv=(\phi_0,\mathbf{s})$, where $\phi_0=\tr(\rho)=1$ and $\mathbf{s}\in\mathbb{R}^{d^2-1}$ is the \emph{traceless (Bloch)} part. 
A maximally mixed state has $\sv=0$, i.e., is in the center of the generalized Bloch sphere. By contrast, a pure state ($\tr(\rho^2)=1$) has $|\sv|^2=d-1$ and thus lies on the hypersphere of radius $\sqrt{d-1}$ (although not all points on it are valid quantum states).

\subsection{Unitary channels}
\label{app_PTM_unitary}

For a unitary channel $\mathcal{U}(\rho)=U\rho U^\dagger$, one has
\beq
T_{U,\alpha\beta}=\frac{1}{2^n}\,\tr \big(P_\alpha\, U P_\beta U^\dagger\big).
\label{eq:PTM_unitary}
\eeq
Trace preservation and unitality imply the form
\beq
\label{eq:block-unitary}
T_U = \begin{pmatrix} 1 & \mathbf{0}^\top \\ \mathbf{0} & \Omega_U \end{pmatrix},
\eeq
i.e., the traceless coefficients undergo a real orthogonal rotation with $\Omega_U\in\mathrm{SO}(d^2-1)$.
For Clifford unitaries, $T_U$ is a signed permutation of Pauli strings.

Consider a \emph{separable} reservoir Hamiltonian $H_R=\sum_{j=1}^n h_R^{(j)}$, such that the reservoir unitary factorizes as $U(t)=\bigotimes_{j=1}^n U_j(t)$. Such a structure emerges, e.g., for a TFIM with $h/J\to \infty$ [\cref{eq_H_TFIM}].
This implies a Kronecker-product structure of the unitary PTM \eqref{eq:PTM_unitary} (and similarly for any other product quantum channel),
\beq
T_U(t)=V^{(1)}(t)\otimes\cdots\otimes V^{(n)}(t),
\label{eq_ptm_sep}
\eeq
where each $V^{(j)}(t)\in\mathrm{SO}(4)$ acts on the single-qubit feature vector $\phiv^{(j)}$.
Consequently, the evolved features remain factorized:
\beq
\phiv'(\xv)=T_U(t)\phiv(\xv)=\bigl(V^{(1)}(t)\phiv^{(1)}(x_1)\bigr)\otimes\cdots\otimes\bigl(V^{(n)}(t)\phiv^{(n)}(x_n)\bigr).
\label{eq_phi_sep_evol}
\eeq
Thus, each Pauli-string feature $\phi_{a_1\ldots a_n}$ evolves only by local mixing among the single-qubit labels $a_j\in\{0,x,y,z\}$, without coupling different qubits and without operator spreading.

Each local unitary $U_k(t)=e^{-\im t h_k}$ rotates the local Pauli components, so $Y$-strings can become nonzero only through this local mixing.
More explicitly, writing the local Pauli feature vector as $\tilde\phiv^{(k)}\equiv(\phi_x^{(k)},\phi_y^{(k)},\phi_z^{(k)})^\top$, any single-qubit unitary induces an $\mathrm{SO}(3)$ rotation
\beq
\tilde\phiv^{'(k)}(u_k)=\Omega^{(k)}(t)\,\tilde\phiv^{(k)}(u_k),\qquad \phi_0^{'(k)}=\phi_0^{(k)}=1,
\eeq
i.e., $\phi_a^{'(k)}=\sum_{b\in\{x,y,z\}} \Omega^{(k)}_{ab}(t)\,\phi_b^{(k)}$, with $\Omega^{(k)}$ the adjoint representation of $U_k(t)$ on the Pauli vector.
For example, for a local $Z$-field $h_k=Z$,
\beq
U_k(t)^\dagger X U_k(t)=\cos(2t)\,X-\sin(2t)\,Y,\qquad
U_k(t)^\dagger Y U_k(t)=\cos(2t)\,Y+\sin(2t)\,X,
\eeq
so that $\phi_y^{'(k)}(u_k)=\sin(2t)\,\phi_x^{(k)}(u_k)$ and, correspondingly, $\phi'_{\ldots y_k\ldots}(\xv)=\sin(2t)\,\phi_{\ldots x_k\ldots}(\xv)$, while $Z$-features remain invariant.
This shows how unitaries act to make specific features accessible to the readout when the latter uses a restricted set of observables.

\subsection{General quantum channels: affine action on the Bloch part}
\label{sec:general-noise}
We discuss now briefly the representation of general noisy channels in the PTM formalism, including some geometrical intuitions \cite{bethruskai_analysis_2002,angrisani_simulating_2025}.
Consider a channel $\Ecal$ that is completely positive and trace-preserving (CPTP) with associated Kraus operators $\{K_\ell\}$. Then
\beq
\label{eq:ptm-kraus}
T_{\Ecal,\alpha\beta}=\frac{1}{2^n}\sum_\ell \tr \big(P_\alpha K_\ell P_\beta K_\ell^\dagger\big),
\eeq
which corresponds in block form (compatible with $\phiv=(1,\mathbf{s})$) to
\beq
\label{eq:affine-form}
T_{\Ecal}=\begin{pmatrix} 1 & \mathbf{0}^\top \\ \mathbf{t} & \Omega \end{pmatrix},
\qquad
\mathbf{s}'=\mathbf{t}+\Omega\mathbf{s},
\eeq
where $\mathbf{t}=\mathbf{0}$ for unital channels. For a single qubit, $\Omega$ is a contracting map (singular values $\leq 1$). 
Non-unitality produces a translation $\mathbf{t}$ of the Bloch part, e.g., a maximally mixed state would be shifted away from the center of the Bloch sphere. 

Consider the following single-qubit PTMs for non-unitary \emph{noise channels} \cite{nielsen_quantum_2010,wilde_quantum_2017,watrous_theory_2018}:
\begin{itemize}
\item Isotropic depolarization $\mathcal{D}_\lambda(\rho)=\lambda \rho+(1-\lambda)\Imat/2$:
\beq
T_{\mathrm{dep}}(\lambda)=\mathrm{diag}(1,\lambda,\lambda,\lambda).
\label{eq_PTM_depol}\eeq
For a global isotropic depolarizing channel on dimension $d=2^n$, $T_{\mathrm{dep}}^{(n)}(\lambda)=\mathrm{diag}\big(1,\lambda \Imat_{d^2-1}\big)$.
\item $Z$-dephasing with probability $p$:
\beq
T_{\mathrm{dephZ}}(p)=\mathrm{diag}(1,1-2p,1-2p,1).
\label{eq_PTM_dephZ}\eeq
\item Amplitude damping with parameter $\gamma\in[0,1]$ and polarization $z_* \in [-1,1]$ (with $z_*=1$ corresponding to the zero-temperature fixed point $|0\rangle\langle 0|$):
\beq
T_{\mathrm{AD}}(\gamma)=
\begin{pmatrix}
1 & 0 & 0 & 0\\
0 & \sqrt{1-\gamma} & 0 & 0\\
0 & 0 & \sqrt{1-\gamma} & 0\\
z_*\gamma & 0 & 0 & 1-\gamma
\end{pmatrix}.
\label{eq_PTM_ampl_damp}\eeq
\end{itemize}
Modeling $n$-qubit noise as tensor products of single-qubit channels $\Ecal^{(n)}=\bigotimes_{i=1}^n \Ecal^{(i)}$ is often used as a baseline in QRC models \cite{kubota_quantum_2022,domingo_taking_2023,monzani_nonunital_2024,martinez-pena_inputdependence_2025}.
Since the PTMs \eqref{eq_PTM_depol} and \eqref{eq_PTM_dephZ} are Pauli-diagonal, they do not lead to Pauli spreading and hence do not enlarge the accessible feature subspace.
Amplitude damping [\cref{eq_PTM_ampl_damp}] produces Pauli-I/Z-mixing, which causes operator weight decay.

A particularly transparent characterization emerges if the noise channel is approximately diagonal. For instance, for the $n$-qubit depolarizing channel $\Dcal^{(n)}=\otimes_{i=1}^n \Dcal_\lambda$ [see \cref{eq_PTM_depol}], any Pauli string $P$ of weight $\nu(P)$ is an eigenoperator with 
\beq \Dcal^{(n)}(P) = \lambda^{\nu(P)} P.
\label{eq_noise_pauli_suppress}\eeq
Accordingly, high-weight Pauli strings are exponentially suppressed by near-diagonal noise channels. In conjunction with unitary-induced operator spreading, this can increase information loss under finite sampling.
On the other hand, amplitude damping noise is often observed to improve QRC performance as the associated superoperator spreads the state’s probability weight into more Pauli directions, thereby enriching the feature map \cite{domingo_taking_2023}. This has been related to the `coherence influx' present for amplitude damping but not depolarization \cite{kobayashi_coherence_2024}.

\section{TFIM symmetries and their impact on temporal multiplexing}
\label{sec_tfim_sector_rank}

We now specialize the general discussion of \cref{sec_op_spread} to the TFIM reservoirs in \cref{eq_H_TFIM} and to the observable families used in the numerics. 
An analytical discussion is possible based on the Jordan--Wigner/Majorana formalism for the Ising chain \cite{prosen_operator_2007,mbeng_quantum_2024}.
Once the Lieb--Robinson lightcone has covered the chain, the support of an initially local operator has saturated. 
The accessible operator subspace controls the achievable rank of the temporally multiplexed PTM.

\subsection{Heisenberg sectors of the relevant observables}
\label{sec_tfim_heisenberg_sectors}

Introducing the Jordan--Wigner Majorana quasiparticle operators
\beq
\gamma_{2j-1}=\Bigl(\prod_{m<j} Z_m\Bigr)X_j,\qquad 
\gamma_{2j}=\Bigl(\prod_{m<j} Z_m\Bigr)Y_j,\qquad j=1,\dots,n,
\label{eq_tfim_majoranas}
\eeq
the TFIM in \cref{eq_H_TFIM_xxz} can be written as
\beq
H\st{TFIM}\ut{xx-z}=
\im J\sum_{j=1}^{n-1}\gamma_{2j}\gamma_{2j+1}+ \im h\sum_{j=1}^{n}\gamma_{2j-1}\gamma_{2j}.
\label{eq_tfim_majorana_h}
\eeq
Since the Hamiltonian is quadratic in the Majoranas, Heisenberg evolution acts linearly on them,
\beq
\gamma_a\ut{(H)}(t)=\sum_{b=1}^{2n} \mathsf{O}_{ab}(t)\,\gamma_b,
\qquad
\mathsf{O}(t)\in \mathrm{SO}(2n),
\label{eq_tfim_majorana_linear}
\eeq
and therefore preserves the \emph{Majorana degree} $r$, i.e., the number of Majorana factors in a monomial. The degree-$r$ operator sector
\beq
\mathcal V_r=\operatorname{span}\{\gamma_{a_1}\cdots \gamma_{a_r}:1\le a_1<\cdots<a_r\le 2n\}
\eeq
with dimension
\beq
\dim \mathcal V_r=\binom{2n}{r},
\label{eq_sector_dim}
\eeq
is thus invariant.

The observables relevant for the QELM fall into simple sectors:
\beq
Z_j=-\im \gamma_{2j-1}\gamma_{2j}\in \mathcal V_2,
\qquad
Z_i Z_j\in \mathcal V_4 \ \ (i\neq j),
\label{eq_tfim_obs_deg_even}
\eeq
while
\beq
X_j=\Bigl[\prod_{m<j}(-\im\gamma_{2m-1}\gamma_{2m})\Bigr]\gamma_{2j-1} \in\mathcal V_{2j-1},
\qquad
Y_j= \Bigl[\prod_{m<j}(-\im\gamma_{2m-1}\gamma_{2m})\Bigr]\gamma_{2j} \in\mathcal V_{2j-1}.
\label{eq_tfim_obs_deg_odd}
\eeq
Thus $Z$-type observables under $H\st{TFIM}\ut{xx-z}$ stay in low-degree Gaussian sectors, whereas $X_j$ and $Y_j$ occupy odd sectors whose degree grows with the site index. In particular, a bulk $X_j$ or $Y_j$ lives in a sector of dimension $\binom{2n}{2j-1} \sim\Ocal(4^n)$.

These sectors have a highly structured Pauli-string content. From the bilinear form of $Z_j$, its Heisenberg evolution under $H\st{TFIM}\ut{xx-z}$ is restricted to
\beq
Z_j\ut{(H)}(t)
=
\sum_{p=1}^{n} a_p(t)\, Z_p
+
\sum_{1\le p<q\le n}\ \sum_{\alpha,\beta\in\{X,Y\}}
b_{pq}^{\alpha\beta}(t)\,
\alpha_p\, Z_{p+1}\cdots Z_{q-1}\, \beta_q ,
\label{eq_z_bilinear_strings}
\eeq
i.e., to strings with two $X/Y$ endpoints and a $Z$-string in between. Hence, after full spatial spreading the number of Pauli strings remains $\sim \mathcal O(n^2)$. Likewise, $(Z_i Z_j)\ut{(H)}(t)$ remains in the quartic sector and becomes a sum of products of two such interval strings. These strings have at most four $X/Y$ endpoints connected by $Z$-segments, and their total number is $\sim \mathcal O(n^4)$.

The second TFIM in \cref{eq_H_TFIM_zzx} follows from the Hadamard equivalence \eqref{eq_Hadamard_equiv}:
\beq
O\ut{(H)}(t)\big|_{\st{zz-x}}
=
U\st{H}^\dagger
\Bigl[(U\st{H} O U\st{H}^\dagger)\ut{(H)}(t)\big|_{\st{xx-z}}\Bigr]
U\st{H}.
\label{eq_tfim_hadamard_orbit}
\eeq
Therefore $Z$-type observables under $H\st{TFIM}\ut{zz-x}$ inherit the same operator-space complexity as $X$-type observables under $H\st{TFIM}\ut{xx-z}$:
\beq
Z_j \ \longleftrightarrow\  X_j\in\mathcal V_{2j-1},
\qquad
Z_i Z_j \ \longleftrightarrow\ X_iX_j\in \mathcal V_{2|i-j|}\ \ (i\neq j).
\label{eq_tfim_hadamard_degrees}
\eeq
Hence $Z_j\ut{(H)}(t)$ under $H\st{TFIM}\ut{zz-x}$ can occupy exponentially large sectors for bulk $j$, and $(Z_iZ_j)\ut{(H)}(t)$ does the same for separations $|i-j|\approx n/2$. 
$Z$-type observables are low-complexity under $H\st{TFIM}\ut{xx-z}$, but high-complexity under $H\st{TFIM}\ut{zz-x}$. By contrast, the full single-site Pauli family $\{X_j,Y_j,Z_j\}_{j=1}^n$ already contains bulk odd-degree operators in either TFIM and therefore has exponential Heisenberg complexity for both variants.

We remark that the Majorana degree $r$ of a Pauli observable is defined only after choosing a Jordan--Wigner/Majorana representation. Here, we use the convention in which the corresponding TFIM Hamiltonian is quadratic in the Majorana operators, so that Heisenberg evolution acts linearly on the Majoranas and preserves the degree-$r$ sectors.

\subsection{Asymptotic rank of the temporally multiplexed PTM}
\label{sec_tfim_asymptotic_rank}

We now connect the above Heisenberg-sector structure to the effective PTM in temporal multiplexing. Denote by $R_\numTM$ the effective PTM in \cref{eq_tmux} evaluated for $\numTM$ multiplexing steps and $\Lcal(\cdot)=\im [H,\cdot]$.
For generic sampling times $\{t_\ell\}$, the large-$\numTM$ rank is the dimension of the Heisenberg orbit, equivalently of the Krylov span \cite{cindrak_engineering_2025}:
\beq\begin{split}
r_\infty(H,\mathcal S) &:=\lim_{\numTM\to\infty}\rank(R_\numTM)
=\dim \operatorname{span}\{e^{t\mathcal{L}}(P):P\in\mathcal S,\ t\in\mathbb R\} \\
&=\dim \operatorname{span}\{\mathcal{L}^m(P):P\in\mathcal S,\ m\ge 0\}.
\label{eq_rank_limit_krylov}
\end{split}\eeq
This is the operator-space saturation rank before feature clipping. The actual effective PTM rank is therefore bounded by
\beq
\rank(R_\numTM)\le \min(\numF,\ r_\infty(H,\mathcal S)),
\label{eq_rank_feature_clip}
\eeq
with $\numF$ the effective number of nonzero encoded features from \cref{sec_classical}.

We consider the three observable families
\beq
\mathcal S_Z=\{Z_j\}_{j=1}^n,
\qquad
\mathcal S_{Z,ZZ}=\{Z_j,\ Z_iZ_j\}_{i,j=1}^n,
\qquad
\mathcal S_{XYZ}=\{X_j,Y_j,Z_j\}_{j=1}^n.
\label{eq_rank_sets}
\eeq
For $\mathcal S_{Z,ZZ}$, ordered pairs do not increase the operator-space dimension since $Z_iZ_j=Z_jZ_i$, while the diagonal $i=j$ contributes only the trivial identity operator.

Using the sector decomposition from \cref{sec_tfim_heisenberg_sectors}, one obtains the following sector-dimension upper bounds for $H\st{TFIM}\ut{xx-z}$:
\begin{align}
r_\infty(H\st{TFIM}\ut{xx-z},\mathcal S_Z)
&\le \binom{2n}{2}
= n(2n-1)
\sim \Ocal(n^2),
\label{eq_rank_xxz_z}\\
r_\infty(H\st{TFIM}\ut{xx-z},\mathcal S_{Z,ZZ})
&\le 1+\binom{2n}{2}+\binom{2n}{4}
\sim \Ocal(n^4).
\label{eq_rank_xxz_zz}
\end{align}
The term $1$ in the second line is the static identity contribution
arising from $Z_iZ_i=\Imat$ and should be omitted if only $i<j$ is
included in $\mathcal S_{Z,ZZ}$.

The sector dimensions are ceilings on the temporal Krylov rank and need not be saturated. In particular, for a fixed quadratic Hamiltonian the rank is further restricted by degeneracies of the Liouvillian frequencies. After Bogoliubov diagonalization, the TFIM can be written as $H=E_0+\sum_{a=1}^n \epsilon_a\left(f_a^\dagger f_a-\frac12\right)$ and every Liouvillian frequency therefore has the form
\beq
\omega_{\mathbf q}=\sum_{a=1}^n q_a\epsilon_a, \qquad q_a\in\{-1,0,1\}.
\label{eq_tfim_liouvillian_frequencies}
\eeq
Consequently, the quadratic TFIM has at most $|\Omega|=3^n$ distinct Liouvillian frequencies. 
With $m_{\mathcal S}:=\dim\operatorname{span}\mathcal S$, the exact frequency-resolved rank formula \eqref{eq_symmetry_hamiltonian_exact_rank} implies
\beq
r_\infty(H,\mathcal S)\le  m_{\mathcal S}\,3^n.
\label{eq_tfim_frequency_rank_bound}
\eeq

An improved bound becomes available when the frequencies are nonresonant and the projections of the seeded operators onto the Liouvillian-frequency subspaces are nonzero. 
Here nonresonance means that distinct vectors $\mathbf q\in\{-1,0,1\}^n$ give distinct sums $\omega_{\mathbf q}=\sum_a q_a\epsilon_a$. 
In a Majorana-degree-$r$ sector, the number of compatible distinct frequencies is
\beq
N_r = \sum_{\substack{0\le s\le \min(r,2n-r)\\
s\equiv r\pmod 2}} \binom ns\,2^s,
\label{eq_tfim_degree_frequency_count}
\eeq
where $s$ is the number of modes with $q_a=\pm1$.
Under the Hadamard equivalence, the observables in $\mathcal S_Z$ for $H\st{TFIM}\ut{zz-x}$ map to $X_j\in\mathcal V_{2j-1}$ for $H\st{TFIM}\ut{xx-z}$. Therefore,
\begin{align}
r_\infty(H\st{TFIM}\ut{zz-x},\mathcal S_Z)
&=\sum_{j=1}^n N_{2j-1} =\sum_{\substack{s=1\\s\ {\rm odd}}}^{n}
\binom ns\,2^s(n-s+1)\nonumber\\
&=\frac{(n+3)3^{n-1}+(n-1)(-1)^n}{2}
=\Theta(n3^n),
\label{eq_rank_zzx_z}
\end{align}
where the notation $\Theta(\cdot)$ indicates a tight bound.
For $H\st{TFIM}\ut{xx-z}$ one has $\Lcal(X_j)=2hY_j$ ($h\neq 0$), so including $Y_j$ does not enlarge the Krylov space already generated by $X_j$. The $Z_j$ observables lie in the separate degree-two sector and it follows that
\begin{equation}
r_\infty(H\st{TFIM}\ut{xx-z},\mathcal S_{XYZ}) =r_\infty(H\st{TFIM}\ut{zz-x},\mathcal S_Z) +r_\infty(H\st{TFIM}\ut{xx-z},\mathcal S_Z)  =\Theta(n3^n).
\label{eq_rank_xxz_xyz}
\end{equation}
Since the full single-site Pauli family is invariant under the global Hadamard transformation up to signs and relabeling,
\beq
r_\infty(H\st{TFIM}\ut{zz-x},\mathcal S_{XYZ})=
r_\infty(H\st{TFIM}\ut{xx-z},\mathcal S_{XYZ})=\Theta(n3^n).
\eeq
Finally, the frequency bound gives
\beq
r_\infty(H\st{TFIM}\ut{zz-x},\mathcal S_Z) \le r_\infty(H\st{TFIM}\ut{zz-x},\mathcal S_{Z,ZZ}) \le \left(1+n+\binom n2\right)3^n \le \Ocal(n^2 3^n).
\label{eq_rank_zzx_zz_bounds}
\eeq

In summary, temporal multiplexing of $Z$-type observables under $H\st{TFIM}\ut{xx-z}$ explores only low-degree Majorana sectors and gives a polynomial rank. Under $H\st{TFIM}\ut{zz-x}$, these observables touch high-degree sectors and the rank is exponential, since the growth is governed by the $3^n$ Liouvillian frequencies. The Majorana-sector dimension characterizes the possible Pauli support of a fully evolved operator. The temporal Krylov rank depends on the multiplicities of the Liouvillian frequencies and can be reduced by resonances or commensurate sampling times.

\section{Influence of general Liouvillian symmetries}
\label{app_sym_kry}

In \cref{sec_tfim_sector_rank} we derived the asymptotic rank of the temporally multiplexed PTM for the TFIM by exploiting a special operator-space sector decomposition. 
We will generalize here these results in terms of symmetries of the Liouvillian $\Lcal$. 
Although we phrase the discussion for unitary evolution generated by a Hamiltonian, the same sector-decomposition arguments apply to any linear Heisenberg generator $\Lcal$.

\subsection{Operator-space formulation of the effective PTM}
\label{app_sym_kry_framework}

Let $\mathcal{H}$ be a $d$-dimensional Hilbert space and let $\mathcal{O}$ be the associated operator space equipped with the Hilbert--Schmidt inner product $\langle A,B \rangle = \frac{1}{d} \mathrm{tr}(A^{\dagger} B)$.
A Hamiltonian $H$ induces the Heisenberg evolution $O\ut{(H)}(t)=e^{t\mathcal{L}} O$ with Liouvillian $\mathcal{L}(O) = \im [H,O]$.
Given a selected observable family $\mathcal{S}$, the ``seed'' space of measurement operators is
\beq
\mathcal{M} = \operatorname{span}\{P_k : k \in \mathcal{S}\} \subseteq \mathcal{O}.
\eeq
Likewise, let $\mathcal{F} \subseteq \mathcal{O}$ denote the feature subspace spanned by those Pauli operators whose input features are not identically zero under the chosen encoding, so that $\dim \mathcal{F} = \numF$ in the notation of \cref{sec_classical}. We write $\Pi_{\mathcal{F}}$ for the orthogonal projector onto $\mathcal{F}$.
For temporal multiplexing at times $t_1,\dots,t_L$, define
\beq
\mathcal{W}_L(H,\mathcal{M};\mathcal{F})=\operatorname{span}\left\{\Pi_{\mathcal{F}} e^{t_{\ell} \mathcal{L}}(M): M \in \mathcal{M},\ \ell=1,\dots,\numTM
\right\}.
\label{eq_symmetry_row_space}
\eeq
In any orthonormal operator basis whose first $\numF$ basis elements span $\mathcal{F}$, the effective PTM $R_\numTM$ in \cref{eq_tmux} is the coordinate matrix of these vectors, hence
\beq
\rank(R_\numTM) = \dim \mathcal{W}_L(H,\mathcal{M};\mathcal{F}).
\label{eq_symmetry_rank_exact}
\eeq
In the absence of feature clipping, one may set $\mathcal{F}=\mathcal{O}$ and obtain the operator-space saturation rank
\beq
r_{\infty}(H,\mathcal{M}):=\lim_{\numTM\to\infty} \rank(R_\numTM)
=
\dim \operatorname{span}\{ e^{t \mathcal{L}}(M) : M \in \mathcal{M},\ t \in \mathbb{R} \}.
\label{eq_symmetry_rank_orbit}
\eeq
For generic sampling times, this coincides with the Krylov dimension [\cref{eq_rank_limit_krylov}],
\beq
r_{\infty}(H,\mathcal{M})= \dim \operatorname{span}\{ \mathcal{L}^m(M) : M \in \mathcal{M},\ m \ge 0 \}.
\label{eq_symmetry_rank_krylov}
\eeq

\subsection{Invariant sectors induced by symmetries}
\label{app_sym_kry_sectors}
Assume a symmetry of the Liouvillian described by the superoperator $\Gamma : \mathcal{O} \to \mathcal{O}$ satisfying
\beq
[\Gamma,\mathcal{L}] = 0.
\label{eq_Liouvillian_symmetry}\eeq
In the physically relevant cases of a unitary or Hermitian $\Gamma$, the spectral theorem yields $\Gamma = \sum_\alpha \lambda_\alpha \Pi_\alpha$, where the $\Pi_\alpha$ are mutually orthogonal spectral projectors. 
These render the orthogonal decomposition
\beq
\mathcal{O} = \bigoplus_{\alpha} \mathcal{O}_{\alpha},
\qquad
\Pi_{\alpha} \mathcal{O} = \mathcal{O}_{\alpha},
\qquad
\Pi_{\alpha} \Pi_{\beta} = \delta_{\alpha\beta} \Pi_{\alpha},
\qquad
\sum_{\alpha} \Pi_{\alpha} = \Imat_{\mathcal{O}} .
\label{eq_symmetry_sector_decomp}
\eeq
Since \cref{eq_Liouvillian_symmetry} implies
\beq
[\Pi_{\alpha},\mathcal{L}] = 0 \qquad \text{for all } \alpha ,
\label{eq_symmetry_sector_commute}
\eeq
every sector is invariant under Heisenberg evolution, i.e.,
\beq
\mathcal{L}(\mathcal{O}_\alpha) \subseteq \mathcal{O}_\alpha,
\qquad e^{t\mathcal{L}} \mathcal{O}_\alpha \subseteq \mathcal{O}_\alpha .
\eeq

When both $\mathcal{M}$ and $\mathcal{F}$ are sector-compatible, i.e. the measured observables and the retained feature basis can each be chosen sector by sector, then
\beq
\mathcal{M} = \bigoplus_{\alpha} \mathcal{M}_{\alpha},
\qquad
\mathcal{F} = \bigoplus_{\alpha} \mathcal{F}_{\alpha},
\label{eq_symmetry_sector_compatible}
\eeq
with $\mathcal{M}_{\alpha}=\Pi_{\alpha} \mathcal{M} \subseteq \mathcal{O}_{\alpha}$ and $\mathcal{F}_{\alpha}=\Pi_{\alpha} \mathcal{F} \subseteq \mathcal{O}_{\alpha}$. In that case the accessible space decomposes orthogonally,
\beq
\mathcal{W}_L(H,\mathcal{M};\mathcal{F})=\bigoplus_{\alpha}\mathcal{W}_{\numTM,\alpha}(H,\mathcal{M}_{\alpha};\mathcal{F}_{\alpha}),
\label{eq_symmetry_rank_sector_sum_space}
\eeq
where
\beq
\mathcal{W}_{\numTM,\alpha}(H,\mathcal{M}_{\alpha};\mathcal{F}_{\alpha})
=\operatorname{span}\left\{\Pi_{\mathcal{F}_{\alpha}} e^{t_{\ell} \mathcal{L}}(M)
:
M \in \mathcal{M}_{\alpha},\ \ell=1,\dots,\numTM\right\}.
\eeq
Hence symmetries limit the accessible operator space by restricting the dynamics to the sectors touched by the measurement subspace.
Consequently,
\beq
\rank(R_\numTM)=\sum_{\alpha} \rank(R_{\numTM,\alpha}),
\label{eq_symmetry_rank_sector_sum}
\eeq
with one effective PTM block $R_{\numTM,\alpha}$ per active sector. Writing
\beq
r_{\infty,\alpha}(H,\mathcal{M}_{\alpha})
=
\dim \operatorname{span}\{ e^{t \mathcal{L}}(M) : M \in \mathcal{M}_{\alpha},\ t \in \mathbb{R} \},
\eeq
one has the sectorwise bound
\beq
\rank(R_{\numTM,\alpha})
\le
\min\bigl(\numTM \dim \mathcal{M}_{\alpha},\ \dim \mathcal{F}_{\alpha},\ r_{\infty,\alpha}(H,\mathcal{M}_{\alpha})\bigr).
\label{eq_symmetry_rank_sector_bound}
\eeq
Since $r_{\infty,\alpha}(H,\mathcal{M}_{\alpha})\leq \dim \mathcal{O}_{\alpha}$, the symmetry-resolved PTM rank bound follows as
\beq
r_{\infty}(H,\mathcal{M};\mathcal{F}):=\lim_{\numTM\to\infty}\dim\mathcal{W}_L(H,\mathcal{M};\mathcal{F})
\le\sum_{\alpha:\mathcal{M}_{\alpha}\neq 0}\min\bigl(\dim\mathcal{F}_{\alpha},r_{\infty,\alpha}(H,\mathcal{M}_{\alpha})\bigr)
\le\sum_{\alpha:\mathcal{M}_{\alpha}\neq 0}\dim\mathcal{F}_{\alpha}.
\label{eq_symmetry_rank_global_bound}
\eeq

If the measured observables are not sector-compatible, the above reasoning yields an upper bound after decomposing them into sector components, but the exact sum rule \eqref{eq_symmetry_rank_sector_sum} can fail because a single observable may contain linear combinations of several sectors.

\subsection{Hamiltonian criteria for underexploration}
\label{app_sym_kry_hamiltonian}

The symmetry decomposition in \cref{eq_symmetry_rank_global_bound} determines only the maximal accessible dimension. Whether a populated sector is fully explored depends on the restricted generator $\mathcal{L}|_{\mathcal{O}_{\alpha}}$ and on the choice of seed subspace $\mathcal{M}_{\alpha}$. 
Let $\mathcal{O}_{\alpha}$ be a symmetry-compatible invariant sector and $\mathcal{L}_{\alpha} := \mathcal{L}|_{\mathcal{O}_{\alpha}}$.
As above, we consider the spectral decomposition
\begin{equation}
\mathcal{L}_{\alpha} = \sum_{\omega \in \Omega_{\alpha}} \im \omega \, \Pi_{\alpha,\omega},
\qquad
\Ocal_{\alpha,\omega} := \Pi_{\alpha,\omega} \mathcal{O}_{\alpha},
\label{eq_symmetry_hamiltonian_gap_decomp}
\end{equation}
where $\Omega_{\alpha} \subset \mathbb{R}$ is the set of Bohr frequencies $\omega$ appearing in sector $\alpha$, with multiplicity $\dim \Ocal_{\alpha,\omega}$.
The Bohr frequencies constitute the set $\{E_n-E_m : E_n, E_m\in \sigma(H)\}$ in terms of the eigenvalues $E_n$ of the Hamiltonian.
We choose now a basis $M_1,\dots,M_{\dim \Mcal_\alpha}$ of $\mathcal{M}_{\alpha}$ and an orthonormal basis $\{F_{\alpha,\omega,s}\}_{s=1}^{\dim \Ocal_{\alpha,\omega}}$ of each eigenspace $\Ocal_{\alpha,\omega}$.
Accordingly, each observable $M_k$ expands as
\beq
\Pi_{\alpha,\omega} M_k
=
\sum_{s=1}^{\dim \Ocal_{\alpha,\omega}} (C_{\alpha,\omega})_{s,k} F_{\alpha,\omega,s},
\label{eq_symmetry_hamiltonian_gap_expansion}
\eeq
with the coefficient matrix $C_{\alpha,\omega} \in \mathbb{C}^{\dim \Ocal_{\alpha,\omega} \times \dim \Mcal_\alpha}$.
Then the asymptotic rank contributed by sector $\alpha$ is exactly
\begin{equation}
r_{\infty,\alpha}(H,\mathcal{M}_{\alpha})
=
\sum_{\omega \in \Omega_{\alpha}} \rank \, C_{\alpha,\omega}.
\label{eq_symmetry_hamiltonian_exact_rank}
\end{equation}
This result shows that underexploration occurs because several operator directions share the same frequency [large $\dim \Ocal_{\alpha,\omega}$ in \cref{eq_symmetry_hamiltonian_gap_expansion}], while the seed subspace supplies too few independent directions inside that degenerate eigenspace (small $\rank\,  C_{\alpha,\omega}$).

A direct consequence of \cref{eq_symmetry_hamiltonian_exact_rank} is the general bound
\begin{equation}
r_{\infty,\alpha}(H,\mathcal{M}_{\alpha})
\le
\sum_{\omega \in \Omega_{\alpha}} \min\bigl(\dim \Ocal_{\alpha,\omega}, \dim \Mcal_\alpha\bigr) .
\label{eq_symmetry_hamiltonian_gap_ceiling}
\end{equation}
For a generic ($\dim \Mcal_\alpha$)-dimensional choice of seed subspace inside $\mathcal{O}_{\alpha}$, one expects equality in \cref{eq_symmetry_hamiltonian_gap_ceiling}. Hence the Hamiltonian alone determines the generic underexploration deficit
\begin{equation}
\dim \Ocal_\alpha - \sum_{\omega \in \Omega_{\alpha}} \min\bigl(\dim \Ocal_{\alpha,\omega}, \dim \Mcal_\alpha\bigr)
=
\sum_{\omega \in \Omega_{\alpha}} \max\bigl(\dim \Ocal_{\alpha,\omega} - \dim \Mcal_\alpha, 0\bigr).
\label{eq_symmetry_hamiltonian_generic_deficit}
\end{equation}
In particular, full sector saturation is generically possible only if the multiplicity of each Bohr frequency does not exceed the dimension of the seed subspace:
\begin{equation}
\dim \Ocal_{\alpha,\omega} \le \dim \Mcal_\alpha
\qquad
\text{for all } \omega \in \Omega_{\alpha}.
\label{eq_symmetry_hamiltonian_saturation_condition}
\end{equation}
The above approach avoids the construction of Krylov spaces and instead relies on the spectral properties of the Hamiltonian or Liouvillian.

It is useful to note that operators diagonal in the energy eigenbasis belong to the $\omega=0$ eigenspace. Hence, temporal multiplexing cannot increase their span and the asymptotic rank equals the dimension of the initially seeded diagonal subspace.

\subsection{Single-observable limit}
We now apply the above formalism to the Krylov grade of a single observable, recovering a key result in \cite{cindrak_engineering_2025}.
We set thus $\mathcal{F}=\mathcal{O}$ and choose a one-dimensional seed space $\mathcal{M}=\operatorname{span}\{O\}$ for a single observable $O$. Then the Krylov grade $M$ is the dimension of the minimal Liouvillian Krylov space generated by $O$ \cite{cindrak_engineering_2025},
\beq
M := \dim \operatorname{span}\{O,\mathcal{L}(O),\mathcal{L}^2(O),\ldots\}
= r_\infty(H,\mathcal{M}).
\label{eq_symmetry_hamiltonian_krylov_grade}\eeq
Suppressing the symmetry label $\alpha$ for simplicity, write the Liouvillian spectral decomposition as
\beq
\mathcal{L}=\sum_{\omega\in\Omega} \im \omega\,\Pi_\omega,
\qquad
\Omega=\{\omega_P\}_{P=0}^{N_\omega-1},
\eeq
where $\omega_P=E_m-E_n$ with $E_m,E_n\in\sigma(H)$ are the pairwise distinct Bohr frequencies of $H$, so that $N_\omega:=|\Omega|$ is the number of distinct transition frequencies. For the single seed $O$, define
\beq
\sigma_P:=\Pi_{\omega_P}O,
\eeq
so that $O(t)=\sum_{P=0}^{N_\omega-1} e^{\im \omega_P t}\sigma_P$. Let
\beq
N_1:=\bigl|\{P\in\{0,\dots,N_\omega-1\}:\sigma_P=0\}\bigr|
\eeq
be the number of vanishing frequency blocks. Since $\dim\mathcal{M}=1$, each coefficient matrix $C_{\omega_P}$ in \cref{eq_symmetry_hamiltonian_gap_expansion} is a single column, and therefore
\beq
\operatorname{rank} C_{\omega_P}
=
\begin{cases}
1, & \sigma_P=\Pi_{\omega_P}O\neq 0,\\
0, & \sigma_P=\Pi_{\omega_P}O=0.
\end{cases}
\eeq
Hence \cref{eq_symmetry_hamiltonian_exact_rank,eq_symmetry_hamiltonian_krylov_grade} give
\beq
M=r_\infty(H,\mathcal{M})=\sum_{P=0}^{N_\omega-1}\operatorname{rank} C_{\omega_P}
=
\bigl|\{P:\sigma_P\neq 0\}\bigr|=N_\omega-N_1,
\eeq
recovering the corresponding theorem in \cite{cindrak_engineering_2025} as the single observable specialization of the present formalism.

\subsection{Examples of symmetry-induced rank bounds}

\subsubsection{Hilbert-space symmetries}

Consider an ordinary Hilbert-space symmetry with $U H U^\dagger = H$. Then the superoperator
\beq
\Gamma_U(O) = U O U^\dagger
\eeq
is unitary on $\mathcal{O}$ and satisfies $[\Gamma_U,\mathcal{L}] = 0$. 
Similarly, for a $\mathbb{Z}_2$ symmetry with $U^2 = \Imat$, the corresponding spectral projectors are
\beq
\Pi_\pm = \frac{1}{2} (\Imat_{\mathcal{O}} \pm \Gamma_U) .
\eeq

If $Q = \sum_q q P_q$ is a conserved charge with $[H,Q]=0$, then
\beq
\Pi_{q q'}(O) = P_q O P_{q'}
\eeq
defines orthogonal projectors commuting with $\mathcal{L}$, which yields the block decomposition
\beq
\mathcal{O} = \bigoplus_{q,q'} P_q \mathcal{O} P_{q'} .
\eeq
If the measured observables commute with $Q$, only the diagonal sectors $\mathcal{O}_{q q}$ can be seeded, and therefore ($d_q=\rank P_q$)
\beq
r_{\infty}(H,\mathcal{M}) \le \sum_q d_q^2.
\label{eq_symmetry_charge_bound}
\eeq
Thus a symmetry reduces the PTM rank because temporal multiplexing can only enlarge the Krylov span inside the sectors allowed by the symmetry and by the measured observables.

\subsubsection{Majorana symmetries of the TFIM}
\label{app_sym_kry_tfim}

The TFIM analysis in \cref{sec_tfim_sector_rank} is a direct instance of the above framework, but with a Liouville-space symmetry that is stronger than a conventional conserved charge. For the quadratic Majorana form of $H_{\mathrm{TFIM}}\ut{xx-z}$ in \cref{eq_tfim_majorana_h}, the operator space decomposes as
\beq
\mathcal{O} = \bigoplus_{r=0}^{2n} \mathcal{V}_r,
\qquad
\dim \mathcal{V}_r = \binom{2n}{r},
\label{eq_symmetry_tfim_majorana_decomp}
\eeq
which is the sector structure used in \cref{sec_tfim_heisenberg_sectors}.
Since the usual observable families are sector-compatible,
\beq
Z_j \in \mathcal{V}_2,
\qquad
Z_i Z_{j\neq i} \in \mathcal{V}_4,
\qquad
X_j,Y_j \in \mathcal{V}_{2j-1},
\label{eq_symmetry_tfim_observable_sectors}
\eeq
the general bound \eqref{eq_symmetry_rank_global_bound} reproduces the polynomial versus exponential behavior found in \cref{sec_tfim_asymptotic_rank}. In particular, for $H_{\mathrm{TFIM}}\ut{xx-z}$, $Z$-type observables seed only the low-degree sectors $\mathcal{V}_2$ and $\mathcal{V}_4$, which yields polynomial asymptotic ranks, while inclusion of $X_j$ or $Y_j$ activates odd sectors whose dimensions are already exponential for bulk sites. By contrast, under the Hadamard-related Hamiltonian $H_{\mathrm{TFIM}}\ut{zz-x}$, the same $Z$-type observables are mapped to the high-degree operators discussed around \cref{eq_tfim_hadamard_orbit,eq_tfim_hadamard_degrees}. 

\subsection{Consequences for PTM analysis and feature decodability}
\label{app_sym_kry_decodability}

In a symmetry-adapted orthonormal operator basis, the unitary PTM becomes block diagonal,
\beq
V(t) = \bigoplus_{\alpha} V_{\alpha}(t).
\label{eq_symmetry_block_ptm}
\eeq
Under the sector-compatibility assumptions in \cref{eq_symmetry_sector_compatible}, the temporally multiplexed effective PTM \eqref{eq_tmux} can be brought into a block form
\beq
R_\numTM \sim \bigoplus_{\alpha} R_{\numTM,\alpha},
\label{eq_symmetry_block_eff_ptm}
\eeq
and the projector defined in \cref{eq_decod_score} decomposes as
\beq
R_\numTM^{+} R_\numTM \sim \bigoplus_{\alpha} \bigl(R_{\numTM,\alpha}^{+} R_{\numTM,\alpha}\bigr).
\label{eq_symmetry_block_projector}
\eeq
Thus, if a feature lies in a sector with $\mathcal{M}_{\alpha}=0$, then it is symmetry-forbidden and cannot be (geometrically) decoded by the chosen observable family, i.e. $\gamma_r^2 = (R_\numTM^{+} R_\numTM)_{rr} = 0$.
Note that, in the raw Pauli basis, symmetry-induced restrictions need not appear as perfectly sharp blocks, because the symmetry-adapted basis is generally different from the Pauli basis.

\bibliography{bibliography_cited}

\end{document}